\documentclass[aps,prd,reprint,superscriptaddress,longbibliography,floatfix]{revtex4-2}
\usepackage{amsmath,amssymb,amsthm,mathtools}
\usepackage{booktabs,array,graphicx}
\newcommand{\dd}{\mathrm{d}}
\newcommand{\rankop}{\operatorname{rank}}
\theoremstyle{definition}
\newtheorem{stabilitytheorem}{Theorem}
\newtheorem{stabilitycorollary}{Corollary}
\usepackage{microtype}
\usepackage[unicode,colorlinks=true,linkcolor=blue,citecolor=blue,urlcolor=blue]{hyperref}
\hypersetup{pdftitle={Equivalence of Radial Stability Criteria for Multifluid Stars},pdfauthor={Tian-Shun Chen, Xiao-Ding Zhou, Hao Feng, Kilar Zhang}}
\usepackage{refcount}

\theoremstyle{plain}
\newtheorem{theorem}{Theorem}[section]
\newtheorem{lemma}[theorem]{Lemma}
\newtheorem{proposition}[theorem]{Proposition}
\newcounter{proofsublemma}
\newcommand{\setproofparent}[1]{%
  \edef\proofparentnumber{\getrefnumber{#1}}%
  \gdef\proofparentanchor{#1}%
  \setcounter{proofsublemma}{0}}
\newcommand{\continueproofparent}[1]{%
  \edef\proofparentnumber{\getrefnumber{#1}}%
  \gdef\proofparentanchor{#1}}

\newtheorem{sublemma}[proofsublemma]{Lemma}
\newtheorem{subproposition}[proofsublemma]{Proposition}
\theoremstyle{definition}
\newtheorem{assumption}[theorem]{Assumption}
\newtheorem{definition}[theorem]{Definition}
\newtheorem*{nameddefinition}{Definition}
\theoremstyle{remark}
\newtheorem{remark}[theorem]{Remark}
\newtheorem{subremark}[proofsublemma]{Remark}

\newcommand{\conditiontarget}[1]{\phantomsection\label{en:condition-#1}}
\newcommand{\conditionref}[1]{%
  \hyperref[en:condition-#1]{\textup{#1}}}
\newcommand{\conditionrefp}[1]{%
  \hyperref[en:condition-#1]{\textup{(#1)}}}
\newcommand{\statementpartref}[2]{%
  \ref{#1}\textup{(\ref{#2})}}
\newcommand{\statementpartrange}[3]{%
  \ref{#1}\textup{(\ref{#2})--(\ref{#3})}}
\usepackage{etoolbox}

\makeatletter
\newwrite\SupplementAuxStream
\newcommand{\ReadSupplementAux}{%
  \begingroup
  \def\@extra@binfo{.supp}%
  \@input{\jobname-supp.aux}%
  \endgroup}
\newcommand{\StartSupplementAux}{%
  \if@filesw
    \immediate\write\@mainaux{\string\ReadSupplementAux}%
    \immediate\openout\SupplementAuxStream=\jobname-supp.aux\relax
    \let\@auxout\SupplementAuxStream
    \immediate\write\@auxout{\string\bibstyle{apsrev4-2}}%
  \fi
  \def\@extra@b@citeb{.supp}}
\newcommand{\PrintSupplementBibliography}{%
  \if@filesw
    \immediate\write\@auxout{\string\citation{REVTEX42Control}}%
    \immediate\write\@auxout{\string\citation{apsrev42Control}}%
    \immediate\write\@auxout{\string\bibdata{\jobname Notes,CZFZ}}%
  \fi
  \begingroup
    \let\@FMN@list\@empty
    \let\SavedSupplementLabel\label
    \def\label##1{\def\TmpLabel{##1}\def\LastLabel{LastBibItem}%
      \ifx\TmpLabel\LastLabel\SavedSupplementLabel{LastSupplementBibItem}\else\SavedSupplementLabel{##1}\fi}%
    \@input@{\jobname-supp.bbl}%
  \endgroup
  \clearpage
  \immediate\closeout\SupplementAuxStream
  \let\@auxout\@mainaux
  \let\@extra@b@citeb\@empty}
\makeatother

\begin{document}
\let\SavedEmail\email
\let\SavedAffiliation\affiliation
\title{Equivalence of Radial Stability Criteria for Multifluid Stars}
\newcommand{\Tdli}{Tsung-Dao Lee Institute, Shanghai Jiao-Tong University, Shanghai 201210, China}
\newcommand{\Shu}{Department of Physics and Institute for Quantum Science and Technology, Shanghai University, Shanghai 200444, China}
\newcommand{\Fudan}{Department of Physics, Fudan University, Shanghai 200438, China}
\newcommand{\KIts}{Kavli Institute for Theoretical Sciences (KITS) and School of Quantum, University of the Chinese Academy of Sciences, Beijing 100190, China}
\newcommand{\Sklt}{Shanghai Key Laboratory of High Temperature Superconductors, Shanghai 200444, China}
\newcommand{\manuscriptauthors}{%
\author{Tian-Shun Chen}
\email{DeltaChen@sjtu.edu.cn}
\affiliation{\Tdli}
\affiliation{\Shu}
\author{Xiao-Ding Zhou}
\email{xdzhou26@m.fudan.edu.cn}
\affiliation{\Fudan}
\affiliation{\Shu}
\author{Hao Feng}
\email{fenghaozi@shu.edu.cn}
\affiliation{\KIts}
\affiliation{\Shu}
\author{Kilar Zhang}
\email[Corresponding author: ]{kilar@shu.edu.cn}
\affiliation{\Shu}
\affiliation{\Sklt}
}
\manuscriptauthors
\begin{abstract}
Neutron stars admixed with dark matter can probe the gravitational effects of a dark component through their masses, radii and tidal responses. Using these observables requires identifying which hydrostatic configurations are radially stable. We prove a static criterion for physical radial zero modes of any finite number of separately conserved, isentropic fluids coupled only through gravity. On a regular family of spherical equilibria, variations that preserve all particle numbers to first order are in one-to-one correspondence with zero-frequency radial displacements. Continuity of the ordered squared frequencies implies that every boundary point of strict radial stability satisfies the static criterion. For a common nuclear equation of state, we compare the static and pulsation boundaries in two-fluid stars containing one dark component and in three-fluid stars containing both bosonic and fermionic dark matter. We project the stable two-fluid models onto the mass--radius and mass--tidal deformability planes, which probe the emitting surface and the distribution of matter through the entire star. By proving that the particle number Jacobian loses rank if and only if a physical radial zero mode exists, we establish a rigorous basis for using the static criterion to trace the radially stable branch.
\end{abstract}
\maketitle

\section{Introduction}\label{sec:intro}

The tidal signal from the binary neutron star merger GW170817~\cite{Abbott2017,Abbott2018} and X-ray pulse profile measurements of PSR J0030+0451, PSR J0740+6620, PSR J0437--4715 and PSR J0614--3329~\cite{Riley2019,Miller2019,Riley2021,Miller2021,Choudhury2024,Mauviard2025} probe complementary properties of relativistic stars. In a star admixed with dark matter, the gravitational mass includes both constituents, whereas surface emission originates in the nuclear component. Dark matter concentrated in a core can alter the pressure support of the nuclear fluid. A dark halo extending beyond the emitting surface also modifies photon propagation and contributes to the tidal response~\cite{Nelson2019,Miao2022,Leung2022Tidal}.

Dark matter capture and accumulation depend on the stellar formation history and on the properties of the particles~\cite{GoldmanNussinov1989,Kouvaris2008}. Two-fluid models of nuclear matter mixed with one dark constituent admit compact dark cores and extended halos~\cite{Leung2011,Leung2012,Kain2021,Karkevandi2022}. Recent work has examined how dark matter modifies the relation between the visible radius and nuclear pressure~\cite{LiJiangSun2026}, limits the maximum mass of a quark star with a compact core of heavy, non-self-interacting dark fermions~\cite{SahooBhuyan2026}, and affects mass--radius sequences in rotating fermion--boson stars in \(R+\alpha R^2\) gravity~\cite{Fakhry2026}.

Three-fluid models motivated by dark matter inherited from the progenitor and dark matter captured later admit equilibria in which one dark component lies within the nuclear surface and the other extends beyond it~\cite{Rezaei2018}. Particle models with both a stable scalar and a stable fermion provide a microscopic motivation for considering two dark constituents~\cite{YagunaZapata2022}. When particle conversion and momentum exchange are negligible on the oscillation timescale, each constituent can undergo its own radial displacement. We study this limit for relativistic perfect fluids coupled only through the common metric.

These effects make neutron stars potential laboratories for dark matter, on condition that the stellar configurations used for comparison with observations are dynamically viable. The classical variational principle for stellar oscillations~\cite{chandrasekhar_general_1964} has a relativistic radial counterpart: for a single barotropic star the sign of the lowest squared radial frequency determines stability~\cite{Chandrasekhar1964}. Along a regular one-parameter sequence connected to stable low-density stars, the first mass turning point provides a familiar criterion for marginal stability~\cite{HarrisonThorneWakanoWheeler1965,BardeenThorneMeltzer1966,HadzicLin2021}. With two independent central pressures, a mass extremum along a chosen sequence is more difficult to interpret. In particular, varying the nuclear central pressure at fixed dark central pressure generally changes both particle numbers. A radial perturbation of an isolated star must preserve each of them. This distinction is important when the same equilibrium survey is used to infer a maximum mass or an allowed range of radii.

A static critical curve criterion was introduced for boson--fermion equilibria~\cite{Henriques1990}. In gravitationally coupled two-fluid models the corresponding critical curves have been compared with the zeros of the fundamental radial mode~\cite{Kain2021,ZhouChenWuZhang2025}. Coupled pulsation equations have been derived for an arbitrary finite number of fluids and solved for a three-fluid equilibrium~\cite{Kain2020}. Direct evolutions, mode calculations and variational methods provide complementary tests of radial stability~\cite{GleasonBrownKain2022,CaballeroRipleyYunes2024,KumarSotani2025}. A separate equilibrium calculation locates mass extrema along sequences of fixed dark particle number~\cite{ScordinoBombaci2025}. General turning point arguments relate constrained extrema to changes in stability under specified thermodynamic and equilibrium family assumptions~\cite{Sorkin1981,sorkin_stability_1982,SchiffrinWald2014}. For a multifluid star, the remaining issue is to establish which regular dynamical perturbations are represented by the static condition, including cases where several neutral displacements coexist.

To answer this question, we consider a regular equilibrium family with any finite number of fluid components. We prove that variations preserving every particle number to first order correspond one to one to physical radial zero modes. The particle number Jacobian loses rank precisely when the star admits a nontrivial zero-frequency radial displacement, and its nullity counts the independent zero modes (Theorem~\ref{thm:neutral}). Every boundary point of strict radial stability satisfies this rank condition. Near a simple transverse zero of the fundamental mode, the condition locates the stability boundary (Corollary~\ref{cor:stable}). We apply it to neutron--dark matter mixtures to identify the stable stars whose radii and tidal responses can be compared with observations.

Section~\ref{sec:family} defines the equilibrium family and its conserved particle numbers. Section~\ref{sec:proof} constructs the correspondence by following material shells, and Sec.~\ref{sec:boundary} relates the radial spectrum to the local stability boundary. Section~\ref{en:sec-numerical} first compares the two-fluid stability boundaries and presents the mass--radius and tidal projections of the stable stars. It then compares the static and pulsation results for the critical surface of a three-fluid star containing both dark species. Section~\ref{sec:discussion} discusses the observational implications. The \hyperref[app:radial-numerics]{Appendix} describes the radial pulsation calculation. The Supplemental Material~\cite{SupplementalMaterial} gives detailed proofs of the radial zero mode criterion and its relation to the local stability boundary.

\section{Equilibrium configurations and conserved particle numbers}\label{sec:family}\label{sec:scope}

Consider a static spherical star in general relativity. Label its components by \(I=1,\ldots,K\), with \(K\) finite, and write \(n_I,\epsilon_I,p_I\) for the rest frame number density, energy density and pressure. We use \(G=c=1\) except when specifying numerical units. Each component is an isentropic perfect fluid with its own conservation equations,
\begin{equation}
\begin{gathered}
 \nabla_\mu(n_I u_I^\mu)=0,\qquad \nabla_\mu T_I^{\mu\nu}=0,\\
 T_I^{\mu\nu}=(\epsilon_I+p_I)u_I^\mu u_I^\nu+p_I g^{\mu\nu}.
 \end{gathered}
 \label{eq:currents}
\end{equation}
All four-velocities are aligned with the static time direction in equilibrium. Under a perturbation their radial components may differ. The common metric responds to the sum of the stress tensors, so the motion of one species can affect another without transferring a single particle between them.

In areal radius \(r\) let
\begin{equation}
\begin{aligned}
 \dd s^2&=-e^{2\Phi}\dd t^2+e^{2\Lambda}\dd r^2+r^2\dd\Omega^2,\\
 A&=e^{-2\Lambda}=1-2m/r .
 \end{aligned}
 \label{eq:metric}
\end{equation}
Here \(m(r)\) is the enclosed gravitational mass. Writing \(h_I=\epsilon_I+p_I\) and \(p=\sum_I p_I\), the coupled Tolman--Oppenheimer--Volkoff equations are~\cite{Tolman1939,OppenheimerVolkoff1939,Kain2020}
\begin{equation}
\begin{gathered}
 m'=4\pi r^2\sum_I\epsilon_I,\qquad p_I'=-h_I\Phi',\\
 \Phi'=\frac{m+4\pi r^3p}{r^2 A}.
 \end{gathered}
 \label{eq:tov}
\end{equation}
A prime denotes differentiation with respect to \(r\). Each component obeys the same potential gradient up to its free surface at radius \(R_I\). While any other species remains, its pressure and the metric continue through \(R_I\). The star meets the Schwarzschild exterior only at \(R_{\rm out}=\max_I R_I\). We normalize time at infinity, so the mass measured there is \(M=m(R_{\rm out})\).

It is helpful to specify the central conditions by the dimensionless enthalpies
\begin{equation}
\begin{aligned}
 H_I(p_I)&=\int_0^{p_I}\frac{\dd\tilde p}{\epsilon_I(\tilde p)+\tilde p},\\
 H_I(r)&=H_I^c-[\Phi(r)-\Phi(0)]\quad (r<R_I).
 \end{aligned}
 \label{eq:H}
\end{equation}
The second relation follows at once from hydrostatic balance. Starting at the center and moving outwards, every active fluid loses the same amount of enthalpy. Its central value determines at which potential difference it reaches vacuum. Two species may therefore have different radii even though they experience exactly the same gravitational field. If their surfaces coincide, their constitutive laws and particle currents still retain their individual labels.

For fixed equations of state, the central enthalpies \(\sigma=(H_1^c,\ldots,H_K^c)\) specify a local family of models. Equivalently one can use the positive central pressures, since \(\dd H_I/\dd p_I=1/h_I>0\) there. Two stars with the same total central pressure can allocate it differently between species and are in general different equilibria. The full family therefore has \(K\) independent central data.

The number inside areal radius \(r\) and its total value at the surface are
\begin{equation}
\begin{gathered}
 g_I(r)=4\pi r^2e^\Lambda n_I(r),\\
 {\cal N}_I(r)=\int_0^r g_I(s)\,\dd s,\qquad N_I={\cal N}_I(R_I).
 \end{gathered}
 \label{eq:N}
\end{equation}
Outside component \(I\) set \(g_I=0\).

For later use, the cold equation of state specifies the chemical potential \(\mu_I=\dd\epsilon_I/\dd n_I\), with
\begin{equation}
\begin{gathered}
 h_I=n_I\mu_I,\qquad \dd p_I=n_I\,\dd\mu_I,\\
 \mu_I(r)e^{\Phi(r)}=\mu_I^\infty.
 \end{gathered}
 \label{eq:mu}
\end{equation}
The last equality follows from Eq.~\eqref{eq:H}; the constant \(\mu_I^\infty\) is the redshifted chemical potential of species \(I\) on this equilibrium, as in the Klein equilibrium condition~\cite{Klein1949}.

We assume regular, asymptotically flat stellar equilibria whose inverse radial metric coefficient \(A(r)=g^{rr}=1-2m(r)/r\) is positive throughout the star. Each fluid has positive central density and a finite free surface where \(n_I,\epsilon_I,p_I\) vanish continuously. Each internal interface is the free surface of one or more constituents. The bulk fields match there without a surface layer or independent interface degrees of freedom.

In the positive-density interior we take \(\epsilon_I(n_I)\) to be locally \(C^3\), with \(\mu_I>0\) and a sound speed \(0<c_{s,I}^2=\dd p_I/\dd\epsilon_I<1\). Near vacuum we assume a controlled polytropic law. The chemical potential \(\mu_I\) tends to a positive finite limit and
\begin{equation}
\begin{aligned}
 n_I&=C_I H_I^{\alpha_I}(1+q_I),&&\alpha_I,C_I>0,\\
 (H_I\partial_{H_I})^k q_I&=O(H_I^{\eta_I}),&&\eta_I>0 .
 \end{aligned}
 \label{eq:surfaceEOS}
\end{equation}
Here \(H_I=\log(\mu_I/\mu_{I,0})\) vanishes at vacuum, \(1+q_I>0\), and the smooth interior remainder obeys the stated bound for nonnegative integer \(k\). The estimate allows a fractional power of the depth below the surface. In physical terms, the outermost shells contain progressively fewer particles. This fact makes it possible to compare stars with slightly different radii, but it also makes the surface a singular point of the radial wave equation. Both features will enter the proof. Global statements about configurations where a component disappears at the center would require extending the analysis beyond this local family. Assumption~\ref{en:ass-regular} and Lemma~\ref{en:lem-equilibrium-package} of the Supplemental Material~\cite{SupplementalMaterial} give the hypotheses and establish differentiability of the labeled radii and particle numbers, including at coincident surfaces.
\nocite{ShapiroTeukolsky1983,BrezisFA,WalterODE,WaldGR,seifert_general_2007,everitt_sturmliouville_1986,niessen_singular_1992,kaper_characterizations_1986,Kato,ReedSimonIV,kong_dependence_1996,zhu_dependence_2017,Mosco1969,KuwaeShioya2003}

For \(K\) fluids, compute the particle number map \(N(\sigma)=(N_1,\ldots,N_K)\) on the equilibrium family and its Jacobian
\begin{equation}
 D_{Ia}(\sigma)=\frac{\partial N_I}{\partial \sigma^a},\qquad
 \delta_v N_I=\sum_{a=1}^{K}D_{Ia}v^a .
 \label{eq:D}
\end{equation}
Each column measures the response of all particle numbers to one central enthalpy. The common gravitational potential produces off-diagonal terms even in the absence of direct interactions. {\bfseries\boldmath A direction \(v\ne0\) preserves the particle number of every component to first order if and only if \(Dv=0\). The corresponding equilibrium is statically critical.}

\section{Equivalence of the static and dynamical criteria}\label{sec:proof}\label{sec:criterion}

\subsection{Particle-preserving equilibrium variations}

We ask which first-order changes between neighboring equilibria can arise solely from redistributing the existing particles of each species, and show that these changes are in one-to-one correspondence with the physical radial zero modes.

The mass and particle number variations can be related directly by comparing neighboring equilibria at fixed equations of state and fixed areal radius \(r\). In the following identity the redshifted chemical potentials are those of the unperturbed star. With \(\delta_E\) denoting variation at fixed radius, one has
\begin{equation}
 \sum_{I=1}^K \mu_I^\infty\,\delta_E{\cal N}_I(r)
      =e^{\Phi(r)+\Lambda(r)}\,\delta m(r).
 \label{eq:enclosedlaw}
\end{equation}
Indeed, the derivative of the left side, by Eqs.~\eqref{eq:N} and \eqref{eq:mu}, is
\begin{align}
 \sum_I \mu_I^\infty\,\delta g_I
 &=4\pi r^2 e^{\Phi+\Lambda}
       \left(\sum_I\delta\epsilon_I+\sum_I h_I\,\delta\Lambda\right) \nonumber\\
 &=e^{\Phi+\Lambda}\big[\delta m'
              +(\Phi'+\Lambda')\,\delta m\big].
 \label{eq:firstlawderivative}
\end{align}
Here \(\delta\Lambda=\delta m/(rA)\) from the metric and \(\Phi'+\Lambda'=4\pi r\sum_I h_I/A\) from hydrostatic balance. Both sides of Eq.~\eqref{eq:enclosedlaw} vanish at the center. At a component surface \(n_I\) and \(\epsilon_I\) vanish continuously, so moving that surface adds no sheet of particles or mass to the first variation. In the Schwarzschild exterior \(e^{\Phi+\Lambda}=1\), and the identity becomes the equilibrium first law
\begin{equation}
 \delta M=\sum_{I=1}^K\mu_I^\infty\,\delta N_I .
 \label{eq:firstlaw}
\end{equation}
The global result agrees with the static, uncharged limit of the multiconstituent mass variation formula [Eqs.~(56) and (57) of Ref.~\cite{Carter1979Variation}]. Section~\ref{en:sec-first-law} of the Supplemental Material~\cite{SupplementalMaterial} derives the multifluid identity across the free surfaces in Eqs.~(\ref{en:eq-enclosed-first-law-derivative})--(\ref{en:eq-global-first-law}) and takes its exterior limit to obtain Eq.~\eqref{eq:firstlaw}.

With one species, a mass extremum along the full one-parameter sequence is a particle number extremum since \(\mu^\infty>0\). With several species a stationary mass along a prescribed sequence only says that the \emph{weighted sum} of their variations vanishes. For example, the mass can remain stationary as one gains baryons and loses dark particles. Such a change cannot be generated by a radial motion of an isolated two-fluid star. A nonzero vector in \(\ker D\) specifies how the central enthalpies can change together while preserving the total particle number of every species to first order. Such a variation redistributes matter inside the star. The equilibrium first law, Eq.~\eqref{eq:firstlaw}, then gives \(\delta M=0\).

Let \(N_D\) and \(N_N\) denote the dark and nuclear particle numbers in a two-fluid star. Suppose \(\nabla N_D\ne0\), and follow the local curve through the model that keeps \(N_D\) fixed. Let \(s\) be a regular parameter on this curve. Then
\begin{equation}
 \left.\frac{\dd M}{\dd s}\right|_{N_D}
   =\mu_N^\infty\left.\frac{\dd N_N}{\dd s}\right|_{N_D}.
 \label{eq:constrainedturn}
\end{equation}
A stationary mass now signals that the tangent preserves the other number too. On a central pressure plot the contours of \(N_D\) and \(N_N\) have a common tangent if both gradients are nonzero. This is the two-fluid critical curve picture of Refs.~\cite{Henriques1990,Kain2021,ZhouChenWuZhang2025}. It remains necessary to determine which radial mode is neutral. The construction locates a stationary point even on a branch that already possesses an unstable mode.

\begin{stabilitytheorem}[Particle number criterion for radial zero modes]\label{thm:neutral}
Fix the equations of state of \(K\) separately conserved, isentropic perfect fluids coupled only through gravity. Let \(\sigma\) be an interior point of their full \(K\)-parameter family of regular spherical equilibria satisfying the center, vacuum and interface hypotheses of Sec.~\ref{sec:scope}. Let \(D(\sigma)\) be the Jacobian of all \(K\) particle numbers in Eq.~\eqref{eq:D}, and \(\mathcal Z_\sigma\) the vector space of regular material radial modes at zero frequency. Following shells of fixed enclosed particle number defines a linear isomorphism
\begin{equation}
  \ker D(\sigma)\simeq\mathcal Z_\sigma,\qquad
  \dim\mathcal Z_\sigma=K-\rankop D(\sigma).
  \label{eq:rank}
\end{equation}
In particular, a nonzero physical radial zero mode exists if and only if \(D(\sigma)\) loses rank.
\end{stabilitytheorem}

Multiplicity here counts linearly independent displacements at a single equilibrium. A vector \(v\) in the null space specifies relative variations of the central data, including their signs. It does not specify the overall oscillation amplitude, which is arbitrary in a linear problem.

The rank in Eq.~\eqref{eq:rank} is independent of the choice of local chart. Changing from central pressures to their logarithms or to central enthalpies multiplies \(D\) on the right by an invertible matrix. Rescaling a particle number by its microscopic normalization rescales one row by a nonzero factor. Both transformations leave the number of independent zero modes unchanged, although a determinant or singular value away from zero changes in magnitude. This distinction matters when a contour is located numerically over several decades in central pressure.

For \(K>2\), the criterion requires linear dependence of the \(K\) particle number gradients. Pairwise parallelism would force the Jacobian to have rank at most one and would miss the generic case of rank \(K-1\).

\subsection{Correspondence with physical radial zero modes}

We prove the correspondence by associating a unique material displacement with each particle-preserving equilibrium variation and then constructing the inverse map.

Inside component \(I\) label each material shell by its enclosed particle number \({\cal N}_I\). This label increases strictly with radius up to \(R_I\), even when a different species ends inside it. Let a shell initially at \(r\) move by \(\xi_I(r)\). Along an equilibrium perturbation \(v\), conservation of its label means
\begin{equation}
 0=\Delta_I{\cal N}_I(r)
     =\delta_{E,v}{\cal N}_I(r)+g_I(r)\xi_I(r).
 \label{eq:materialN}
\end{equation}
The Eulerian variation on the right holds \(r\) fixed. If the neighboring model contains more particles inside that fixed radius, the same material shell lies at a smaller radius, which explains the minus sign in
\begin{equation}
 \xi_I(r)=-\frac{\delta_{E,v}{\cal N}_I(r)}{4\pi r^2 e^\Lambda n_I(r)}.
 \label{eq:xi}
\end{equation}
There is one displacement per species at a common radius. Dark particles may move inward there while baryons move outward, or the displacements may have the same sign with different amplitudes.

Differentiating Eq.~\eqref{eq:materialN} gives the linearized number conservation law, \(\delta g_I=-(g_I\xi_I)'\). Equivalently, two displaced shell boundaries carry the same number of particles between them as before~\cite{FriedmanSchutz1978,Kain2020}. The regular center fixes the integration constant. There \(g_I=O(r^2)\), \(\delta_E{\cal N}_I=O(r^3)\), and Eq.~\eqref{eq:xi} gives \(\xi_I=O(r)\). Any other displacement producing the same equilibrium perturbation would differ by a constant divided by \(g_I\), singular at the center unless that constant vanishes. This establishes uniqueness.

The surface is where the charge constraint becomes essential. For a generic equilibrium variation with \(\delta_v N_I\ne0\), the numerator of Eq.~\eqref{eq:xi} tends to the excess number of particles as \(r\) approaches \(R_I\), while the denominator tends to zero. A material displacement cannot account for an excess of particles at the edge of an isolated star. If \(\delta_vN_I=0\), the numerator is determined instead by the thin tail of the profile between \(r\) and \(R_I\). Let \(x=R_I-r\). The enthalpy drops linearly to zero at the surface, so \(g_I=b\,x^{\alpha_I}[1+o(1)]\), with \(b>0\). The number in the tail is
\begin{equation}
 N_I-{\cal N}_I(r)=\frac{b}{\alpha_I+1}x^{\alpha_I+1}[1+o(1)] .
 \label{eq:tail}
\end{equation}
Varying \(R_I\) at fixed \(r\) gives \(g_I(r)\,\delta_v R_I\) as the leading variation of the tail. The variation of its prefactor is of higher order. The controlled remainder in Eq.~\eqref{eq:surfaceEOS} permits this differentiation. Equation~\eqref{eq:xi} therefore has the finite limit \(\xi_I(R_I)=\delta_v R_I\). The last shell follows its own material surface. If two constituent surfaces coincide on the background this argument still applies separately to each label. The differentiated tail estimate and its displacement limit are derived in Sec.~\ref{sup:sec-surface-tail}, Eq.~(\ref{en:app-tail-quotient-check}), of the Supplemental Material~\cite{SupplementalMaterial}.

At the surface one can express the same condition as
\begin{equation}
 \Delta_I H_I(R_I)=\delta H_I(R_I)+\xi_I(R_I)H_I'(R_I)=0.
 \label{eq:materialsurface}
\end{equation}
It follows by differentiating \(H_I(R_I)=0\). The usual statement that the Lagrangian surface pressure vanishes follows from \(\Delta_I p_I=h_I\Delta_I H_I\). At an ordinary vacuum surface \(h_I\) itself vanishes. Keeping track of \(H_I\) retains the displacement of the last material shell even where \(p_I\) is unusually insensitive to its motion. Across an inner surface, the pressure and the displacement of a surviving species match continuously, and the metric has no surface source.

Differentiating the hydrostatic force balance for a neighboring equilibrium gives, in every region where species \(I\) is present,
\begin{equation}
 \delta p_I'+(\delta\epsilon_I+\delta p_I)\Phi'
             +h_I\delta\Phi'=0 .
 \label{eq:zeroforce}
\end{equation}
The three terms are the change in the pressure gradient, the change in the inertial density acted on by the original gravitational field, and the change in the field acting on the original fluid. Their sum is precisely the radial force perturbation at zero frequency. At a nonzero frequency the force imbalance accelerates the shell. Here the change in pressure and gravity balances at every radius.

The gravitational perturbation can be checked without dividing a time-dependent equation by the frequency. Substituting Eq.~\eqref{eq:materialN} into Eq.~\eqref{eq:enclosedlaw} gives
\begin{equation}
 \delta m(r)=-4\pi r^2\sum_I h_I(r)\xi_I(r),\qquad
 \delta\Lambda=\frac{\delta m(r)}{rA}.
 \label{eq:deltam}
\end{equation}
An outward displacement of material reduces the mass inside its former radius with the relativistic weight \(h_I\). Where a component ends its contribution tends continuously to zero. At the outer edge the result is \(\delta M=0\), consistent with Eq.~\eqref{eq:firstlaw}. The lapse perturbation follows from the radial Einstein equation,
\begin{equation}
 \delta\Phi'=\frac{4\pi r}{A}\sum_I\delta p_I+
                  (2\Phi'+1/r)\delta\Lambda ,
 \label{eq:lapse}
\end{equation}
with the time normalization fixed by the vacuum exterior. Equations~\eqref{eq:materialN}, \eqref{eq:zeroforce}--\eqref{eq:lapse}, and the barotropes give the material and metric perturbations together. A null vector of \(D\) has therefore supplied a regular solution of the physical zero-frequency equations, including the boundary conditions.

Conversely, start from a regular radial mode with zero frequency. Integrating its local conservation law outwards from the center gives Eq.~\eqref{eq:materialN}. Finiteness of the displacement and \(g_I(R_I)=0\) give \(\delta N_I=0\) at \emph{each} free surface. To infer that the mode must appear in the equilibrium survey we must still show that its matter and metric perturbations are the derivative of one of the modeled stars. This is why the use of a complete local family of central data matters.

In the interior the barotrope fixes \(\delta\epsilon_I=\mu_I\delta n_I\) and \(\delta p_I=c_{s,I}^2\delta\epsilon_I\). Equation~\eqref{eq:zeroforce} reduces to
\begin{equation}
\begin{aligned}
 (\delta H_I)'&=-\delta\Phi',\\
 \delta H_I(r)&=\delta H_I^c-[\delta\Phi(r)-\delta\Phi(0)] .
 \end{aligned}
 \label{eq:dH}
\end{equation}
The linearized mass equation, Eq.~\eqref{eq:lapse} and Eq.~\eqref{eq:dH} are exactly the equations obtained by differentiating the equilibrium sequence in the direction \(v=(\delta H_I^c)\). Both perturbations have zero enclosed mass at the center. To see why there is no extra freedom, start with the center expansions \(m=O(r^3)\) and \(H_I=H_I^c+O(r^2)\). Regularity fixes the center solution for the specified \(v\). At each successive material surface the vanishing of the ending enthalpy fixes its shift, and the enclosed mass, metric and surviving pressures give the initial data for continuation. A derivative of the density can behave as \(x^{\alpha_I-1}\) at an inner edge and remains integrable for every \(\alpha_I>0\). The linearized mass equation therefore carries no free jump constant across that edge. The final additive constant in \(\delta\Phi\) is fixed by matching the normalized exterior. This yields the same static perturbation throughout the star.

There is no residual mode with vanishing matter fields. If \(v=0\), uniqueness gives all Eulerian field variations zero. Equation~\eqref{eq:materialN} then gives every displacement zero where the component is present. Thus a nontrivial physical zero mode defines a nonzero \(v\) with \(Dv=0\). Starting from this \(v\) and applying Eq.~\eqref{eq:xi} recovers the original displacements shell by shell. These inverse constructions prove Theorem~\ref{thm:neutral}, including its multiplicity statement. For a simple zero, equilibrium data determine the neutral displacement profile up to normalization. Sections~\ref{en:app-proof-static-package}--\ref{en:app-proof-zero-kernel-package} of the Supplemental Material~\cite{SupplementalMaterial} establish, in order, uniqueness of the static variation, recovery of the displacement, reconstruction and matching of the fields, and the inverse map. Theorem~\ref{en:thm-main} combines these results and gives the nullity in Eq.~(\ref{en:eq-nullity}).

\section{Radial spectrum and the local stability boundary}\label{sec:boundary}

Consider a harmonic perturbation \(\xi_I(t,r)=\xi_I(r)e^{i\omega t}\). In terms of \(\zeta_I=r^2 e^{-\Phi}\xi_I\), an oscillation of the isolated star satisfies a coupled radial eigenvalue equation~\cite{Chandrasekhar1964,Kain2020,CaballeroRipleyYunes2024},
\begin{equation}
\begin{gathered}
 ({\cal L}\zeta)_I=\omega^2 W_I\zeta_I,\qquad 0<r<R_I,\\
 W_I=\frac{e^{3\Lambda+\Phi}}{r^2}h_I .
 \end{gathered}
 \label{eq:mode}
\end{equation}
The positive weight \(W_I\) is the inertia of component \(I\) in this variable. The highest radial derivative in \({\cal L}\) is \(-(P_I\zeta_I')'\), with \(P_I=e^{\Lambda+3\Phi}h_I c_{s,I}^2/r^2\). The sound speed thus controls the local cost of compression. The other terms contain the hydrostatic and metric response, including the coupling to displacements of other species.

The radial variational principle identifies the fundamental mode by minimizing the ratio of restoring energy to inertia~\cite{Chandrasekhar1964,CaballeroRipleyYunes2024},
\begin{equation}
 \omega_0^2=\min_{\zeta\ne0}
 \frac{a[\zeta,\zeta]}{\displaystyle\sum_I\int_0^{R_I}W_I|\zeta_I|^2\,\dd r}.
 \label{eq:variational}
\end{equation}
Here \(a\) is the symmetric quadratic form of Eq.~\eqref{eq:mode}, defined for regular finite-energy trial displacements and their energy norm limits. It includes both pressure and gravity and is bounded from below. Reference~\cite{CaballeroRipleyYunes2024} derives the two-fluid variational principle in Sec.~IV and the symmetric pulsation equations for any finite number of fluids in Sec.~V. If the minimum is positive, all small radial normal modes oscillate. When it is negative, a displacement along its eigenfunction has an exponentially growing solution with growth rate \(\sqrt{-\omega_0^2}\) in the time coordinate normalized at infinity.

There is a physical reason to check the use of Eq.~\eqref{eq:variational} for a multifluid star. The integrals extend to a different radius for each fluid. At a material surface both inertia and pressure support vanish, at different rates, and that surface can be inside another moving fluid. An inadmissible singular displacement there could solve the local differential equation and be counted as a mode if only the interior equations were used. For \(x=R_I-r\) the ordinary vacuum law gives \(W_I\) proportional at leading order to \(x^{\alpha_I}\) and \(P_I\) to \(x^{\alpha_I+1}\). A singular homogeneous solution with \(P_I\zeta_I'\) tending to a nonzero constant would have divergent compressional energy \(\int P_I|\zeta_I'|^2\,\dd r\). The finite-energy surface branch has finite displacement and satisfies the material condition~\eqref{eq:materialsurface}. At an inner surface the continuing species transmits pressure and displacement, while the ending fluid carries no sheet of stress. These conditions eliminate unphysical boundary work in the variational argument.

The detailed endpoint estimates show that the self-adjoint radial problem has discrete squared frequencies, bounded below, with no finite accumulation point. Its finite-energy eigenfunctions obey the same center, surface and matching conditions used in Sec.~\ref{sec:proof}. The zero counted by the static criterion therefore belongs to the spectrum of the physically matched star. Further, the ordered squared frequencies \(\omega_0^2\leq\omega_1^2\leq\cdots\), counted with multiplicity, vary continuously along the equilibrium family. One may compare models with different radii by measuring distance in each fluid in units of its own radius. The part of the energy integral contributed by an overlap that is disappearing is controlled by the vanishing surface weights. Continuity consequently extends to backgrounds where two radii coincide. Differentiability of a simple frequency is used below only along a smooth sector on which the ordering of all labeled radii is fixed. Theorem~\ref{en:thm-fixed-spectral-package} of the Supplemental Material~\cite{SupplementalMaterial} identifies the physical eigenspaces of the self-adjoint operator. Theorem~\ref{en:thm-ordered-spectral-continuity} in Sec.~\ref{en:sec-spectral-parameter-dependence} proves continuity of the ordered spectrum. Section~\ref{sup:sec-kato-riesz} gives the projection argument for differentiability of a simple eigenvalue on a fixed-order sector.

\begin{stabilitycorollary}[Boundary of strict radial stability]\label{cor:stable}
Under the hypotheses of Theorem~\ref{thm:neutral}, let \(U\) be a regular neighborhood in the equilibrium family and let \(U_{\rm st}=\{\sigma\in U:\omega_0^2(\sigma)>0\}\). Then

\begin{equation}
 \partial_U U_{\rm st}
 \ \subseteq\ \{\sigma:\omega_0^2=0\}
 \ \subseteq\ \{\sigma:\operatorname{rank}D<K\}.
 \label{eq:inclusions}
\end{equation}
If a simple fundamental eigenvalue has a nonzero tangential derivative on a smooth sector of fixed radius ordering, all three sets coincide locally at its zero, relative to that sector.
\end{stabilitycorollary}
By continuity, either strict sign of \(\omega_0^2\) persists in a neighborhood, so \(\omega_0^2=0\) on the boundary of the strictly stable set. Equation~\eqref{eq:rank} then gives the second inclusion.

At a simple fundamental zero, the first overtone has \(\omega_1^2>0\) and remains positive nearby. On the smooth sector \(\mathcal U\) of the corollary, the nonzero derivative of \(\omega_0^2\) makes its zero set a hypersurface across which the sign changes. In a sufficiently small relative neighborhood \(V\subset\mathcal U\), positivity on one side and growth on the other give
\begin{equation}
\begin{aligned}
 \{\sigma\in V:\operatorname{rank}D<K\} & =\{\sigma\in V:\omega_0^2=0\}\\
 &=V\cap\partial_{\mathcal U}(U_{\rm st}\cap\mathcal U).
\end{aligned}
 \label{eq:local}
\end{equation}
Proposition~\ref{en:prop-spectral-boundary} of the Supplemental Material~\cite{SupplementalMaterial} proves the inclusions and the local equality in Eqs.~(\ref{en:eq-boundary-inclusions})--(\ref{en:eq-local-stable-rank-equality}).

An overtone can reach zero in a star that already has \(\omega_0^2<0\). The fundamental squared frequency can also touch zero without changing sign along a sequence, as can happen when the sequence is tangent to a regular stability boundary.

Along a regular sequence starting from a known stable model, continuity keeps \(\omega_0^2>0\) until \(D\) first loses rank. At each candidate boundary a pulsation calculation identifies the neutral modes and checks the sign of \(\omega_0^2\) on either side along the sequence. Section~\ref{en:sec-numerical} uses this procedure to select the configurations for the mass--radius and tidal projections.

\section{Radial stability and observables of neutron--dark matter mixtures}\label{en:sec-numerical}

\subsection{Two-fluid stability boundaries and observables}\label{sec:two-fluid-application}

\begin{figure*}[!ht]
 \centering
 \includegraphics[width=0.90\textwidth]{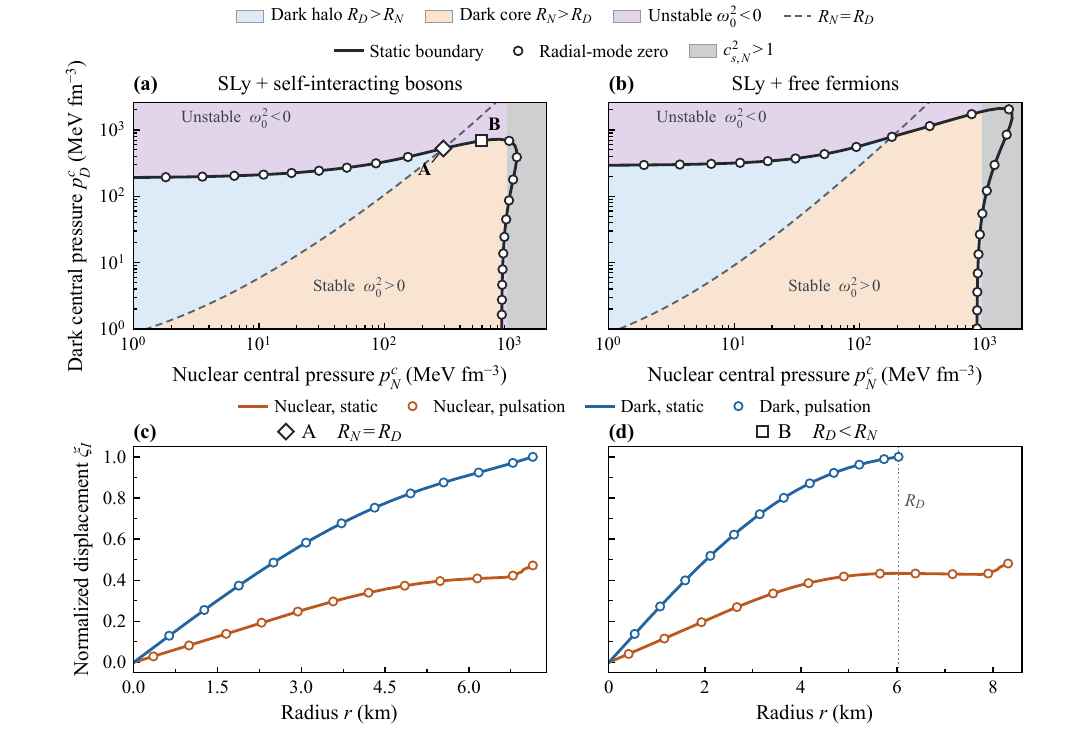}
 \caption{Radial stability and the associated neutral displacements. (a,b) SLy matter mixed with bosons and fermions, respectively. Within the stable region, blue marks dark halos (\(R_D>R_N\)) and orange dark cores (\(R_N>R_D\)). Lavender marks radial instability (\(\omega_0^2<0\)). The dashed curve marks equal central enthalpies, where the surfaces coincide. Gray denotes \(c_{s,N}^2>1\). The black curve tracks loss of rank of the particle number Jacobian. Open circles mark zeros of the radial fundamental mode. The diamond A \((295,524)\) and square B \((598,686)\) in (a) select the profiles in (c) and (d), respectively. Coordinates are \((p_N^c,p_D^c)\) in \(\mathrm{MeV\,fm^{-3}}\). In (c,d), the dark matter displacement is blue and the nuclear displacement orange. Lines give the reconstruction from Eq.~\eqref{eq:xi} and open circles give the pulsation solutions. For each star the two solutions are aligned in sign and inertia norm, then scaled by the largest absolute static displacement across both fluids. The dotted line in (d) marks the dark surface.}
 \label{en:fig-static-dynamic-validation}
\end{figure*}

We apply the static criterion to two families of cold, nonrotating stars in which nuclear and dark matter interact through gravity and conserve their particle numbers separately. For the nuclear fluid we use the unified SLy equation of state (EoS), which uses the same SLy4 effective interaction for the inner crust and core~\cite{DouchinHaensel2001}. We use the analytic pressure fit of Ref.~\cite{HaenselPotekhin2004}.
We smoothly join the SLy fit to a cold envelope with \(p_N\propto\rho^\gamma\) for \(\rho=\epsilon_N/c^2\leq10^5~\mathrm{g\,cm^{-3}}\) and \(\gamma\simeq1.756\), so that \(n_N\sim C_N H_N^{1/(\gamma-1)}\to0\) at the zero-pressure surface as required by Eq.~\eqref{eq:surfaceEOS}.
We restrict the observable projections to the causal part of the EoS. In the adopted analytic representation the nuclear sound speed reaches \(c\) at a central pressure \(p_N^c=943~\mathrm{MeV\,fm^{-3}}\). Keeping this nuclear EoS fixed lets us compare how two different sources of dark pressure affect stability and tidal response.

In the first family, the bosonic dark fluid is the strong coupling limit of a massive scalar with a repulsive quartic self-interaction~\cite{ColpiShapiroWasserman1986}. In solar units its barotrope is~\cite{ZhangLin2020Hybrid}
\begin{equation}
 \bar\epsilon_D=3\bar p_D+B_4\sqrt{\bar p_D},\qquad B_4=0.1 ,
\end{equation}
where bars denote energy density and pressure in units of \(M_\odot c^2/(GM_\odot/c^2)^3\). In the second family, degeneracy pressure supports a zero-temperature gas of free spin-\(\tfrac12\) fermions with two spin states and particle mass \(m_\chi=1~\mathrm{GeV}\)~\cite{Kain2021}. With \(x=k_F/m_\chi\), where \(k_F\) is the Fermi momentum and \(\hbar=c=1\), the relativistic Fermi gas equation of state reads~\cite{OppenheimerVolkoff1939}
\begin{equation}
 \begin{aligned}
 \epsilon_D &= \frac{m_\chi^4}{\pi^2}\int_0^x t^2\sqrt{1+t^2}\,\dd t,\\
 p_D &= \frac{m_\chi^4}{3\pi^2}\int_0^x \frac{t^4}{\sqrt{1+t^2}}\,\dd t .
 \end{aligned}
\end{equation}

In Figs.~\ref{en:fig-static-dynamic-validation}(a) and \ref{en:fig-static-dynamic-validation}(b) the first loss of rank of the particle number Jacobian coincides with a change of sign of the fundamental radial eigenvalue \(\omega_0^2\). The first overtone stays positive at these crossings. For example, on the equal-central-enthalpy sequences [dashed curves in Figs.~\ref{en:fig-static-dynamic-validation}(a,b)], the two fluids share a surface. The critical central pressures \((p_N^c,p_D^c)\), in \(\mathrm{MeV\,fm^{-3}}\), are \((295,524)\) for bosons (point A) and \((181,786)\) for fermions. On the causal sequences tested, the two boundary determinations differ by less than \(2\times10^{-5}\) in the plane of base-ten logarithmic pressures. The fermionic boundary also reproduces the published critical curve for the same SLy--Fermi mixture~\cite{Kain2021}. Across all 7,182 equilibria, the static and dynamical criteria give the same classification into stable and unstable configurations.

At the critical points A and B in Fig.~\ref{en:fig-static-dynamic-validation}(a), \(D\) loses rank, so there is a nonzero direction \(v\) in central enthalpy space that preserves both particle numbers to first order (\(Dv=0\)). Equation~\eqref{eq:xi} gives each shell's displacement by holding its enclosed particle number fixed along \(v\). Since this variation preserves hydrostatic balance to first order, the reconstructed displacements have no linear restoring force and together form a radial zero mode (Theorem~\ref{thm:neutral}). We approximate \(v\) by the right singular vector of \(D\) associated with its smallest singular value. The neighboring equilibria used for this directional derivative may lie off the critical curve. For each star, the two fluids share one overall normalization and sign, so comparison with an independent pulsation solution tests both the shapes of the displacement profiles and their relative amplitude.

\nocite{Hinderer2008Love,Hinderer2010Tides,J0614Data,Dittmann2024,GW170817Data,Abbott2020GW190425,GW190425Data}
\begin{figure*}[!t]
 \centering
 \includegraphics[width=0.92\textwidth]{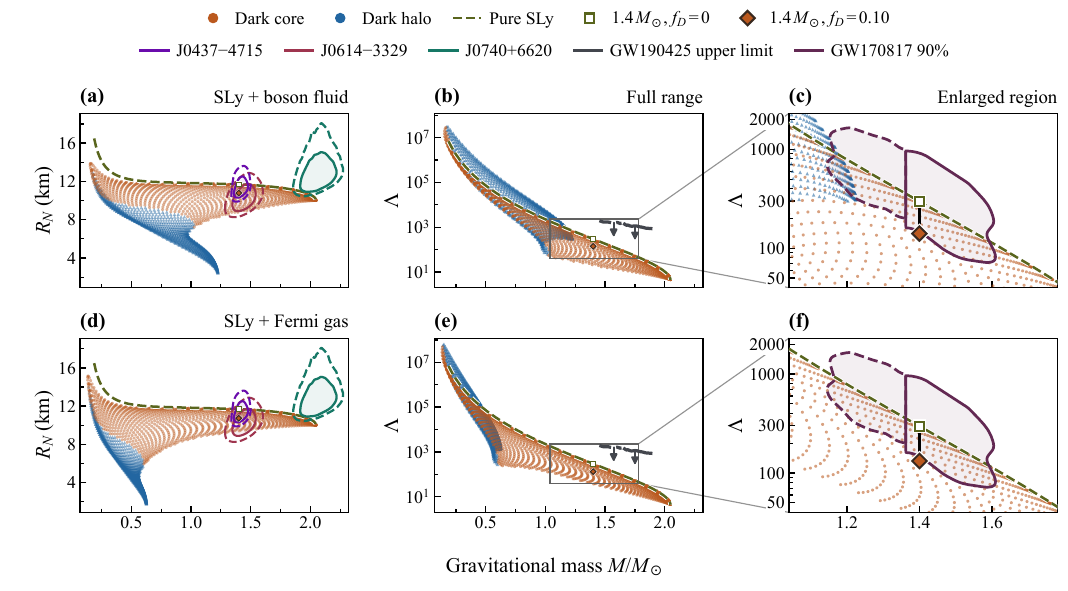}
 \caption{Stable mixtures with bosonic dark matter (top row) or fermionic dark matter (bottom row). Blue triangles denote dark halos, orange dots denote dark cores, and the dark olive dashed curve is the pure SLy sequence. Squares and diamonds mark the pure SLy and 10\% dark core stars at \(1.4\,M_\odot\) in Table~\ref{tab:matched}. (a,d) Nuclear emitting radius and joint mass--equatorial radius credible regions for PSR J0437$-$4715 (68.3\%, 95.4\%), J0614$-$3329 (68\%, 95\%) and J0740$+$6620 (68\%, 90\%)~\cite{Choudhury2024,J0437Data,Mauviard2025,J0614Data,Dittmann2024,J0740Data}. (b,e) Full tidal projection. Arrows mark the direction of the GW190425 upper limits, given by the 90th percentile of each component's \(\Lambda_i\) in source mass bins under the low-spin prior~\cite{Abbott2020GW190425,GW190425Data}. (c,f) Enlargements of the boxed regions with the GW170817 90\% joint contours in \((M_i,\log_{10}\Lambda_i)\), assuming a common EoS~\cite{Abbott2018,GW170817Data}. Solid and dashed GW curves denote the heavier and lighter component, respectively.}
 \label{en:fig-two-dimensional-projections}
\end{figure*}

At A, the equal-enthalpy bosonic model, both surfaces lie at \(7.14\) km. Figure~\ref{en:fig-static-dynamic-validation}(c) shows that the surfaces move in the same direction, with \(\xi_D(R_D)/\xi_N(R_N)=2.12\). Two initially coincident material surfaces therefore separate to first order along this mode. At B, with \((p_N^c,p_D^c)=(598,686)~\mathrm{MeV\,fm^{-3}}\), the dark core ends at \(6.03\) km while the nuclear fluid extends to \(8.31\) km [Fig.~\ref{en:fig-static-dynamic-validation}(d)]. The ratio of their surface displacements is \(2.08\). The nuclear displacement continues through the edge of the dark core, where the two one-sided values of the nuclear Eulerian pressure perturbation agree to better than one part in \(10^6\).

In both stars \(|\omega_0^2|/\omega_1^2<2\times10^{-9}\). The overlap mismatch, \(1-|\mathcal O_W|\), is below \(10^{-6}\) for both stars. Here \(\mathcal O_W\) is the normalized overlap of the static and pulsation profiles of \(\zeta_I=r^2e^{-\Phi}\xi_I\), evaluated with the inner product \(\langle a,b\rangle_W=\sum_I\int_0^{R_I}W_I a_I b_I\,\dd r\) of Eq.~\eqref{eq:mode}. As a separate surface test, we compare both the limiting static displacement and the pulsation displacement with the independent variation \(D_vR_I\) of the equilibrium radius. Their relative differences are below \(10^{-5}\) for each fluid in both stars.

Using the sign of the fundamental squared frequency to select the stable side, we retain 2,348 causal bosonic models and 2,559 causal fermionic models for the observable projections in Fig.~\ref{en:fig-two-dimensional-projections}. The two radii \(R_N\) and \(R_D\) locate the surfaces of nuclear and dark matter. Surface X-rays originate at \(R_N\)~\cite{Miao2022}, while the tidal field probes matter out to \(R_t=\max(R_N,R_D)\)~\cite{Leung2022Tidal}. The quadrupolar Love number \(k_2\) sets the dimensionless tidal deformability through the relation~\cite{Hinderer2008Love,Hinderer2010Tides,Leung2022Tidal}
\begin{equation}
 \label{eq:tidal-love}\Lambda=\frac{2}{3}k_2\left(\frac{c^2 R_t}{GM}\right)^5 .
\end{equation}
In a dark core configuration the same radius enters both observables. A dark halo extends the tidal response beyond the emitting surface and also modifies the propagation of surface photons.

\begin{table}[tb]
 \caption{Stable stars of gravitational mass \(1.4\,M_\odot\). Both mixed models have \(M_D=0.14\,M_\odot\). The nuclear and dark radii are in km.}
 \label{tab:matched}
 \centering
 \setlength{\tabcolsep}{4pt}
 \begin{tabular}{lrrrrr}
 \toprule
 Dark component & \(f_D\) & \(R_N\) & \(R_D\) & \(k_2\) & \(\Lambda\)\\
 \midrule
 None & 0 & 11.70 & \dots & 0.0767 & 297.0\\
 Boson fluid & 0.10 & 10.78 & 6.52 & 0.0547 & 140.7\\
 Fermi gas & 0.10 & 10.74 & 5.44 & 0.0525 & 132.2\\
 \bottomrule
 \end{tabular}
\end{table}

Table~\ref{tab:matched} isolates the effect of composition by holding the gravitational mass fixed at \(1.4\,M_\odot\). We compare a pure SLy star with one star from each mixed family at \(f_D=M_D/M=0.10\), where \(M_D\) is the dark contribution to the gravitational mass. Both mixtures form dark cores and have radial fundamental frequencies of about \(3.06~\mathrm{kHz}\). The emitting radius shrinks by \(7.8\%\) for bosons and \(8.2\%\) for fermions. The tidal response changes more strongly. Relative to \(\Lambda=297.0\) for pure SLy, the bosonic and fermionic values fall to \(140.7\) and \(132.2\), reductions of \(52.6\%\) and \(55.5\%\), respectively. Both \(k_2\) and the radius decrease, and the radius enters \(\Lambda\) to the fifth power at fixed mass. At this composition the two dark models differ in \(\Lambda\) by only \(8.5\), much less than either shift from the pure nuclear star.

The X-ray contours in Fig.~\ref{en:fig-two-dimensional-projections} provide two comparisons near the benchmark mass. PSR~J0437$-$4715 has an inferred equatorial radius of \(11.36^{+0.95}_{-0.63}\) km~\cite{Choudhury2024}. For PSR~J0614$-$3329 the preferred hot spot model gives \(M=1.44^{+0.06}_{-0.07}\,M_\odot\) and \(R_{\rm eq}=10.29^{+1.01}_{-0.86}\) km~\cite{Mauviard2025}. At \(1.4\,M_\odot\), both dark core stars fall inside the published 68\% joint contour for J0614$-$3329, and the 95\% contour also contains the pure SLy star~\cite{J0614Data}. Near \(2.1\,M_\odot\), PSR~J0740$+$6620 provides a separate radius measurement of \(12.92^{+2.09}_{-1.13}\) km~\cite{Dittmann2024}. These rotating star inferences give a scale for comparing the spherical models.

Binary inspirals primarily measure a mass-weighted combination of the two tidal deformabilities. The public samples retain each source frame component mass paired with its own \(\Lambda\). For GW170817 we show the component posteriors inferred with a common parametrized EoS that supports \(1.97\,M_\odot\)~\cite{Abbott2018,GW170817Data}. GW190425 carries much less tidal information. Its low-spin analysis with independent component deformabilities gives broad, prior-sensitive upper bounds on each component. For example, binning the released posterior samples by source frame secondary mass, we find that in the bin with median mass \(1.56\,M_\odot\), the 90th percentile is \(\Lambda_2\simeq1570\)~\cite{Abbott2020GW190425,GW190425Data}. The overlays illustrate the scale of current measurements. Inferring a dark fraction would require a binary likelihood that allows the two stars to have different compositions.

\subsection{Stability surface of three-fluid stars}\label{sec:three-fluid-application}

To model a star containing both dark species discussed in Sec.~\ref{sec:intro}, we combine the EoSs of Sec.~\ref{sec:two-fluid-application} for SLy nuclear matter (\(N\)), self-interacting bosons (\(B\), \(B_4=0.1\)), and free fermions (\(F\), \(m_\chi=1~\mathrm{GeV}\)). Each fluid conserves its own particle number and interacts with the other two only through gravity.

For each fixed ratio of central enthalpies, we follow the compact star branch from a stable configuration and increase \(s=\sum_I H_I^c\) until the particle number Jacobian first loses rank. Figure~\ref{fig:three-fluid-boundary}(a) shows the resulting surface traced along 5,995 directions. Its edges recover two-fluid critical curves when one species is absent.

We independently locate the fundamental mode zero along 703 directions, including 595 with all three species present, using the pulsation equations in Appendix~\ref{app:radial-numerics}. For a given direction, \(s_{\rm stat}\) denotes the total central enthalpy at the first loss of rank, and \(s_{\rm mode}\) its independently determined value at \(\omega_0^2=0\). On every tested ray, \(\omega_0^2\) changes sign as \(s\) crosses \(s_{\rm mode}\), while \(\omega_1^2\) remains positive at the root.

\begingroup
\ifdim\columnwidth<\textwidth
  \onecolumngrid
  \vspace*{\fill}
  \makeatletter
  \renewenvironment{figure*}[1][]
    {\noindent\begin{minipage}{\textwidth}\def\@captype{figure}}
    {\end{minipage}\par\newpage\twocolumngrid}
  \makeatother
\fi
\begin{figure*}[!t]
 \centering
 \includegraphics[width=0.90\textwidth]{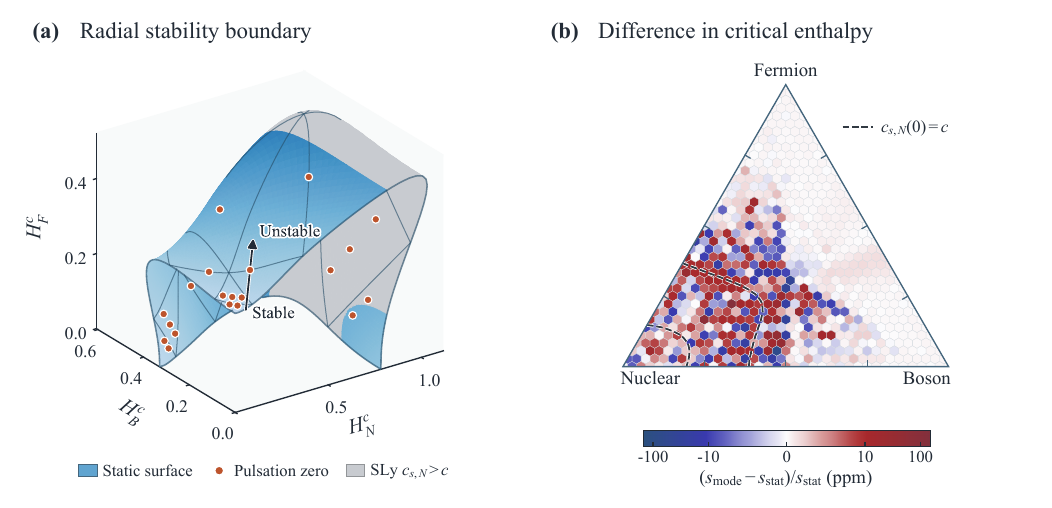}
 \caption{Radial stability of three-fluid stars. (a) The axes are the dimensionless central enthalpies [Eq.~\eqref{eq:H}]. Increasing \(s=H_N^c+H_B^c+H_F^c\) at fixed enthalpy ratios takes the compact star sequence from the radially stable side facing the origin (\(\omega_0^2>0\)), through the critical surface (\(\omega_0^2=0\)), into the unstable region (\(\omega_0^2<0\)). Orange points mark independent pulsation roots. (b) Each cell gives \((s_{\rm mode}-s_{\rm stat})/s_{\rm stat}\) for one direction \(q_I=H_I^c/s\), with pure fluids at the vertices. The scale is in parts per million (ppm). Both dashed curves mark the same central SLy sound speed threshold, \(c_{s,N}(0)=c\). The band between them has \(c_{s,N}(0)>c\) and corresponds to the gray part of (a). On either side, models containing nuclear matter have \(c_{s,N}(0)<c\).}
 \label{fig:three-fluid-boundary}
\end{figure*}
\endgroup

Figure~\ref{fig:three-fluid-boundary}(b) compares the roots on each common ray. Positive values place the pulsation root farther from the origin than the static root. The median absolute relative difference is \(0.71\) parts per million. The maximum is \(0.0074\%\) on the causal part of the surface and \(0.0123\%\) over the full grid. The two dashed curves arise because the nuclear central enthalpy at marginal stability varies nonmonotonically with composition. For example, following \(q_B=q_F\) away from the pure nuclear vertex, one first enters and then leaves the band where the SLy fit gives a central sound speed above \(c\).

With three fluids, a pulsation shooting method must determine two relative central amplitudes along with the frequency. The increased cost led Ref.~\cite{Kain2020} to present the pulsation solution for a single three-fluid equilibrium. The static criterion requires only neighboring equilibria and their particle numbers. For the same 703 directions, with identical bracket searches and root tolerances on a single thread of an Intel Core i9-14900KF processor, the static calculation took \(5.25\) s and our finite element fundamental mode calculation took \(29.18\) s. This corresponds to a \(5.6\)-fold speedup for the static method at the resolutions used here.

\section{Discussion and conclusions}\label{sec:discussion}

For any equilibrium satisfying the assumptions of Sec.~\ref{sec:family}, the number of independent physical radial zero modes equals the nullity of its particle number Jacobian (Theorem~\ref{thm:neutral}). Near a simple transverse zero of the fundamental mode, Corollary~\ref{cor:stable} identifies the local stability boundary.

In both stable \(1.4\,M_\odot\) models with a dark core containing 10\% of the gravitational mass, the emitting radius decreases by about 8\% and the tidal deformability by more than half relative to the pure SLy star of the same mass (Table~\ref{tab:matched}). The X-ray emitting surface follows the nuclear fluid. The quadrupolar tidal response depends on matter out to the outermost radius~\cite{Miao2022,Leung2022Tidal}. Even after imposing stability, distinct compositions can yield similar projected masses and radii~\cite{Hippert2023,LuWen2025}.

Higher signal-to-noise observations of future binary neutron star inspirals could reduce the statistical uncertainty in the effective binary tidal deformability \(\tilde\Lambda\). Combining events across a range of masses would then test the predicted mass dependence of the stellar deformability \(\Lambda\) more precisely~\cite{Walker2024Tides}. If the inferred tidal response at well-measured masses consistently lay outside the range predicted by purely baryonic EoSs compatible with independent nuclear and astrophysical constraints, dark matter admixture could provide a compelling physical explanation. In simulated signals from stars with a small fermionic dark component, omitting that component from the inference can bias the inferred nuclear EoS toward softer models~\cite{Koehn2024Tides}. Testing the dark core interpretation would therefore require a joint analysis of tidal and radius measurements that allows for uncertainty in the nuclear EoS and the inspiral waveform.

\begin{acknowledgments}
The authors thank Feng-Li Lin, Dong Lai, Qiyuan Pan, Shao-Feng Ge, Zhoujian Cao, Nobutoshi Yasutake, Chian-Shu Chen, Alessandro Parisi, Chen Zhang, Zihao Zhang, Si-Man Wu, and Jing-Yi Wu for helpful discussions. K.Z. (Hong Zhang) is supported by a classified fund from Shanghai city.

OpenAI Codex is applied to polish the manuscript and to assist with some of the mathematical derivations in the Supplemental Material. The authors reviewed and verified all such output and take full responsibility for the content. 
\end{acknowledgments}

\section*{Data availability}
The pulse profile posteriors for PSR~J0437$-$4715, PSR~J0614$-$3329 and PSR~J0740$+$6620 are available in Refs.~\cite{J0437Data,J0614Data,J0740Data}. The component posteriors for GW170817 and GW190425 are available in Refs.~\cite{GW170817Data,GW190425Data}. The model data and code supporting the numerical analysis are available from the corresponding author upon reasonable request.

\begingroup
\appendix
\section{Numerical solution of the radial pulsation equations}\label{app:radial-numerics}

We use the coupled radial pulsation equations for gravitationally coupled fluids [Eq.~(55) of Ref.~\cite{Kain2020}]. Their symmetric form for an arbitrary number of fluids is given in Sec.~V and Appendix~A of Ref.~\cite{CaballeroRipleyYunes2024}. Here we group the terms of the weak equation for numerical integration. The fluid labels are \(N,B\) or \(N,F\) in Sec.~\ref{sec:two-fluid-application} and \(N,B,F\) in Sec.~\ref{sec:three-fluid-application}. On each equilibrium of Eq.~\eqref{eq:tov}, let \(\zeta_I=r^2e^{-\Phi}\xi_I\), \(h_I=\epsilon_I+p_I\), \(b_I=h_Ic_{s,I}^2\), and \(R_*=\max_I R_I\). Sums at radius \(r\) run over the fluids present there. With primes denoting radial derivatives, define
\begin{align}
 \Gamma &= 4\pi r e^{2\Lambda},\qquad
 \Theta=\Gamma\sum_I h_I,\qquad \chi=\Phi',\notag\\
 \Xi &= e^{\Lambda+3\Phi}/r^2,\qquad
 V=\chi'+\chi^2-(\Theta+2/r)\chi .
 \label{eq:num-radial-background}
\end{align}
For a real trial displacement \(u\) in the \(\zeta\) variables, write \(S[u]=\sum_Jh_Ju_J\) and
\begin{equation}
 C_I[u]=u_I'+\Theta u_I-\Gamma S[u].
 \label{eq:num-radial-compression}
\end{equation}
This combination gives the Lagrangian pressure perturbation, \(\Delta_I p_I=-e^\Phi b_I C_I[\zeta]/r^2\). For each real test displacement \(v\), the equation we discretize is
\begin{align}
 a[\zeta,v]&=\omega^2\sum_I\int_0^{R_I} W_I\zeta_I v_I\,\dd r,
 \label{eq:num-radial-eigen}\\
 a[u,v]&=\int_0^{R_*}\!\Xi\bigg\{\sum_I b_I C_I[u]C_I[v]\notag\\
 &\hspace{2.0em}+V\sum_Ih_Iu_Iv_I-\frac{\Gamma}{r}S[u]S[v]\bigg\}\,\dd r .
 \label{eq:num-radial-form}
\end{align}
Here \(W_I=e^{3\Lambda+\Phi}h_I/r^2\). Expanding Eq.~\eqref{eq:num-radial-form} yields the diagonal and interfluid terms in Eq.~(\ref{en:eq-full-spectral-form-expanded}) of the Supplemental Material~\cite{SupplementalMaterial}, with the coefficients given in Sec.~\ref{en:app-spectral-realization}.

We expand \(\zeta_I/r^3\) in continuous piecewise quadratic elements, which enforces \(\xi_I=O(r)\) at the center. Every \(R_I\) is a mesh node. The endpoint displacement of the ending fluid is free, giving the natural condition \(\Delta_I p_I(R_I)=0\). For a continuing fluid the displacement is continuous and the weak equation matches its pressure perturbation across the interface. Four-point Gauss quadrature on each element gives the symmetric generalized eigenproblem \(\mathbf K\mathbf x=\omega^2\mathbf M\mathbf x\). We compute the lowest eigenvalue by shift-and-invert iteration. At fixed central enthalpy ratios we increase the overall enthalpy scale from a model with \(\omega_0^2>0\), bracket the first sign change, and locate its root by Brent's method. We check that \(\omega_1^2>0\) there. This procedure uses only the equilibrium profiles and their equations of state.

To test the finite element spectrum, we also compute the fundamental mode by outward shooting following Ref.~\cite{Kain2020}. At each trial frequency we combine two regular central solutions to satisfy the inner fluid's free surface condition, then adjust \(\bar\omega^2\) until the outer fluid also satisfies its surface condition. We use two free Fermi gases with particle masses \(m_2=7m_1\). At each of four fixed central energy densities \(\bar\epsilon_{2,c}=1,20,100,300\), we vary the central pressure \(\bar p_{1,c}\) over 35 equilibria that span the first loss of stability. Bars denote the dimensionless variables of Ref.~\cite{Kain2020}.

\begin{figure}[!t]
 \centering
 \includegraphics[width=\columnwidth]{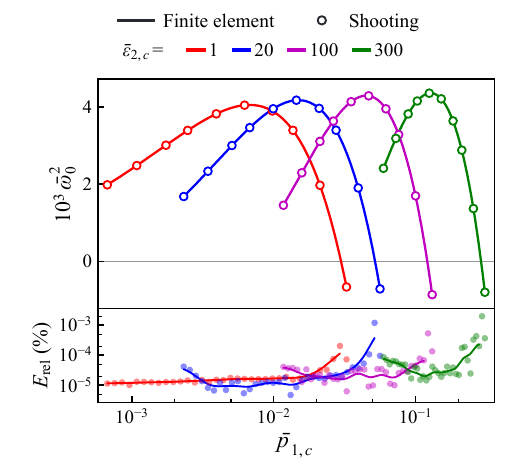}
 \caption{Fundamental radial mode versus the central pressure of fluid 1. Colors denote the fixed central energy density of fluid 2. Solid curves give the finite element eigenvalues and open circles mark selected shooting solutions. The lower strip gives the relative error \(E_{\mathrm{rel}}=\lvert\bar\omega_{0,\mathrm{FEM}}^2/\bar\omega_{0,\mathrm{shoot}}^2-1\rvert\).}
 \label{fig:kain-two-fluid}
\end{figure}

The two methods give the same sign of \(\bar\omega_0^2\) for all 140 equilibria in Fig.~\ref{fig:kain-two-fluid}. The median relative error is \(2.1\times10^{-5}\%\) and the maximum is \(2.0\times10^{-3}\%\). The relative error is undefined where the shooting eigenvalue vanishes, so we compare the critical pressures directly. Their largest relative difference across the four sequences is \(1.7\times10^{-7}\). Refining both radial grids reduces the maximum absolute eigenvalue difference from \(5.0\times10^{-9}\) to \(4.9\times10^{-10}\).

\endgroup
\onecolumngrid\twocolumngrid
\setlength{\emergencystretch}{1em}
\bibliography{CZFZ}
\onecolumngrid
\clearpage
\setlength{\emergencystretch}{0pt}

\StartSupplementAux
\renewcommand{\citenumfont}[1]{s#1}
\renewcommand{\bibnumfmt}[1]{[s#1]}
\makeatletter
\onecolumn@grid@setup
\let\set@footnotewidth\set@footnotewidth@one
\let\compose@footnotes\compose@footnotes@one
\makeatother

\begingroup
\makeatletter
\frontmatter@init
\patchcmd{\addcontents@abstract}{section*.2}{\getrefbykeydefault{sec:included-supplement}{anchor}{section*.2}}{}{}
\let\maketitle\frontmatter@maketitle
\let\author\frontmatter@author
\let\affiliation\SavedAffiliation
\let\email\SavedEmail
\let\abstract\frontmatter@abstract
\let\endabstract\endfrontmatter@abstract
\def\title@column#1{\minipagefootnote@init#1\minipagefootnote@foot}
\patchcmd{\titleblock@produce}{\label{FirstPage}}{}{}{\errmessage{Cannot patch FirstPage}}
\patchcmd{\frontmatter@title@produce}{\phantomsection}{\phantomsection\label{sec:included-supplement}}{}{\errmessage{Cannot insert supplement label}}
\patchcmd{\frontmatter@footnotemark}{frontmatter.}{suppfrontmatter.}{}{\errmessage{Cannot patch footnote link}}
\patchcmd{\frontmatter@makefntext}{frontmatter.}{suppfrontmatter.}{}{\errmessage{Cannot patch footnote destination}}
\makeatother
\title{Supplemental Material for Equivalence of Radial Stability Criteria for Multifluid Stars}
\manuscriptauthors
\begin{abstract}This supplement gives the precise hypotheses and full proofs for
the particle number criterion for neutral radial modes of relativistic
multifluid stars. The static argument constructs the shell displacement
associated with a particle-preserving equilibrium variation. The spectral
argument treats the vanishing of the pressure and the inertia of each
component at its own free surface, where the radial equations become
singular. Sections, equations, and numbered mathematical statements use the prefix S.
\end{abstract}
\maketitle
\thispagestyle{empty}
\endgroup

\setcounter{section}{0}
\renewcommand{\thesection}{S\arabic{section}}
\renewcommand{\thesubsection}{\thesection.\Alph{subsection}}
\makeatletter
\renewcommand{\p@subsection}{}
\makeatother
\numberwithin{equation}{section}
\renewcommand{\theHsection}{supp.\arabic{section}}
\renewcommand{\theHequation}{supp.\arabic{section}.\arabic{equation}}
\renewcommand{\theHtheorem}{supp.\arabic{section}.\arabic{theorem}}
\setcounter{table}{0}
\renewcommand{\thetable}{\Roman{table}}
\renewcommand{\theHtable}{supp.\arabic{table}}

\noindent Sections~\ref{en:sec-physical-setting} and
\ref{en:app-equilibrium-structure} state the hypotheses and prove the
equilibrium and surface properties used in Sec.~\ref{sec:family} of the main text.
Section~\ref{en:sec-static-solution-spaces} derives the equilibrium first
law and constructs the correspondence between particle-preserving variations
and physical displacements. Section~\ref{en:sec-main-proof} completes the
proof of the particle number criterion and its multiplicity statement
(Theorem~\ref{thm:neutral} of the main text). Section~\ref{en:sec-spectral} treats the
radial spectrum and proves the local boundary result (Corollary~\ref{cor:stable} of the
main text). Section~\ref{sup:sec-spectral-estimates} gives the supporting
endpoint and operator estimates.

\section{Physical setting, notation, and hypotheses}
\label{en:sec-physical-setting}
We consider static, spherically symmetric, asymptotically flat relativistic
stars composed of \(K\) separately conserved isentropic barotropic perfect
fluids, coupled only through the common spacetime metric and allowed to have
distinct free surfaces.

We use geometrized units \(G=c=1\), where \(G\) is Newton's constant and \(c\)
is the speed of light.  Here and below \(I\in\{1,\ldots,K\}\) labels a fluid
component sharing the common metric.  The constitutive, conservation, and
coupling hypotheses are collected below.

\subsection{Equilibrium configurations and conserved charges}
\label{en:sec-equilibrium-notation}

The symbols \(n_I,u_I^\mu,J_I^\mu\) denote the rest frame particle number
density, four-velocity, and particle current of component \(I\), respectively;
\(\nabla\) is the Levi--Civita covariant derivative of the common metric.  The
energy density, pressure, chemical potential, and adiabatic sound speed are
denoted by \(\epsilon_I,p_I,\mu_I,c_{s,I}\), respectively, with
\(c_{s,I}^2:=\dd p_I/\dd\epsilon_I\); in the density parametrization below,
\(c_{s,I}^2=n_I\mu_I'(n_I)/\mu_I(n_I)\).

In areal radius gauge, \(\dd s^2\) denotes the spacetime line element,
\(t\) is the static time coordinate, \(r\) is the areal radius, and
\(\dd\Omega^2\) is the unit sphere metric:
\begin{equation}
  \dd s^2=-e^{2\Phi(r)}\dd t^2+A(r)^{-1}\dd r^2+r^2\dd\Omega^2,
  \qquad
  A(r)=1-\frac{2m_{\mathrm{tot}}(r)}r=e^{-2\Lambda(r)}.
  \label{en:eq-metric}
\end{equation}
The functions \(\Phi\) and \(\Lambda\) are the temporal and radial metric
potentials, \(m_{\mathrm{tot}}(r)\) is the total mass enclosed by radius \(r\), and \(A\) is
the radial metric coefficient defined in \eqref{en:eq-metric}.  A prime
denotes \(\dd/\dd r\).
For later use, set
\begin{equation}
  \begin{gathered}
    m_{\mathrm{tot}}:=\sum_I m_I,
    \qquad p_{\mathrm{tot}}:=\sum_I p_I,
    \qquad \epsilon_{\mathrm{tot}}:=\sum_I\epsilon_I,\\
    h_I:=\epsilon_I+p_I,
    \qquad h_{\mathrm{tot}}:=\sum_Ih_I
      =\epsilon_{\mathrm{tot}}+p_{\mathrm{tot}},
    \qquad b_I:=h_Ic_{s,I}^2,
    \qquad b_{\mathrm{tot}}:=\sum_Ib_I.
  \end{gathered}
  \label{en:eq-total-matter-notation}
\end{equation}
Here \(m_I\) is the mass contribution of component \(I\), \(h_I\) is its
enthalpy density, and \(b_I\) is its adiabatic bulk coefficient.  The
equilibrium equations are~\cite{Tolman1939,OppenheimerVolkoff1939,ShapiroTeukolsky1983,Kain2020}
\begin{align}
  m_I'&=4\pi r^2\epsilon_I,
  \label{en:eq-tov-mass}\\
  p_I'&=-h_I
  \frac{m_{\mathrm{tot}}+4\pi r^3p_{\mathrm{tot}}}
       {r(r-2m_{\mathrm{tot}})},
  \label{en:eq-tov-pressure}\\
  \Phi'&=\frac{m_{\mathrm{tot}}+4\pi r^3p_{\mathrm{tot}}}
              {r(r-2m_{\mathrm{tot}})}.
  \label{en:eq-tov-phi}
\end{align}
Component \(I\) occupies \(0\le r<R_I\); it is active on \([0,R_I)\), with
\(R_I\) its component radius.  The total mass and particle
numbers are
\begin{align}
  M&=m_{\mathrm{tot}}(R_{\mathrm{out}})
  =4\pi\sum_I\int_0^{R_I}r^2\epsilon_I(r)\,\dd r,
  \label{en:eq-mass}\\
  N_I&=4\pi\int_0^{R_I}r^2A^{-1/2}n_I\,\dd r,
  \qquad R_{\mathrm{out}}=\max_I R_I.
  \label{en:eq-number}
\end{align}

Define the normalized relativistic enthalpy of component \(I\) by
\begin{equation}
  H_I(p):=\int_0^p\frac{\dd\widetilde p}
  {\epsilon_I(\widetilde p)+\widetilde p},
  \qquad H_I^c:=H_I\bigl(p_I(r=0)\bigr).
  \label{en:eq-enthalpy-definition}
\end{equation}
For the generated barotropic EoS in
Assumption~\ref{en:ass-regular}\conditionrefp{H1}, let
\(\mu_{I,0}:=\lim_{n\to0^+}\mu_I(n)\), whose existence and positivity are
required in \conditionref{H2}.  Writing
\(H_I(n):=H_I(p_I(n))\) for the enthalpy pulled back along the EoS, under
the vacuum normalization imposed in
\conditionref{H2}, the lower endpoint \(p_I=0\) is the density endpoint
\(n=0\), where
\(\left.\epsilon_I\right|_{n=0}=\left.p_I\right|_{n=0}=0\).
Since \(\dd p_I=n\,\dd\mu_I\) and
\(\epsilon_I+p_I=n\mu_I\), integration from this endpoint gives
\begin{equation}
  H_I(n)=\log\frac{\mu_I(n)}{\mu_{I,0}}.
  \label{en:eq-enthalpy-chemical-potential}
\end{equation}
At a vacuum surface, \(\mu_I=\mu_{I,0}\), so
\eqref{en:eq-enthalpy-chemical-potential} gives \(H_I=0\).  The central enthalpies
\begin{equation}
  \sigma=(H_I^c)_{I=1}^K\in(0,\infty)^K
  \label{en:eq-central-chart}
\end{equation}
serve as local equilibrium coordinates near the background point
\(\sigma_0\) fixed in \conditionref{H3} below.  The symbol
\(U\Subset(0,\infty)^K\) denotes the open neighborhood of \(\sigma_0\)
whose size is fixed in
Lemma~\ref{en:lem-equilibrium-package}: \(U\) is taken sufficiently small
that the equilibrium family and the regularity properties stated there hold
for every \(\sigma\in U\).  Relative compactness by itself is not being used
to infer those properties.  Define the
particle number map
\begin{equation}
  N=(N_I)_{I=1}^K:U\longrightarrow\mathbb R^K.
  \label{en:eq-number-map}
\end{equation}
For later use, set
\begin{equation}
  g_I(r)=4\pi r^2A^{-1/2}n_I(r),
  \qquad
  \mathcal N_I(r)=\int_0^r g_I(s)\,\dd s,
  \qquad N_I=\mathcal N_I(R_I).
  \label{en:eq-enclosed-number}
\end{equation}
Here \(g_I(r)\) is the particle number density per unit areal radius,
\(\mathcal N_I(r)\) is the particle number enclosed within radius \(r\), and
\(N_I=\mathcal N_I(R_I)\) is the total particle number of component \(I\).

\subsection{Perturbation variables and static criticality}
\label{en:sec-perturbation-notation}

Throughout the perturbation analysis, \(\delta\) denotes a first-order
Eulerian variation.  The symbol \(\delta R_I\) is the corresponding variation
of the labeled component radius.  For a cumulative quantity with such a moving
endpoint, \(\delta_E\mathcal N_I(r)\) denotes the Eulerian variation of the enclosed
number at fixed areal radius \(r\).  Along a family direction
\(v=v^a\partial_a\in T_\sigma U\), where \(\partial_a=\partial/\partial\sigma^a\),
define the directional derivative of any scalar family quantity \(F\) by
\(D_v F:=DF_\sigma[v]=v^a\partial_aF\), and write \(\delta_vF:=D_v F\) when the
same derivative is regarded as a perturbation.  In particular, we write
\(\delta_{E,v}\mathcal N_I(r):=D_v \mathcal N_I(r;\sigma)\) with
\(r\) fixed, and \(\delta_vR_I:=D_v R_I\) for the corresponding
labeled radius variation.  The total surface variation also contains the endpoint term:
\(\delta N_I=\delta_E\mathcal N_I(R_I)+g_I(R_I)\delta R_I\).

Let \(\xi_I(t,r)\) be the radial Lagrangian displacement and define the reduced
displacement vector \(\zeta=(\zeta_I)_{I=1}^K\) by
\begin{equation}
  \zeta_I=r^2e^{-\Phi}\xi_I.
  \label{en:eq-zeta}
\end{equation}
For a normal mode proportional to \(e^{i\omega t}\), \(\omega\) denotes the
mode frequency.  We denote by \(\mathcal L_\sigma\) the radial differential
expression obtained from the Einstein--Euler pulsation equations and by
\(\mathcal W_\sigma\) the positive multiplication operator defined in
\eqref{en:eq-generalized-mode}.
The full dynamical equations reduce to
\begin{equation}
  \mathcal L_\sigma\zeta
  =\omega^2\mathcal W_\sigma\zeta,
  \qquad
  (\mathcal W_\sigma\zeta)_I
  =\frac{e^{3\Lambda+\Phi}(\epsilon_I+p_I)}{r^2}\zeta_I.
  \label{en:eq-generalized-mode}
\end{equation}
The explicit recovery formulas and the resulting differential expression for
\(\mathcal L_\sigma\) are given in
Section~\ref{en:app-proof-reduction-package}.

\begin{definition}[Static criticality]
\label{en:def-static-critical}
An equilibrium \(\sigma\in U\) is statically critical when
\begin{equation}
  \rankop DN_\sigma<K.
  \label{en:eq-rank-deficiency}
\end{equation}
Since \(\dim T_\sigma U=K\), this is equivalent to
\(\ker DN_\sigma\ne\{0\}\).
Here \(T_\sigma U\) is the tangent space of the equilibrium chart.  In local
equilibrium coordinates, the Fr\'echet differential of \(N\) at
\(\sigma\) is the unique linear map characterized by
\begin{equation}
  \begin{gathered}
    DN_\sigma:T_\sigma U\longrightarrow\mathbb R^K,\\
    N(\sigma+v)
    =N(\sigma)+DN_\sigma[v]+o(\lVert v\rVert),
    \qquad v\to0.
  \end{gathered}
\end{equation}
\end{definition}

\subsection{Physical and EoS assumptions}
\label{en:sec-physical-hypotheses}

\begin{assumption}[Independent physical hypotheses]
\label{en:ass-regular}
Fix a background equilibrium configuration \(\sigma_0\), and assume the
following.
\begin{enumerate}
\renewcommand{\theenumi}{H\arabic{enumi}}
\renewcommand{\labelenumi}{(\theenumi)}
  \item \conditiontarget{H1}
  \emph{Constitutive and coupling model.}
  The \(K\) components are isentropic barotropic perfect fluids that are
  independently minimally coupled to the common metric, with no
  nongravitational exchange of particles or stress--energy between
components.  Accordingly, for
  \(T_I^{\mu\nu}=(\epsilon_I+p_I)u_I^\mu u_I^\nu+p_Ig^{\mu\nu}\),
  \begin{equation}
    J_I^\mu=n_Iu_I^\mu,\qquad
    \nabla_\mu J_I^\mu=0,\qquad
    \nabla_\mu T_I^{\mu\nu}=0,\qquad I=1,\ldots,K.
    \label{en:eq-current}
  \end{equation}
  For component \(I\), a fixed energy function
  \(\epsilon_I=\epsilon_I(n)\), with \(n\) its density argument, defines
  \begin{equation}
    \mu_I(n)=\frac{\dd\epsilon_I}{\dd n},\qquad
    p_I(n)=n\mu_I(n)-\epsilon_I(n).
    \label{en:eq-generated-eos}
  \end{equation}
  Choose \(n_{I,\max}>n_I(0;\sigma_0)\) for each label and restrict the
  local family so that all of its central densities lie below these bounds.
  On the density interval sampled by this construction assume
  \begin{equation}
    \begin{gathered}
      \left.\epsilon_I\right|_{[n_-,n_+]}
      \in C^3([n_-,n_+])
      \quad\text{for every }0<n_-<n_+<n_{I,\max},\\
      \mu_I(n)>0,\qquad
      0<\frac{n\mu_I'(n)}{\mu_I(n)}=c_{s,I}^2<1,
      \qquad 0<n<n_{I,\max},
    \end{gathered}
    \label{en:eq-microphysical-stability}
  \end{equation}
  where the prime in this item denotes differentiation with respect to \(n\).
  All positive-density assertions below refer to this interval.

  \item \conditiontarget{H2}
  \emph{Vacuum-scale regular polytropic law.}
  The EoS is vacuum-normalized at zero density: \(\epsilon_I\) and \(p_I\),
  viewed as functions of the density argument, extend continuously to
  \(n=0\) with
  \(\left.\epsilon_I\right|_{n=0}=\left.p_I\right|_{n=0}=0\), and
  the previously defined limit \(\mu_{I,0}\) belongs to \((0,\infty)\).
  In the law below, \(H_I\) is the enthalpy normalized at this vacuum
  endpoint: the lower limit \(p=0\) in
  \eqref{en:eq-enthalpy-definition} corresponds to \(n=0\), so
  \(H_I(n)\to0\) as \(n\to0^+\).  This choice fixes the additive constant
  in the enthalpy coordinate.
  Together with H1, this normalization implies
  \(\epsilon_I(n)>0\) and \(p_I(n)>0\) for \(n>0\), and \(n=0\) is the
  unique zero-pressure state.
  For some \(H_{I,*},C_I,\alpha_I,\eta_I>0\) fixed by the low-density
  microphysics,
  \begin{equation}
    n_I(H_I)=C_IH_I^{\alpha_I}\{1+q_I(H_I)\},
    \qquad 0<H_I\le H_{I,*},
    \label{en:eq-eos-vacuum-asymptotic}
  \end{equation}
  where \(1+q_I>0\),
  \(q_I\in C^\infty((0,H_{I,*}])\cap C^0([0,H_{I,*}])\), \(q_I(0)=0\), and,
  for every integer \(k\ge0\),
  \begin{equation}
    \left|(H_I\partial_{H_I})^kq_I(H_I)\right|
    \le C_{I,k}H_I^{\eta_I}.
    \label{en:eq-eos-conormal-regularity}
  \end{equation}
  These bounds express smoothness of the normalized EoS on the logarithmic
  vacuum scale; the
  estimates used below require only \(0\le k\le3\).  We call a component
  endpoint satisfying this vacuum law an \emph{ordinary vacuum
  endpoint}; there the particle number density, energy density, pressure, and
  enthalpy density \(\epsilon_I+p_I\) all tend to zero according to the
  controlled power laws derived below.  Fractional positive-power corrections
  are allowed, so ordinary \(C^\infty\) extendibility at \(H_I=0\) is not
  required.

  \item \conditiontarget{H3}
\emph{Regular background and junction model.}
The background is a regular, static, spherically symmetric, asymptotically
flat, horizonless equilibrium, with all matter contained inside a finite
outer radius.  The model has no internal phase transition surfaces and
assigns no independent degrees of freedom to an interface.  Each component
terminates at its own labeled surface \(R_I\).  Across each
such surface \(\Sigma_I:=\{r=R_I\}\), the no-thin-shell Darmois--Israel
junction conditions are
\begin{equation}
  [\gamma_{ab}]_{\Sigma_I}=0,
  \qquad
  [K_{ab}]_{\Sigma_I}=0.
  \label{en:eq-background-darmois-junction}
\end{equation}
Here \(\gamma_{ab}\) and \(K_{ab}\) are the induced metric and extrinsic
curvature of \(\Sigma_I\), respectively, and \([X]_{\Sigma_I}\) denotes the
difference between the two one-sided limits of \(X\).  Equivalently, the
Israel surface stress tensor vanishes, so the surface carries no thin shell
or delta function source.
\end{enumerate}
\end{assumption}

Assumption~\ref{en:ass-regular}\conditionrefp{H3} is imposed at the
background point \(\sigma_0\).  Lemma~\ref{en:lem-equilibrium-package}
shows that \(U\) may be taken sufficiently small about \(\sigma_0\) so that
all equilibria in \(U\) retain the stated regularity, finite labeled
surfaces, and no-horizon property.  Whenever a proof below decreases \(U\)
further, it means replacing \(U\) by a smaller open neighborhood of
\(\sigma_0\) with compact closure in \((0,\infty)^K\).

\section{Regularity of the equilibrium family}
\label{en:app-equilibrium-structure}
\subsection{Labeled surfaces and fixed-domain state space}
\label{en:sec-equilibrium-state-space}

\phantomsection
\label{en:def-surface-decomposition}
\smallskip\noindent\textbf{Surface decomposition and piecewise
regularity.}
For a fixed equilibrium star, let
\begin{equation}
  0<\mathcal R_1<\cdots<\mathcal R_q=R_{\mathrm{out}},\qquad
  \mathcal B_a:=\{I:R_I=\mathcal R_a\},\qquad
  \mathcal A_a^+:=\{J:R_J>\mathcal R_a\}
  \label{en:eq-surface-blocks}
\end{equation}
be the distinct component radii, the labels ending there, and the components
surviving that radius, respectively.  We call \(\mathcal B_a\) the
\emph{ending block} and each distinct radius \(\mathcal R_a\) a
\emph{surface node}.  Put \(\mathcal R_0:=0\) and call
\(\mathcal I_a:=(\mathcal R_{a-1},\mathcal R_a)\), \(a=1,\ldots,q\),
the open active cells, and define their active label sets by
\begin{equation}
  \mathcal K_a:=\{I:\mathcal I_a\subset(0,R_I)\}
  =\{I:R_I\ge\mathcal R_a\}.
  \label{en:eq-active-label-set}
\end{equation}
A component \(I\) is present on the cells for which
\(I\in\mathcal K_a\).  Unless stated otherwise, fluid sums on \(\mathcal I_a\)
run over \(\mathcal K_a\); equivalently, local matter fields and matter-weighted
products are zero-extended, but bare displacements are not.  A field is
\emph{piecewise classical} when it has the stated regularity on each active
cell; derivatives need not match at an ending block.  Whenever a field is
evaluated at an endpoint of an active cell, its value means the finite
one-sided limit taken from within that cell.  At an internal ending block,
the left and right one-sided limits are required to agree whenever a matching
condition is imposed.

\phantomsection
\label{en:def-fixed-state-space}
\smallskip\noindent\textbf{Fixed-domain equilibrium state space.}
The local equilibrium state space \(\mathcal X_{\mathrm{eq}}\) consists of tuples
\begin{equation}
  x=
  \bigl((H_I,m_I)_{I=1}^K,\Phi;(R_I)_{I=1}^K,M\bigr).
  \label{en:eq-state-tuple}
\end{equation}
On the neighborhood constructed in Lemma
~\ref{en:lem-equilibrium-package}, choose
\(R_*>\sup_{\overline U}R_{\mathrm{out}}\).  For each \(I\), define the constant
extension
\begin{equation}
  \widetilde m_I(r):=m_I(\min\{r,R_I\}),
  \qquad 0\le r\le R_*.
\end{equation}
Restrict the normalized \(\Phi\) to \([0,R_*]\) and zero-extend the matter
fields.  We represent a state by the fixed-domain chart
\begin{equation}
  J_{\mathrm{fd}}(x):=\Bigl(
  \bigl(H_I(R_Is)\bigr)_{I=1}^K,
  \bigl(\widetilde m_I\bigr)_{I=1}^K,
  \Phi|_{[0,R_*]};
  \bigl(R_I\bigr)_{I=1}^K,M,
  \bigl(n_I^0,\epsilon_I^0,p_I^0,g_I^0\bigr)_{I=1}^K
  \Bigr)
  \label{en:eq-fixed-state-chart}
\end{equation}
in the Banach space
\begin{equation}
 \mathcal X_{\mathrm{fd}}:=[C^1([0,1])]^K\times[C^0([0,R_*])]^K\times C^1([0,R_*])
 \times\mathbb R^{K+1}\times[L^1_{\mathrm{fd}}(0,R_*)]^{4K}.
\end{equation}
Here the superscript \(0\) denotes the zero extensions just specified, and
\(\|f\|_{L^1_{\mathrm{fd}}}:=\int_0^{R_*}(1+r^2)|f(r)|\,\dd r\).  The subscript
\({\mathrm{fd}}\) indicates the fixed-domain norm.  The space
\(\mathcal X_{\mathrm{eq}}\) carries the subspace topology induced by \(J_{\mathrm{fd}}\), and
all assertions that a family is \(C^1\) are assertions in this fixed Banach
chart.  Equivalently, enthalpies are compared after the labelwise pullback
\(r=sR_I(\sigma)\), \(0\le s\le1\), in uniform \(C^1\), while the extended
mass and lapse profiles are compared in the fixed-domain norms defined above.
The fixed-domain \(L^1\) components allow the endpoint behavior
\(O((R_I-r)^{\alpha_I-1})\).
The central data map \(\pi_{\mathrm{cen}}:\mathcal X_{\mathrm{eq}}\to\mathbb R^K\) is
\(\pi_{\mathrm{cen}}(x)=(H_I(0))_{I=1}^K\).  In this chart, the physical
center values encode the parameter \(\sigma\).

\begin{lemma}[Equilibrium structure, endpoints, and parameter regularity]
\label{en:lem-equilibrium-package}
Under Assumption~\ref{en:ass-regular}, \(U\) can be taken sufficiently small
about \(\sigma_0\), with \(U\Subset(0,\infty)^K\), so that the
following hold for every \(\sigma\in U\).
\begin{enumerate}
\renewcommand{\theenumi}{\alph{enumi}}
\renewcommand{\labelenumi}{(\theenumi)}
  \item \label{en:item-eos-consequences}
Put \(\beta_I=\min\{1,\eta_I\}\).  As \(H_I\to0^+\),
  \begin{align}
    h_I(H_I)=\epsilon_I(H_I)+p_I(H_I)
      &=\mu_{I,0}C_IH_I^{\alpha_I}[1+O(H_I^{\beta_I})],
    \label{en:eq-enthalpy-density-H-asymptotic}\\
    \epsilon_I(H_I)&=\mu_{I,0}C_IH_I^{\alpha_I}[1+O(H_I^{\beta_I})],
    \label{en:eq-epsilon-H-asymptotic}\\
    p_I(H_I)&=\frac{\mu_{I,0}C_I}{\alpha_I+1}
      H_I^{\alpha_I+1}[1+O(H_I^{\beta_I})],
    \label{en:eq-pressure-H-asymptotic}\\
    c_{s,I}^2(H_I)&=\frac{H_I}{\alpha_I}[1+O(H_I^{\eta_I})].
    \label{en:eq-sound-speed-H-asymptotic}
  \end{align}
  Combining the preceding asymptotics and setting \(b_I=h_Ic_{s,I}^2\) gives
  \begin{equation}
    b_I(H_I)=\frac{\mu_{I,0}C_I}{\alpha_I}
      H_I^{\alpha_I+1}[1+O(H_I^{\beta_I})].
    \label{en:eq-b-H-asymptotic}
  \end{equation}
  After factoring out the leading powers in
  \eqref{en:eq-enthalpy-density-H-asymptotic}--\eqref{en:eq-b-H-asymptotic},
  the corresponding remainders and all of their Euler derivatives
  \((H_I\partial_{H_I})^k\) obey the analogous positive-power bounds
  (with exponent \(\beta_I\), or \(\eta_I\) for the relative sound speed
  remainder).
  Moreover, \(n_I,\epsilon_I,p_I\to0\),
  \(\mu_I=\mu_{I,0}e^{H_I}\ge\mu_{I,0}\), and
  \(n_I/(\epsilon_I+p_I)=1/\mu_I\) is bounded.  The vacuum normalization
  and \conditionref{H1} also give \(\epsilon_I,p_I>0\) whenever \(H_I>0\), while
  \(p_I=0\) only at \(H_I=0\).

  \item \label{en:item-monotone-surfaces}
  At the center and on every active cell,
  \begin{equation}
    H_I'(0)=0,
    \qquad
    H_I'(r)=\frac{p_I'}{\epsilon_I+p_I}=-\Phi'(r)<0
    \quad(0<r<R_I).
    \label{en:eq-enthalpy-monotonicity}
  \end{equation}
  Each support is \([0,R_I)\), and
  \begin{equation}
    H_I'(R_I)=-\Phi'(R_I)<0.
    \label{en:eq-transverse-surface}
  \end{equation}
  Every labeled \(R_I(\sigma)\) is \(C^1\).

  \item \label{en:item-surface-tail}
  With \(x=R_I-r\), \(\kappa_I=\Phi'(R_I)>0\), and
  \(g_I=4\pi r^2A^{-1/2}n_I\), uniformly on \(\overline U\),
  \begin{align}
    H_I&=\kappa_Ix+O(x^2),
    \label{en:eq-H-surface-expansion}\\
    n_I&=C_I\kappa_I^{\alpha_I}x^{\alpha_I}
      [1+O(x^{\beta_I})],
    \label{en:eq-n-surface-expansion}\\
    g_I&=a_Ix^{\alpha_I}+g_I^{\mathrm{rem}},
    \label{en:eq-g-derived-expansion}
  \end{align}
  where
  \begin{equation}
    a_I=4\pi R_I^2A(R_I)^{-1/2}C_I\kappa_I^{\alpha_I}>0,
    \label{en:eq-surface-leading-coefficient}
  \end{equation}
  and, with the parameter derivatives below taken at fixed \(r\),
  \begin{equation}
    g_I^{\mathrm{rem}}=O(x^{\alpha_I+\beta_I}),\qquad
    \partial_{\sigma^a}g_I^{\mathrm{rem}}
    =O(x^{\alpha_I+\beta_I-1}).
    \label{en:eq-derived-remainder-control}
  \end{equation}
  For \(N_I^{\mathrm{tail}}(r)=\int_r^{R_I}g_I\,\dd u\),
  \begin{equation}
    D_v N_I^{\mathrm{tail}}=g_I D_v R_I+o(g_I),
    \qquad
    D_v N_I=0\Longrightarrow
    -\frac{\delta_{E,v}\mathcal N_I}{g_I}\to D_v R_I.
    \label{en:eq-derived-surface-quotient}
  \end{equation}

  \item \label{en:item-crossing-regularity}
  Zero extension to \([0,R_*]\) defines \(C^1\) maps
  \begin{equation}
    \sigma\longmapsto f_I(\cdot;\sigma)
    \quad\text{into weighted }L^1,
    \qquad f_I\in\{n_I,\epsilon_I,p_I,g_I\},
    \label{en:eq-zero-extension-map}
  \end{equation}
  with parameter derivatives continuous in the corresponding fixed-domain norms.

  \item \label{en:item-map-regularity}
  There is a unique \(C^1\) equilibrium embedding
  \(X:U\to\mathcal X_{\mathrm{eq}}\), \(\pi_{\mathrm{cen}}\circ X=\operatorname{Id}_U\), and \(M,N\)
  are \(C^1\) on \(U\).  Moreover,
  \begin{equation}
    \inf_{\substack{\sigma\in\overline U\\0\le r\le R_{\mathrm{out}}(\sigma)}}
    A(r;\sigma)>0,
    \qquad rA^{-3/2}n_I\in L^1(0,\infty),
    \label{en:eq-derived-A-and-kernel}
  \end{equation}
\end{enumerate}
\end{lemma}

\setproofparent{en:lem-equilibrium-package}

\subsection{Vacuum thermodynamics: proof of part (a)}

Fix one component and suppress its label.  By \conditionref{H2}, the density endpoint
\(n=0\) is the zero-pressure vacuum and
\(\lim_{n\downarrow0}\epsilon(n)=\lim_{n\downarrow0}p(n)=0\).
Since \(\mu(n)\to\mu_0\) there,
\begin{equation*}
  \epsilon(n)=\int_0^n\mu(s)\,\dd s,
  \qquad
  p(n)=n\mu(n)-\int_0^n\mu(s)\,\dd s
      =\int_0^n s\mu'(s)\,\dd s.
\end{equation*}
The integral representations imply that \(\epsilon(n)>0\) and \(p(n)>0\)
for \(n>0\), and that \(p=0\) only at \(n=0\).  For \(n>0\),
\[
  \frac{\dd p}{\epsilon+p}
  =\frac{n\,\dd\mu}{n\mu}=\dd\log\mu.
\]
Integrating from the vacuum endpoint and using \(\mu(n)\to\mu_0\) gives
\(H=\log(\mu/\mu_0)\).  Together with the thermodynamic relations, this yields
\begin{equation}
  \dd\epsilon=\mu\,\dd n,\qquad
  \mu=\mu_0e^H,
  \qquad \frac{\dd p}{\dd H}=n\mu=\epsilon+p,
  \label{en:eq-dp-dH}
\end{equation}
\begin{align*}
 p(H)&=\mu_0C\int_0^H s^\alpha e^s[1+q(s)]\,\dd s,
 & h&=n\mu,
 & \epsilon&=h-p,\\
 c_s^2&=\frac{n}{\partial_Hn}
 =\frac{H(1+q)}{\alpha(1+q)+H\partial_Hq}.
\end{align*}
Using the conormal bounds in \conditionref{H2}, the product rules, and the rescaling identity
\(\int_0^H s^\gamma F(s)\,\dd s
 =H^{\gamma+1}\int_0^1t^\gamma F(tH)\,\dd t\),
we obtain \eqref{en:eq-enthalpy-density-H-asymptotic}--
\eqref{en:eq-b-H-asymptotic}, including all stated Euler derivative bounds.
Finally, \(n/(\epsilon+p)=1/\mu\le1/\mu_0\).  This proves part~\textup{(a)}.

As the equilibrium parameters vary, the vacuum surfaces of the individual
components move.  Near each ordinary vacuum endpoint, the matter profiles
vanish to positive order.  Combined with the conormal bounds above, this
makes their zero extensions \(C^1\) in the parameters as \(L^1\)-valued maps
on a fixed interval.  The parameter dependence of the associated Volterra
integral equations then transfers this regularity to the full equilibrium
solution.  The following lemma makes these two steps precise.
\begin{sublemma}[Shifted conormal profiles and Volterra dependence]
\label{en:lem-shifted-conormal-dependence}
Let \(J\) be compact, let \(\mathcal O\) be a finite-dimensional open
parameter set, let \(\ell\in C^1(\mathcal O)\), and
\(F_\theta(y)=y^\gamma a(y,\theta)\) for \(y>0\), \(\theta\in\mathcal O\),
extended by zero for
\(y\le0\), where \(\gamma>0\).  Assume that \(a,y\partial_ya\) and their
first parameter derivatives are continuous and locally bounded in the
corresponding conormal \(L^\infty\) norms, meaning the joint sup norms of
these listed functions on the relevant compact \(y\)-ranges.  Write
\([x]_+:=\max\{x,0\}\).  Then
\(\theta\mapsto F_\theta([\ell(\theta)-\cdot]_+)\) is \(C^1\) into
\(L^1(J)\), with
\begin{equation}
 D_v[F_\theta([\ell-s]_+)]=(D_v F_\theta)([\ell-s]_+)
 +F_\theta'(\ell-s)D_v \ell\,\mathbf 1_{\{s<\ell\}}.
  \label{en:eq-shifted-conormal-derivative}
\end{equation}
The assertion is stable under bounded \(C^1\) prefactors and finite sums.
Moreover, for a fixed \(t_0\in J\), a Carath\'eodory--Volterra family
\[
Z(t;\theta)=Z_0(\theta)+\int_{t_0}^{t}
F_{\mathrm{V}}(s,Z(s;\theta),\theta)\,\dd s
\]
is \(C^1\) in
\(C^0(J)\) whenever \(Z_0\) and the associated Nemytskii map into \(L^1(J)\)
are \(C^1\), and \(|D_ZF_{\mathrm{V}}|\le a(t)\) on the relevant tube for some
\(a\in L^1(J)\).
Here the relevant tube is a compact neighborhood in \((t,Z,\theta)\)-space
containing the reference solution graph and the nearby solution graphs under consideration.
\end{sublemma}

\begin{proof}
After a harmless cutoff, \(F_\theta\in W^{1,1}(\mathbb R)\) because
\(F_\theta'=O(y^{\gamma-1})\), and
\(D_yF_\theta=F_\theta'\mathbf 1_{\{y>0\}}\) has no boundary delta since
\(F_\theta(0)=0\).  The conormal assumptions give \(C^1\) dependence in
\(W^{1,1}\); differentiability of translations
\(W^{1,1}\times\mathbb R\to L^1\)
\cite{BrezisFA} yields \eqref{en:eq-shifted-conormal-derivative}.  For the
Volterra claim, integration maps \(L^1\) to \(C^0\).  Differentiating the
difference quotients and applying Gronwall's inequality~\cite{WalterODE}
yield the unique variational equation and continuous dependence on the
parameters.
\end{proof}

\subsection{Monotonicity and labeled surfaces: proof of part (b)}

By part~\textup{(a)}, every matter profile is nonnegative and is strictly
positive on its active region.  Hence the total pressure is nonnegative;
the positive central enthalpies also give \(m_{\mathrm{tot}}(r)>0\) for \(r>0\).
From \eqref{en:eq-tov-pressure}, \eqref{en:eq-tov-phi}, and
\eqref{en:eq-dp-dH},
\(H_I'=p_I'/(\epsilon_I+p_I)=-\Phi'\).  Center regularity gives
\(H_I'(0)=0\); for \(r>0\), the no-horizon condition
\(r-2m_{\mathrm{tot}}>0\) then implies \(\Phi'>0\), hence \(H_I'<0\).  Strict constitutive
monotonicity therefore gives the connected supports.  At every endpoint the
one-sided fields are finite and the same formula gives
\(H_I'(R_I)=-\Phi'(R_I)<0\), proving transversality.

Put \(\psi=\Phi-\Phi(0)\).  The fixed-domain construction in
Lemma~\ref{en:lem-fixed-domain-parameter-regularity} first proves
\(\psi\in C^1(U;C^1)\), independently of parts~\textup{(c)}--\textup{(d)}.
Since \(\widehat H_I=H_I^c-\psi\), it is jointly \(C^1\); the labelwise
implicit function theorem then gives \(R_I\in C^1\), also at coincident zeros.

\subsection{Surface and tail expansions: proof of part (c)}\label{sup:sec-surface-tail}

Using this \(C^1\) conclusion, put \(x=R_I-r\), \(B_I=4\pi r^2A^{-1/2}\), and
\(\kappa_I=\Phi'(R_I)>0\).  From \(H_I'=-\Phi'\),
\(H_I=\kappa_Ix+O(x^2)\), while fixed-radius differentiation of
\(H_I(R_I(\sigma);\sigma)=0\) gives
\begin{equation}
\begin{aligned}
  0&=(D_v H_I)(R_I)+H_I'(R_I)D_v R_I,\\
  D_v H_I(r)&=\kappa_I D_v R_I+O(x).
\end{aligned}
\label{en:app-fixed-r-enthalpy-derivative}
\end{equation}
With \(a_I\) from \eqref{en:eq-surface-leading-coefficient},
\conditionref{H2} gives
\begin{align*}
  \frac{g_I}{a_Ix^{\alpha_I}}
  &=\frac{B_I(r)}{B_I(R_I)}
   \left(\frac{H_I}{\kappa_Ix}\right)^{\alpha_I}
   [1+q_I(H_I)]
  =1+\vartheta_I,\\
  \vartheta_I&=O(x^{\beta_I}),
  & D_v \vartheta_I&=O(x^{\beta_I-1}),
\end{align*}
where \(\beta_I=\min\{1,\eta_I\}\); the last estimate follows from the
conormal chain rule and the preceding fixed-radius identity.  Setting
\(g_I^{\mathrm{rem}}=a_Ix^{\alpha_I}\vartheta_I\), we obtain
\[
 D_v g_I^{\mathrm{rem}}=O(x^{\alpha_I+\beta_I-1}),\qquad
 D_v g_I=\alpha_Ia_Ix^{\alpha_I-1}D_v R_I
 +(D_v a_I)x^{\alpha_I}+D_v g_I^{\mathrm{rem}},
\]
This proves \eqref{en:eq-g-derived-expansion}--
\eqref{en:eq-derived-remainder-control}; the expansions for
\(n_I,\epsilon_I,p_I\) are already \eqref{en:eq-n-surface-expansion} and
part~\textup{(a)}.

For \(N_I^{\mathrm{tail}}=\int_r^{R_I}g_I\), integration and the
moving-endpoint rule give
\[
\begin{aligned}
 N_I^{\mathrm{tail}}
   &=\frac{a_I}{\alpha_I+1}x^{\alpha_I+1}
     +O(x^{\alpha_I+\beta_I+1}),\\
 D_v N_I^{\mathrm{tail}}
   &=g_I D_v R_I+O(x^{\alpha_I+\beta_I})
     =g_I D_v R_I+o(g_I).
\end{aligned}
\]
Since \(N_I=\mathcal N_I+N_I^{\mathrm{tail}}\), \(D_v N_I=0\) yields
\begin{equation}
 \delta_{E,v}\mathcal N_I=-g_I D_v R_I+o(g_I),
 \qquad -\frac{\delta_{E,v}\mathcal N_I}{g_I}\longrightarrow D_v R_I,
 \label{en:app-tail-quotient-check}
\end{equation}
which is \eqref{en:eq-derived-surface-quotient}.  All estimates are uniform on
\(\overline U\), including coincident surfaces.

\subsection{Crossing-compatible zero extension: proof of part (d)}

It suffices to treat one endpoint.  Every profile is a finite sum of terms of
the form obtained below.  Write \(f_{\mathrm{rem}}^+\) for the zero extension
of a remainder satisfying the estimates in part~\textup{(c)}:
\begin{equation}
  f(r;\sigma)=b(r;\sigma)[R(\sigma)-r]_+^\gamma
  +f_{\mathrm{rem}}^+(r;\sigma),
  \qquad \gamma>0,
  \label{en:app-crossing-profile}
\end{equation}
Here \(b\) is \(C^1\), and \(\gamma=\alpha_I+1\) for pressure while
\(\gamma=\alpha_I\) for
\(n_I,\epsilon_I,g_I\).  Its fixed-domain \(L^1\) parameter derivative is
\begin{equation}
  \partial_af=(\partial_ab)[R-r]_+^\gamma
  +\gamma b(R-r)^{\gamma-1}\mathbf 1_{\{r<R\}}\partial_aR
  +\partial_af_{\mathrm{rem}}^+.
  \label{en:app-zero-extension-derivative}
\end{equation}
Here the parameter derivative is taken with \(r\) held fixed.
The potentially singular term is integrable exactly because \(\gamma>0\),
and the lemma proves continuity of this derivative.  There is no surface
delta because the zero-extended profile itself vanishes at \(R\).  Set
\(R_\tau:=R(\sigma+\tau v)\) and \(R_{\mathrm{bg}}:=R(\sigma)\).  Since
\(R_\tau=R_{\mathrm{bg}}+O(\tau)\), the noncommon layer satisfies
\begin{equation}
  \frac1{|\tau|}
  \int_{\min\{R_\tau,R_{\mathrm{bg}}\}}^{\max\{R_\tau,R_{\mathrm{bg}}\}}
  |R_\tau-r|^\gamma\,\dd r
  =O(|\tau|^\gamma)\longrightarrow0.
  \label{en:app-thin-layer-estimate}
\end{equation}
The remainder is smaller, so the layer is \(o(|\tau|)\);
\eqref{en:app-zero-extension-derivative} then proves Fr\'echet
differentiability.
Applying the argument to each label proves part (d), also at coincidences and
order exchanges.

\subsection{Independent fixed-domain construction: proof of part (e)}

Independently of parts~\textup{(c)}--\textup{(d)}, set
\(\psi=\Phi-\Phi(0)\) and
\begin{equation}
 H_I=H_I^c-\psi,\qquad
 P(\psi;\sigma):=\sum_Ip_I([H_I^c-\psi]_+),\qquad
 E(\psi;\sigma):=\sum_I\epsilon_I([H_I^c-\psi]_+).
 \label{en:app-global-positive-part-matter}
\end{equation}
Evaluating \(H_I=H_I^c-\psi\) at \(r=R_I\) gives
\(\psi(R_I)=H_I^c\).  Writing \(m_{\mathrm{tot}}=r^3\bar m_{\mathrm{tot}}\)
at the center gives
\begin{align}
 \bar m_{\mathrm{tot}}(r)&=4\pi\int_0^1x^2E(\psi(rx);\sigma)\,\dd x,
 \label{en:eq-center-Volterra-mass}\\
 \psi'(r)&=r\frac{\bar m_{\mathrm{tot}}+4\pi P(\psi;\sigma)}{1-2r^2\bar m_{\mathrm{tot}}},
 \qquad \psi(0)=0.
 \label{en:eq-center-Volterra-H}
\end{align}
All components are active near the center, so standard parameter-dependent
Volterra theory gives a unique \(C^1\) family there.  Since \(\psi'>0\) for
\(r>0\), inversion gives
\begin{align}
 \frac{\dd r}{\dd\psi}
 &=\frac{r(r-2m_{\mathrm{tot}})}{m_{\mathrm{tot}}+4\pi r^3P(\psi;\sigma)},
 \label{en:app-psi-radius-ode}\\
 \frac{\dd m_{\mathrm{tot}}}{\dd\psi}
 &=4\pi r^2E(\psi;\sigma)\frac{\dd r}{\dd\psi}.
 \label{en:app-psi-mass-ode}
\end{align}

For the reference solution, Schwarzschild matching gives
\[
 \psi_{\infty,0}=\max_IH_I^c(\sigma_0)
 -\frac12\log\!\left(1-\frac{2M_0}{R_{{\mathrm{out}},0}}\right)
 >\max_IH_I^c(\sigma_0).
\]
Choose \(\max_IH_I^c(\sigma_0)<\psi_*<\psi_{\infty,0}\).  The reference
solution therefore reaches \(\psi_*\) at a finite vacuum radius.  Then choose
\(0<\psi_0<\min_IH_I^c(\sigma_0)\).  On \([\psi_0,\psi_*]\), the
  shifted-profile lemma gives the positive-part sources \(P,E\).  The reference
trajectory lies in the open region
\(r>0\), \(m_{\mathrm{tot}}+4\pi r^3P>0\), \(r-2m_{\mathrm{tot}}>0\); hence the parameter-dependent
 Carath\'eodory theorem on a compact tube yields solutions on this common
interval when \(U\) is taken sufficiently small about \(\sigma_0\), with
\[
 0<\psi_0<\min_{\overline U,I}H_I^c
 \le\max_{\overline U,I}H_I^c<\psi_*.
\]
At \(\psi_*\) all sources vanish.  If
\(r_*(\sigma)=r(\psi_*;\sigma)\) and
\(M_*(\sigma)=m_{\mathrm{tot}}(\psi_*;\sigma)\), exterior matching gives
\[
 \psi_\infty(\sigma)=\psi_*
 -\frac12\log\!\left(1-\frac{2M_*(\sigma)}{r_*(\sigma)}\right)>\psi_*.
\]
For this choice of \(U\), continuity on the compact set \(\overline U\)
makes this inequality uniform.

\begin{sublemma}[Fixed-domain parameter regularity]
\label{en:lem-fixed-domain-parameter-regularity}
For \(\sigma\in U\), the equilibrium profiles satisfy
\begin{align}
 \sigma&\longmapsto p_I^0(\,\cdot\,;\sigma)
   \in C^1\bigl(U;C^0([0,R_*])\bigr),
 \label{en:eq-fixed-pressure-regularity}\\
 \sigma&\longmapsto(n_I^0,\epsilon_I^0,g_I^0)(\,\cdot\,;\sigma)
   \in C^1\bigl(U;[L^1_{\mathrm{fd}}(0,R_*)]^3\bigr),
 \label{en:eq-fixed-matter-regularity}\\
 \sigma&\longmapsto\widetilde m_I(\,\cdot\,;\sigma)
   \in C^1\bigl(U;C^0([0,R_*])\bigr),\qquad
 \sigma\longmapsto\Phi(\,\cdot\,;\sigma)
   \in C^1\bigl(U;C^1([0,R_*])\bigr).
 \label{en:eq-fixed-metric-regularity}
\end{align}
Moreover, for the pulled-back enthalpy profile in
\eqref{en:eq-pulled-enthalpy-profile},
\begin{equation}
  \sigma\longmapsto H_I^{\mathrm{pb}}(\,\cdot\,;\sigma)
  \in C^1\bigl(U;C^1([0,1])\bigr).
  \label{en:eq-pulled-enthalpy-regularity}
\end{equation}
These regularity conclusions imply that \(J_{\mathrm{fd}}\circ X\), \(M\),
and \(N\) are \(C^1\).
\end{sublemma}

\begin{proof}
By part~\textup{(a)}, the positive-part pressure is \(C^1\) at \(H_I=0\),
whereas \(n_I,\epsilon_I\) and their first \(H_I\)-derivatives have the
integrable orders \(O(H_I^{\alpha_I})\) and
\(O(H_I^{\alpha_I-1})\).  Lemma~\ref{en:lem-shifted-conormal-dependence}
therefore gives \(P\in C^1(U;C^0)\) and \(E\in C^1(U;L^1)\) on
\([\psi_0,\psi_*]\), including moving thresholds.

Let \(Z=(r,m_{\mathrm{tot}})\), let \(F_1\) denote the right side of
\eqref{en:app-psi-radius-ode}.  On the common tube its denominator is bounded
away from zero, multiplication \(C^0\times L^1\to L^1\) is bounded, and
 with \(F_{\mathrm{eq}}(Z,\sigma):=(F_1,4\pi r^2EF_1)\),
\begin{equation}
 \|D_ZF_{\mathrm{eq}}(Z,\sigma)[Z_1]\|_{L^1}
 \le C(1+\|E(\cdot;\sigma)\|_{L^1})\|Z_1\|_{C^0}.
 \label{en:eq-fixed-domain-state-Lipschitz}
\end{equation}
The Volterra part of Lemma~\ref{en:lem-shifted-conormal-dependence} gives
\(Z\in C^1(U;C^0)\), and the first equation gives
\(r\in C^1(U;C^1)\).  On the common compact tube
\(\overline U\times[\psi_0,\psi_*]\), the formula
\eqref{en:app-psi-radius-ode}, the no-horizon bound, and positivity of
\(m_{\mathrm{tot}}+4\pi r^3P\) give
\[
  0<c_0\le \partial_\psi r(\psi;\sigma)\le C_0.
\]
The derivative bounds and the inverse function theorem show that
\(r_\sigma(\psi):=r(\psi;\sigma)\) is a strictly increasing \(C^1\)
diffeomorphism onto its image.  If necessary, replace \(U\) by a
smaller open neighborhood of \(\sigma_0\) with compact closure; the images
then contain a common compact \(r\)-interval, so the inverses form
a single parameter-dependent chart there.  If
\(\psi_\sigma=r_\sigma^{-1}\), differentiating
\(r_\sigma(\psi_\sigma(r))=r\) gives
\[
  \partial_r\psi_\sigma
    =\bigl(\partial_\psi r_\sigma\bigr)^{-1}\!\circ\psi_\sigma,
  \qquad
  D_v \psi_\sigma
    =-\left(\frac{D_v r_\sigma}{\partial_\psi r_\sigma}\right)\!\circ\psi_\sigma.
\]
To differentiate this inverse in \(C^1\), note that
\(\partial_\psi^2 r_\sigma\) is jointly continuous. Indeed,
differentiating \eqref{en:app-psi-radius-ode} once in \(\psi\) uses only
\(\partial_\psi m_{\mathrm{tot}}=4\pi r^2 E\,\partial_\psi r\) and
\(\partial_\psi P=-\sum_I h_I^0\), both jointly continuous across all
moving thresholds. Thus the formula for \(D_v\psi_\sigma\) above has a
jointly continuous radial derivative. Together with the uniform lower
bound on \(\partial_\psi r\), this gives the parameter-dependent inverse
in \(C^1\) on the common chart.  It is used only away from the center, where \(\psi'(0)=0\);
the center Volterra construction gives the required regularity there, and
uniqueness identifies the two descriptions on their overlap.  The radial TOV
identity yields the additional spatial derivative, while the vacuum equations
extend the same family to the fixed interval \([0,R_*]\).  Hence
\[
 \psi\in C^1(U;C^1([0,R_*]))\cap C^0(U;C^2([0,R_*])).
\]

Since \(\psi\in C^1(U;C^1)\) and \(\partial_r\psi(R_I;\sigma)>0\), the
labelwise implicit function theorem gives \(R_I\in C^1(U)\), also at coincidences.
For the pressure, composition with the positive-part pressure law is
\(C^1\) in the uniform norm. For \(n_I,\epsilon_I\), let \(F\) be the
locally cut-off, zero-extended constitutive profile in enthalpy. It belongs
to \(W^{1,1}\) since \(F'=O(H^{\alpha_I-1})\) above vacuum.
On a fixed collar of \(R_I\), the signed coordinate
\(u_\sigma(r)=H_I^c-\psi_\sigma(r)\) satisfies
\(\lvert\partial_r u_\sigma\rvert\ge c>0\).
Approximate \(F\) in \(W^{1,1}\) by smooth profiles \(F_j\).
Changing variables from \(r\) to \(u_\sigma\) bounds the \(L^1\) errors
in \(F_j(u_\sigma)\) and in the parameter derivatives
\(F_j'(u_\sigma)D_vu_\sigma\) by a uniform constant times
\(\lVert F_j-F\rVert_{W^{1,1}}\) for \(\lvert v\rvert\le1\).
Passing to the limit gives \(C^1\) dependence in \(L^1\) on that collar.
Away from it ordinary composition applies, proving
\eqref{en:eq-fixed-pressure-regularity}--\eqref{en:eq-fixed-matter-regularity}.
Since
\[
 \widetilde m_I(r)=4\pi\int_0^r s^2\epsilon_I^0(s)\,\dd s,\qquad
 g_I^0=4\pi r^2A^{-1/2}n_I^0,
\]
integration and multiplication by the \(C^1\) metric factor prove the
remaining mass and \(g_I^0\) claims.

Put
\begin{equation}
 F_\Phi(r;\sigma):=
 \frac{m_{\mathrm{tot}}(r;\sigma)+4\pi r^3p_{\mathrm{tot}}(r;\sigma)}
 {r[r-2m_{\mathrm{tot}}(r;\sigma)]},\qquad F_\Phi(0;\sigma):=0.
\end{equation}
The center orders and the uniform no-horizon bound give
\(F_\Phi\in C^1(U;C^0([0,R_*]))\).  Since \(M=m_{\mathrm{tot}}(R_*;\sigma)\) is
\(C^1\), normalized exterior matching gives
\begin{equation}
 \Phi(r;\sigma)=\frac12\log\!\left(1-\frac{2M(\sigma)}{R_*}\right)
 -\int_r^{R_*}F_\Phi(s;\sigma)\,\dd s,
 \label{en:app-global-normalized-lapse}
\end{equation}
and proves \eqref{en:eq-fixed-metric-regularity}.  Finally,
\[
 H_I^{\mathrm{pb}}(s;\sigma)
 =H_I^c-\psi(R_I(\sigma)s;\sigma),\qquad
 D_v H_I^{\mathrm{pb}}=v^I-D_v \psi(R_Is)-s\psi'(R_Is)D_v R_I.
\]
The \(C^1\) regularity of \(\psi\) established above proves
\eqref{en:eq-pulled-enthalpy-regularity}.  Integration of \(g_I^0\) gives
\(N_I\in C^1(U)\), while \(M=\sum_I\widetilde m_I(R_*)\) is \(C^1\).
\end{proof}

These facts show that \(J_{\mathrm{fd}}\circ X\) is \(C^1\); since
\(\pi_{\mathrm{cen}}\circ X=\operatorname{Id}_U\), the map \(X\) is an
embedding.
The uniform horizonless bound on the chosen \(\overline U\), together with
\(n_I=O((R_I-r)^{\alpha_I})\), gives the uniform \(A\)-bound and
\(rA^{-3/2}n_I\in L^1\), completing part~\textup{(e)}.

\section{Particle-preserving variations and the radial zero mode equations}
\label{en:sec-static-solution-spaces}\label{en:app-zero-frequency}

\subsection{Enclosed particle numbers and the equilibrium first law}
\label{en:sec-first-law}
Consider a tangent \(v\) to the equilibrium family at fixed equations of
state, and write \(\delta\) for its first variation. Since
\(H_I'=-\Phi'\) and \(H_I=\log(\mu_I/\mu_{I,0})\), the background
redshifted chemical potential \(\mu_I^\infty=e^\Phi\mu_I\) is constant
throughout component \(I\). Extend \(\mathcal N_I(r)\) by \(N_I\) for
\(r\ge R_I\). Variations of these functions are taken at fixed areal
radius. Sums of enclosed masses and particle numbers include all \(K\)
labels: beyond a component's surface its enclosed charge is constant in
\(r\), but its parameter variation can be nonzero. Only the derivatives
of these charges and the local matter densities are summed over the active
labels. In each open active cell, with the \(\mu_I^\infty\) held at their
background values, differentiation of \eqref{en:eq-enclosed-number} gives
\begin{align}
 \left(\sum_{I=1}^K\mu_I^\infty\,\delta_E\mathcal N_I\right)'
 &=4\pi r^2 e^{\Phi+\Lambda}
   \Bigl(\sum_I\delta\epsilon_I+h_{\rm tot}\delta\Lambda\Bigr) \nonumber\\
 &=e^{\Phi+\Lambda}\bigl[\delta m_{\rm tot}'
                 +(\Phi'+\Lambda')\delta m_{\rm tot}\bigr]
   =\bigl(e^{\Phi+\Lambda}\delta m_{\rm tot}\bigr)' .
 \label{en:eq-enclosed-first-law-derivative}
\end{align}
Here \(\delta\epsilon_I=\mu_I\delta n_I\), \(h_I=\mu_I n_I\),
\(\delta\Lambda=\delta m_{\rm tot}/(rA)\), and
\(\Phi'+\Lambda'=4\pi r h_{\rm tot}/A\) follows from
\eqref{en:eq-metric} and \eqref{en:eq-tov-mass}--\eqref{en:eq-tov-phi}.
The integration constant at the center vanishes. At any labeled surface the
moving-endpoint terms in the mass and particle number integrals vanish
because \(\epsilon_I,n_I\to0\). The remaining cellwise variations are
locally integrable by Lemma~\ref{en:lem-equilibrium-package}, including at
coincident surfaces. Thus the identity integrates across every surface to
\begin{equation}
 \sum_{I=1}^K\mu_I^\infty\,\delta_E\mathcal N_I(r)=
 e^{\Phi(r)+\Lambda(r)}\delta m_{\rm tot}(r).
 \label{en:eq-enclosed-first-law}
\end{equation}
At a fixed point in the normalized Schwarzschild exterior,
\(e^{\Phi+\Lambda}=1\), so
\begin{equation}
 \delta M=\sum_{I=1}^K\mu_I^\infty\,\delta N_I .
 \label{en:eq-global-first-law}
\end{equation}
This is the static, uncharged limit of the global mass variation formula in Ref.~\cite{Carter1979Variation}, Eqs.~(56)--(57). In particular \(DN_\sigma[v]=0\) implies \(\delta M=0\). This is a
first-order condition on a tangent. In the two-fluid case, along any regular
curve with \(N_D\) constant, \(\dd M/\dd s=\mu_N^\infty\,\dd N_N/\dd s\),
where \(s\) is a curve parameter and \(N,D\) label nuclear and dark matter.
If \(\dd N_D\ne0\) as a covector, its kernel is the tangent line to this
curve. Since \(\mu_N^\infty>0\), a stationary mass on the curve then gives
a nonzero tangent preserving both numbers.

We continue to use the central enthalpies in
\eqref{en:eq-central-chart} as local coordinates.  Since
\(\dd p_I/\dd H_I=\epsilon_I+p_I>0\) at the center, the central enthalpy
and central pressure charts are related by a local \(C^1\) diffeomorphism.
Rank and nullity are invariant under this coordinate change, so all such
statements below are independent of the chosen chart.

\subsection{Spaces, domains, and maps}
\label{en:sec-spaces-and-maps}

\begin{definition}[Static perturbations, recovery, and the radial domain]
\phantomsection\label{en:def-solution-classes}
\label{en:def-static-perturbation-spaces}
\smallskip\noindent\textbf{(I) Static Eulerian perturbation spaces.}
An element of \(\mathcal S_\sigma^{\mathrm{stat}}\) is a real tuple
\begin{equation}
  \delta X
  =\bigl((\delta H_I,\delta m_I)_{I=1}^K,\delta\Phi;
    (\delta R_I)_{I=1}^K,\delta M\bigr).
  \label{en:eq-static-element-type}
\end{equation}
Here \(\delta H_I\) is defined on \((0,R_I)\), each \(\delta m_I\) is
continued constantly to \((R_I,R_*)\), \(\delta\Phi\) is defined on the
interior and normalized exterior, and \(\delta R_I,\delta M\in\mathbb R\).
The variations \(\delta n_I,\delta\epsilon_I,\delta p_I,\delta g_I\) and
\(\delta\Lambda\) are the fields determined from this tuple by the
linearized EoS relations and the metric identity.  The tuple must satisfy:
\begin{description}
  \item[\textup{(S1)}]\conditiontarget{S1}
  On every active cell on which a component is present, the listed Eulerian fields
  are \(C^1\) and satisfy the equations obtained by linearizing
  \eqref{en:eq-tov-mass}--\eqref{en:eq-tov-phi} at the fixed equilibrium.
  Independently of any equilibrium family representation, their center
  behavior satisfies, for finite constants \(\delta H_I^c\),
  \begin{equation}
    \begin{aligned}
      \delta H_I(r)&=\delta H_I^c+O(r^2),&
      \delta m_I(r)&=O(r^3),\\
      \delta\Lambda(r)&=O(r^2),&
      \delta\Phi'(r)&=O(r)
    \end{aligned}
    \qquad(r\to0^+).
    \label{en:eq-static-center-regularity}
  \end{equation}

  \item[\textup{(S2)}]\conditiontarget{S2}
  After zero-extending \(\delta\epsilon_I\), every component mass satisfies
\begin{equation}
  \delta m_I\in W^{1,1}(0,R_*),\qquad
  \delta m_I(0)=0,\qquad
  \delta m_I'=4\pi r^2\delta\epsilon_I
  \quad\hbox{in }\mathcal D'(0,R_*).
  \label{en:eq-component-mass-global-domain}
\end{equation}
Here \(W^{1,1}(0,R_*)\) denotes the Sobolev space of integrable functions
on \((0,R_*)\) with integrable weak derivative.
The \(W^{1,1}\) regularity therefore forces each component mass variation
to have matching one-sided values at every internal ending block
\(\mathcal R_a\):
\[
 \lim_{r\to\mathcal R_a^-}\delta m_I(r)
 =\lim_{r\to\mathcal R_a^+}\delta m_I(r).
\]

  \item[\textup{(S3)}]\conditiontarget{S3}
  Every label has its own scalar variation \(\delta R_I\) and satisfies
  \begin{equation}
    \delta H_I(R_I)+H_I'(R_I)\delta R_I=0.
    \label{en:eq-static-labeled-surface}
  \end{equation}

  \item[\textup{(S4)}]\conditiontarget{S4}
  At every internal ending block, \(\delta m_{\mathrm{tot}}=\sum_I\delta m_I\)
  (equivalently \(\delta\Lambda\)) and \(\delta\Phi\) have matching one-sided
  limits.  For every surviving component \(J\), \(\delta H_J\)
  (equivalently \(\delta p_J\)) also has matching one-sided limits.  The metric conditions follow from no-layer matching. The pressure condition
  holds for each surviving label by its separate Euler equation
  \eqref{en:app-zf-Euler}: the other terms are locally integrable, so
  the distributional derivative of \(\delta p_J\) has no surface atom.
  Since \(h_J(R)>0\), this gives the same condition on \(\delta H_J\).

  \item[\textup{(S5)}]\conditiontarget{S5}
  Beyond \(R_{\mathrm{out}}\), the perturbation is the linearized Schwarzschild
  exterior with mass variation \(\delta M\):
  \begin{equation}
    \delta m_{\mathrm{tot}}=\delta M,\qquad
    \delta\Lambda=\frac{\delta M}{rA},\qquad
    \delta\Phi=-\frac{\delta M}{rA}.
  \end{equation}
  Here normalized Schwarzschild matching means that the inner one-sided mass
  and metric perturbations at \(R_{\mathrm{out}}\) agree with these asymptotically
  normalized vacuum values:
  \begin{equation}
    \delta m_{\mathrm{tot}}(R_{\mathrm{out}}^-)=\delta M,\qquad
    \delta\Lambda(R_{\mathrm{out}}^-)
      =\frac{\delta M}{R_{\mathrm{out}}A(R_{\mathrm{out}})},\qquad
    \delta\Phi(R_{\mathrm{out}}^-)
      =-\frac{\delta M}{R_{\mathrm{out}}A(R_{\mathrm{out}})}.
    \label{en:eq-static-outer-matching}
  \end{equation}
  In particular, \(\delta\Phi(\infty)=0\).

  \item[\textup{(S6)}]\conditiontarget{S6}
  At every labeled component surface, the one-sided estimates are
\begin{equation}
  \delta H_I=O(1),\qquad
  \delta n_I,\delta\epsilon_I=O((R_I-r)^{\alpha_I-1}),\qquad
  \delta p_I=O((R_I-r)^{\alpha_I}).
  \label{en:eq-static-weighted-limits}
\end{equation}
\end{description}
The particle-number-preserving static subspace is
\begin{equation}
  \mathcal S_{\sigma,N}^{\mathrm{stat}}
  :=\{\delta X\in\mathcal S_\sigma^{\mathrm{stat}}:
  \delta N_I=0\ \forall I\}.
  \label{en:eq-constrained-static-space}
\end{equation}
This is a vector subspace.  The results below show that it is nontrivial if
and only if \(\sigma\) is statically critical.

The linearized EoS relations and the extension rules define an injective
linear map
\begin{equation}
  \iota_\sigma:\mathcal S_\sigma^{\mathrm{stat}}\longrightarrow\mathcal X_{\mathrm{fd}}.
  \label{en:eq-static-chart-embedding}
\end{equation}
Its pulled-back enthalpy component is
\(\delta H_I(R_Is)+sH_I'(R_Is)\delta R_I\); its matter components are the
zero-extended \((\delta n_I,\delta\epsilon_I,\delta p_I,\delta g_I)\), and
the remaining components are the mass, lapse, radius, and total mass
variations in the fixed-domain chart.

We next define the equilibrium family derivative as a field tuple; its
membership in \(\mathcal S_\sigma^{\mathrm{stat}}\) will be proved below.  Put
\begin{equation}
  H_I^{\mathrm{pb}}(s;\sigma):=H_I(R_I(\sigma)s;\sigma),
  \qquad 0\le s\le1.
  \label{en:eq-pulled-enthalpy-profile}
\end{equation}
For \(v\in T_\sigma U\), set
\begin{equation}
\begin{aligned}
 D_v H_I^{\mathrm{pb}}&:=D(H_I^{\mathrm{pb}})_\sigma[v],&
 D_v \widetilde m_I&:=D(\widetilde m_I)_\sigma[v],\\
 D_v \Phi&:=D\Phi_\sigma[v],&
 D_v R_I&:=DR_{I,\sigma}[v],\qquad
 D_v M:=DM_\sigma[v].
\end{aligned}
\label{en:eq-raw-chart-derivatives}
\end{equation}
Lemma
~\statementpartref{en:lem-equilibrium-package}{en:item-map-regularity}
proves that these Banach-valued derivatives exist.  On \(0<r<R_I\), define
\begin{align}
  \delta_vH_I(r)
   &:=(D_v H_I^{\mathrm{pb}})(r/R_I)
     -\frac r{R_I}H_I'(r)D_v R_I,
 \label{en:eq-raw-eulerian-enthalpy}\\
  \delta_vm_I&:=D_v \widetilde m_I\quad\hbox{on }[0,R_*],
  &\delta_v\Phi&:=D_v \Phi\quad\hbox{on }[0,R_*],
 \label{en:eq-raw-fixed-fields}
\end{align}
and assemble
\begin{equation}
 B_\sigma v
 :=\bigl((\delta_vH_I,\delta_vm_I)_I,\delta_v\Phi;
          (D_v R_I)_I,D_v M\bigr).
 \label{en:eq-raw-equilibrium-derivative}
\end{equation}
The first formula is the inverse of the moving-radius chain rule and recovers
the Eulerian derivative at fixed \(r\).  Beyond \(R_*\), continue
\(\delta_v\Phi\) by differentiating the normalized Schwarzschild exterior.
This construction defines \(B_\sigma v\) independently of
\(\mathcal S_\sigma^{\mathrm{stat}}\).  Lemma
~\ref{en:lem-static-representation} shows that it belongs to this space and
represents the fixed-domain differential.

\phantomsection
\label{en:def-full-zero-mode-space}
\smallskip\noindent\textbf{(II) Full zero-frequency perturbations.}
An element of \(\mathcal S_{\sigma,0,N}^{\mathrm{full}}\) is a tuple
\begin{equation}
  \delta X
  =\bigl(\delta X_{\mathrm{stat}},(\xi_I)_{I=1}^K\bigr),
  \qquad
  \delta X_{\mathrm{stat}}\in\mathcal S_{\sigma,N}^{\mathrm{stat}},
  \label{en:eq-full-element-type}
\end{equation}
in which the Eulerian static fields are augmented by one radial Lagrangian
displacement \(\xi_I:(0,R_I)\to\mathbb R\) for each fluid.  For any background
scalar \(f_I\) carried by component \(I\), write
\(\Delta f_I:=\delta f_I+\xi_I f_I'\) for its Lagrangian variation.  The tuple
must satisfy:
\begin{description}
  \item[\textup{(F1) Static and global constraints.}]\conditiontarget{F1}
  Forgetting the displacements leaves an element of
  \(\mathcal S_{\sigma,N}^{\mathrm{stat}}\); in particular, all
  \conditionrefp{S1}--\conditionrefp{S6}
  conditions and \(\delta N_I=0\) hold.

  \item[\textup{(F2) Displacement regularity.}]\conditiontarget{F2}
  Each \(\xi_I\) is \(C^2\) on every active cell contained in
  \((0,R_I)\), has the center expansion
  \begin{equation}
    \xi_I=a_I^{\xi,\mathrm{cen}}r+O(r^3),\qquad
    \xi_I'=a_I^{\xi,\mathrm{cen}}+O(r^2),
  \end{equation}
  and has every finite one-sided limit required below.  No one-sided
  derivative limit is required across an ending block.

  \item[\textup{(F3) Zero-frequency bulk equations.}]\conditiontarget{F3}
  On every active cell, the particle number conservation, EoS, Hamiltonian, static
  \(rr\), and Euler equations
  \eqref{en:app-zf-number}--\eqref{en:app-zf-Euler} hold.

  \item[\textup{(F4) Internal transmission.}]\conditiontarget{F4}
  At every internal ending block, the fields \(\xi_J\) and \(\delta p_J\) of
  each surviving component \(J\), as well as \(\delta\Lambda\) and
  \(\delta\Phi\), have matching one-sided limits across the block.  Every ending label retains its own surface
  condition.

  \item[\textup{(F5) Labeled surfaces and exterior.}]\conditiontarget{F5}
  At every labeled component surface,
  \begin{equation}
    \delta R_I=\xi_I(R_I),\qquad
    \Delta H_I(R_I)
    :=\delta H_I(R_I)+H_I'(R_I)\delta R_I=0.
    \label{en:eq-full-labeled-surface}
  \end{equation}
  The normalized exterior conditions in \conditionrefp{S5} hold.
\end{description}
We call these elements \emph{full zero-frequency perturbations}: they retain
the matter and metric variations, all fluid displacements, the labeled
surface variations, and the global mass variation.  The forgetful map is
\begin{equation}
  F_\sigma:\mathcal S_{\sigma,0,N}^{\mathrm{full}}
  \longrightarrow\mathcal S_{\sigma,N}^{\mathrm{stat}},
  \qquad
  F_\sigma(\delta X_{\mathrm{stat}},\xi)=\delta X_{\mathrm{stat}}.
  \label{en:eq-forgetful-map}
\end{equation}

\phantomsection
\label{en:def-recovery-map}
\smallskip\noindent\textbf{(III) Recovery map.}
A reduced displacement is a real vector
\(\zeta=(\zeta_I)_{I=1}^K\), with
\(\zeta_I:(0,R_I)\to\mathbb R\), related to the physical displacement by
\(\zeta_I=r^2e^{-\Phi}\xi_I\).  For such a vector, first form the local
algebraic expressions in \eqref{en:app-zf-xi} and
\eqref{en:app-recover-lambda}--\eqref{en:app-recover-matter}.  When those
expressions have the cellwise regularity and integrability specified in
\conditionref{C1}--\conditionref{C2}, the integral and exterior formulas
\eqref{en:app-recover-mass}--\eqref{en:app-recover-exterior} define the
remaining fields.  If the global traces specified in \conditionref{C3} also
exist, these fields assemble the tuple
\begin{equation}
  \begin{aligned}
  \widehat R_\sigma\zeta:=\bigl(&
  (\xi_I)_{I=1}^K,(\delta H_I)_{I=1}^K,
  (\delta n_I)_{I=1}^K,(\delta\epsilon_I)_{I=1}^K,\\
  &(\delta p_I)_{I=1}^K,(\delta m_I)_{I=1}^K,
  \delta m_{\mathrm{tot}},\delta\Lambda,\delta\Phi;
  (\delta R_I)_{I=1}^K,\delta M\bigr).
  \end{aligned}
  \label{en:eq-recovery-map-tuple}
\end{equation}
We call this linear operation the \emph{recovery map}, and its output a
\emph{recovered full field tuple}.  The exterior formula builds in the
asymptotic Schwarzschild normalization.  Membership in
\(\mathcal S_{\sigma,0,N}^{\mathrm{full}}\) is established below from the
reduced equation and the domain conditions.

\phantomsection
\label{en:def-reduced-domain}
\smallskip\noindent\textbf{(IV) Reduced ambient class and domain.}
To accommodate the integrable coefficient derivatives generated when another
component terminates inside \((0,R_I)\), define the piecewise codomain
\begin{equation}
	\mathcal Y_\sigma:=\bigoplus_I
	\left\{\begin{array}{@{}l@{}}
		f\in L^1(0,R_I):\
		\forall a,\ 
		\mathcal I_a\subset(0,R_I)
		\Longrightarrow
		f|_{\mathcal I_a}\in C^0(\mathcal I_a),\\[-2pt]
		f=O(r)\quad(r\to0^+),\qquad
		f=O((R_I-r)^{\alpha_I-1})\quad(r\to R_I^-)
	\end{array}\right\},
	\label{en:eq-reduced-target-space}
\end{equation}
with the center and labeled endpoint bounds understood one-sidedly.  At an
  internal ending block \(\mathcal R_a<R_I\), the \(L^1\) requirement permits the
  integrable coefficient-forcing order
  \(O(|r-\mathcal R_a|^{\min_{J\in\mathcal B_a}\alpha_J-1})\).
The ambient class \(\widehat{\mathcal C}_\sigma\) consists of real vectors
\(\zeta=(\zeta_I)_{I=1}^K\) satisfying the following regularity and
definability requirements:
\begin{description}
  \item[\textup{(C1)}]\conditiontarget{C1} \(\zeta_I|_{\mathcal I_a}\in C^2(\mathcal I_a)\)
  whenever component \(I\) is present on \(\mathcal I_a\);
  \item[\textup{(C2)}]\conditiontarget{C2} the local expressions in
  \eqref{en:app-zf-xi},
  \eqref{en:app-recover-lambda}--\eqref{en:app-recover-matter}, and
  \eqref{en:app-recover-phi-prime} are classical on every such cell,
  where \(\delta\Phi'_{\mathrm{int}}[\zeta]\) denotes the interior lapse
  derivative in \eqref{en:app-recover-phi-prime},
  \begin{equation}
    r^2\bigl(\delta\epsilon_I[\zeta]\bigr)^0\in L^1(0,R_*)
    \quad\text{for every \(I\)},\qquad
    \delta\Phi'_{\mathrm{int}}[\zeta]
      \in L^1_{\mathrm{loc}}((0,R_{\mathrm{out}}]),
    \label{en:eq-recovery-integrability}
  \end{equation}
  and the matter fields given by these expressions obey the labeled endpoint
  weights \eqref{en:eq-static-weighted-limits}.  The stated integrability
  guarantees that the integrals
  in \eqref{en:app-recover-mass} and
  \eqref{en:app-recover-phi-interior} are well defined;
  \item[\textup{(C3)}]\conditiontarget{C3} for every label \(I\), the finite
  labeled surface trace
  \[
    \zeta_I(R_I^-):=\lim_{r\to R_I^-}\zeta_I(r)\in\mathbb R
  \]
  exists, and all one-sided traces entering
  \eqref{en:app-B-internal} and \eqref{en:app-B-exterior} exist.  Since
  \(R_I>0\) and \(e^\Phi\) has a finite nonzero surface limit, this is
  equivalent to the existence of a finite trace \(\xi_I(R_I^-)\);
  \item[\textup{(C4)}]\conditiontarget{C4} the cellwise differential expression
  \(\mathcal L_\sigma\zeta\), formed from the already defined recovered local
  expressions, belongs to \(\mathcal Y_\sigma\).
\end{description}
The reduced domain \(\widehat{\mathcal D}_\sigma\) is the subspace of
\(\widehat{\mathcal C}_\sigma\) satisfying the following four groups of additional
conditions:
\begin{description}
  \item[\textup{(D1)}]\conditiontarget{D1} At the center, for some coefficient \(a_I^{\mathrm{cen}}\),
  \(\zeta_I=a_I^{\mathrm{cen}}r^3+O(r^5)\) and
  \(\zeta_I'=3a_I^{\mathrm{cen}}r^2+O(r^4)\).  In particular,
  \(g_I\xi_I=4\pi e^{\Lambda+\Phi}n_I\zeta_I=O(r^3)\), so the
  center integration constant in the particle number equation vanishes.
  \item[\textup{(D2)}]\conditiontarget{D2} At every labeled component
  surface, the one-sided limit defining
  \(\widehat B_{I,\mathrm{s}}(\zeta)\) in
  \eqref{en:app-B-surface} exists and vanishes:
  \[
    \widehat B_{I,\mathrm{s}}(\zeta)=0.
  \]
  No finite one-sided trace of \(\zeta_I'\) itself is required.
  \item[\textup{(D3)}]\conditiontarget{D3} At every internal ending block \(\mathcal R_a\), \(a<q\), the common
  transmission conditions \(B_{a,\mathrm{int}}(\zeta)=0\), namely
  \eqref{en:app-B-internal}, hold.
  \item[\textup{(D4)}]\conditiontarget{D4} At the outermost surface, the normalized Schwarzschild exterior
  conditions \(B_\infty(\zeta)=0\), namely
  \eqref{en:app-B-exterior}, hold.
\end{description}
Taken together, \conditionref{C1}--\conditionref{C4} specify regularity,
integrability, trace existence, and the target space regularity of the
differential expression.  Conditions
\conditionref{D1}--\conditionref{D4} impose, respectively, the center,
  material surface, internal transmission, and normalized exterior boundary
  conditions.  In particular, \conditionref{C3} supplies only the finite
  displacement trace used to define \(\delta R_I\); the existence and
  vanishing of the normalized Lagrangian enthalpy residual are imposed
  separately by \conditionref{D2}.
On this domain, \(\mathcal L_\sigma\) is the cellwise classical
expression \eqref{en:app-reduced-operator}.  Accordingly,
\(\ker\mathcal L_\sigma\) means that the expression vanishes on every active
cell, together with all conditions in \(\widehat{\mathcal D}_\sigma\).  The
equivalence with the distributional formulation is established in
Theorem~\ref{en:thm-reduction-package}.

\phantomsection
\label{en:def-residual-and-boundary-operator}
\smallskip\noindent\textbf{(V) Residual and boundary condition.}
For any candidate or recovered full field tuple \(\delta X\) with the stated
cellwise regularity, define its static Euler residual by
\begin{equation}
  \mathcal E_I(\delta X)
  :=\delta p_I'+(\epsilon_I+p_I)\delta\Phi'
  +(\delta\epsilon_I+\delta p_I)\Phi'.
  \label{en:eq-Euler-residual-definition}
\end{equation}
We organize the remaining residuals and boundary data into the following
blocks.  Let
\(\mathcal X_{\mathrm{full}}^{\mathrm{tr}}\) be the class of candidate full tuples with
the stated cellwise regularity for which every distributional residual, germ,
trace, improper integral, and limit listed below exists and belongs to its
indicated component space.  All endpoint values are one-sided traces from the
indicated active cell.  Scalar traces and jumps take values in finite products
of \(\mathbb R\), while the component mass residual is taken in
\(\mathcal D'(0,R_*)^K\).  The zero-boundary-data condition includes
membership in \(\mathcal X_{\mathrm{full}}^{\mathrm{tr}}\) and componentwise vanishing
of all the data listed below.

\begin{enumerate}
\renewcommand{\theenumi}{\alph{enumi}}
\renewcommand{\labelenumi}{(\theenumi)}
\item \emph{Center block.}
Let \(\mathcal X_{\mathrm{cen}}\) be the product of the one-sided center germs of
\((\xi_I,\xi_I',\delta H_I,\delta m_I)_{I=1}^K\), together with the common
metric germs \(\delta\Lambda\) and \(\delta\Phi'\).  Let
\(\mathcal X_{\mathrm{cen}}^{\mathrm{reg}}\) be the linear subspace consisting of the
germs for which there are constants \(a_I^{\xi,\mathrm{cen}}\) and
\(\delta H_I^c\) such that
\begin{equation}
  \begin{gathered}
    \xi_I=a_I^{\xi,\mathrm{cen}}r+O(r^3),\qquad
    \xi_I'=a_I^{\xi,\mathrm{cen}}+O(r^2),\\
    \delta H_I=\delta H_I^c+O(r^2),\qquad
    \delta m_I=O(r^3),\qquad
    \delta\Lambda=O(r^2),\qquad
    \delta\Phi'=O(r).
  \end{gathered}
  \label{en:eq-full-center-block}
\end{equation}
Set \(\mathcal T_{\mathrm{cen}}:=\mathcal X_{\mathrm{cen}}/
\mathcal X_{\mathrm{cen}}^{\mathrm{reg}}\), and let
\(\mathcal B_{\mathrm{cen}}(\delta X)\in\mathcal T_{\mathrm{cen}}\) be the corresponding
quotient class.  By construction, it vanishes exactly when
\eqref{en:eq-full-center-block} holds.

\item \emph{Component mass block.}
After zero extension of the matter fields, define
\begin{equation}
  \mathcal B_{\mathrm{mass}}(\delta X)
  :=\bigl(\partial_r\delta m_I-4\pi r^2(\delta\epsilon_I)^0\bigr)_{I=1}^K
  \in\mathcal D'(0,R_*)^K.
  \label{en:eq-full-mass-block}
\end{equation}

\item \emph{Particle number block.}
For particle number, put
\(\delta_E\mathcal N_I(r):=\int_0^r\delta g_I(s)\,\dd s\) and
\begin{equation}
  \mathcal B_N(\delta X):=(\delta N_I)_{I=1}^K,\qquad
  \delta N_I
  :=\delta_E\mathcal N_I(R_I^-)+g_I(R_I)\delta R_I
  \in\mathbb R.
  \label{en:eq-full-number-block}
\end{equation}
\item \emph{Labeled surface block.}
At each labeled component surface, set
\begin{equation}
  \mathcal B_{I,\mathrm{s}}(\delta X)
  :=
  \left(
    \delta R_I-\xi_I(R_I^-),
    \delta H_I(R_I^-)+H_I'(R_I)\delta R_I
  \right)\in\mathbb R^2.
  \label{en:eq-full-surface-block}
\end{equation}

\item \emph{Internal node block.}
At every internal ending block \(\mathcal R_a\), \(a<q\), set
\begin{equation}
  \mathcal B_{a,\mathrm{int}}(\delta X)
  :=
  \left(
    ([\xi_J]_{\mathcal R_a},[\delta p_J]_{\mathcal R_a})_{J\in\mathcal A_a^+},
    [\delta m_{\mathrm{tot}}]_{\mathcal R_a},
    [\delta\Lambda]_{\mathcal R_a},
    [\delta\Phi]_{\mathcal R_a}
  \right)
  \in\mathbb R^{2|\mathcal A_a^+|+3}.
  \label{en:eq-full-internal-block}
\end{equation}

\item \emph{Normalized exterior block.}
At the outermost surface, collect the three inner matching traces and the
three exterior field residuals:
\begin{equation}
  \begin{aligned}
  \mathcal B_\infty(\delta X):=\Bigl(&
    \delta m_{\mathrm{tot}}(R_{\mathrm{out}}^-)-\delta M,\,
    \delta\Lambda(R_{\mathrm{out}}^-)
      -\frac{\delta M}{R_{\mathrm{out}}A(R_{\mathrm{out}})},\\
   &\delta\Phi(R_{\mathrm{out}}^-)
      +\frac{\delta M}{R_{\mathrm{out}}A(R_{\mathrm{out}})};\,
    \left.\left(\delta m_{\mathrm{tot}}-\delta M\right)\right|_{[R_{\mathrm{out}},\infty)},\\
   &\left.\left(\delta\Lambda-\frac{\delta M}{rA}\right)
      \right|_{[R_{\mathrm{out}},\infty)},\,
    \left.\left(\delta\Phi+\frac{\delta M}{rA}\right)
      \right|_{[R_{\mathrm{out}},\infty)}
  \Bigr).
  \end{aligned}
  \label{en:eq-full-exterior-block}
\end{equation}
Its target is
\(\mathbb R^3\times[C^0_{\mathrm{loc}}([R_{\mathrm{out}},\infty))]^3\).

\item \emph{Integration constant block.}
The constants left by the cellwise first-order equations are collected as
\begin{equation}
  \begin{gathered}
    C_{\mathrm{full}}(\delta X)
    :=\left(
      (C_{N,I})_{I=1}^K,(C_{m,I})_{I=1}^K,
      C_\Lambda,C_{\Phi,\infty},C_M
    \right)\in\mathbb R^{2K+3},\\
    C_{N,I}:=\lim_{r\to0^+}
       [\delta_E\mathcal N_I(r)+g_I(r)\xi_I(r)],
    \qquad C_{m,I}:=\delta m_I(0),\\
    C_\Lambda:=\lim_{r\to0^+}rA(r)\delta\Lambda(r),\qquad
    C_{\Phi,\infty}:=\lim_{r\to\infty}\delta\Phi(r),\qquad
    C_M:=\delta M-\delta m_{\mathrm{tot}}(R_{\mathrm{out}}^-).
  \end{gathered}
  \label{en:eq-full-integration-constant-block}
\end{equation}
\end{enumerate}

The distributional mass block together with \(C_{m,I}=0\) fixes the
component mass constant on the first cell and its continuation across every
later cell.  Likewise, the conditions
\([\delta\Phi]_{\mathcal R_a}=0\), \(a<q\), identify all cellwise lapse
constants, and \(C_{\Phi,\infty}=0\) fixes their common normalized value.

\smallskip\noindent\emph{Assembly of the boundary data operator.}
The target trace space is
\begin{equation}
  \begin{aligned}
  \mathcal T_{\mathrm{full}}:={}&
  \mathcal T_{\mathrm{cen}}\times\mathcal D'(0,R_*)^K
  \times\mathbb R^K\times\mathbb R^{2K}
  \times\prod_{a<q}\mathbb R^{2|\mathcal A_a^+|+3}\\
  &{}\times
  \left(\mathbb R^3\times
  [C^0_{\mathrm{loc}}([R_{\mathrm{out}},\infty))]^3\right)
  \times\mathbb R^{2K+3},
  \end{aligned}
  \label{en:eq-full-boundary-trace-space}
\end{equation}
The boundary data operator is
\begin{equation}
  \begin{aligned}
  \operatorname{BC}_{\mathrm{full}}:\mathcal X_{\mathrm{full}}^{\mathrm{tr}}
  &\longrightarrow\mathcal T_{\mathrm{full}},\\
  \delta X&\longmapsto\left(
    \mathcal B_{\mathrm{cen}},\mathcal B_{\mathrm{mass}},\mathcal B_N,
    (\mathcal B_{I,\mathrm{s}})_{I=1}^K,
    (\mathcal B_{a,\mathrm{int}})_{a<q},
    \mathcal B_\infty,C_{\mathrm{full}}
  \right).
  \end{aligned}
  \label{en:eq-full-boundary-operator}
\end{equation}
Here \(\operatorname{BC}_{\mathrm{full}}(\delta X)=0\) means componentwise
vanishing in \(\mathcal T_{\mathrm{full}}\).
\end{definition}

\noindent\textbf{Relation among the spaces.}
The nonlinear space \(\mathcal X_{\mathrm{eq}}\) contains equilibrium configurations.  At
a fixed equilibrium \(\sigma\), the real vector space
\(\mathcal S_\sigma^{\mathrm{stat}}\) contains regular Eulerian solutions of the
linearized static equations, and
\(\mathcal S_{\sigma,N}^{\mathrm{stat}}\) is its particle-number-preserving
subspace.  The space \(\mathcal S_{\sigma,0,N}^{\mathrm{full}}\) augments those
constrained perturbations by the compatible Lagrangian displacements, whereas
\(\widehat{\mathcal D}_\sigma\) is a domain of reduced displacement vectors
rather than a space of full field tuples.  The constrained static and full
zero-frequency solution spaces may be trivial at a noncritical equilibrium.
Section~\ref{en:sec-spectral} uses the corresponding complexified spaces,
denoted by the same symbols.

\subsection{Static perturbations as equilibrium tangents}
\label{en:app-proof-static-package}\label{en:sec-auxiliary-results}

\begin{lemma}[Static tangent representation]
\label{en:lem-static-representation}
Under Assumption~\ref{en:ass-regular}, the field derivative defined in
\eqref{en:eq-raw-equilibrium-derivative} satisfies
\begin{equation}
  B_\sigma v\in\mathcal S_\sigma^{\mathrm{stat}},
  \qquad
  \iota_\sigma(B_\sigma v)
  =D(J_{\mathrm{fd}}\circ X)_\sigma[v]
  \quad(v\in T_\sigma U).
  \label{en:eq-chart-field-differential}
\end{equation}
The differential of the equilibrium map is therefore the linear isomorphism
\begin{equation}
  B_\sigma:T_\sigma U\xrightarrow{\ \cong\ }
  \mathcal S_\sigma^{\mathrm{stat}}.
  \label{en:eq-B-isomorphism}
\end{equation}
Equivalently, every regular static perturbation in
\(\mathcal S_\sigma^{\mathrm{stat}}\) is uniquely determined by its central
enthalpy variations.  
\end{lemma}

\setproofparent{en:lem-static-representation}

\begin{proof}
On each open active cell let
\begin{equation}
  Y_a=\bigl((H_I)_{I\in\mathcal K_a},(m_I)_{I=1}^K,\Phi\bigr),
  \qquad m_I'=0\quad(I\notin\mathcal K_a).
  \label{en:app-equilibrium-cell-state}
\end{equation}
Writing \(m_{\mathrm{tot}}=r^3\bar m_{\mathrm{tot}}\) at the center gives
\begin{align}
  \bar m_{\mathrm{tot}}(r)
  &=4\pi\int_0^1x^2
  \epsilon_{\mathrm{tot}}\bigl((H_I(rx))_{I=1}^K\bigr)\,\dd x,
  \label{en:app-center-mass}\\
  H_I'(r)
  &=-r
  \frac{\bar m_{\mathrm{tot}}+4\pi p_{\mathrm{tot}}}{1-2r^2\bar m_{\mathrm{tot}}},
  \qquad H_I(0)=H_I^c.
  \label{en:app-center-pressure}
\end{align}
Away from the center the active fields satisfy a \(C^1\) system
\begin{equation}
  Y_a'=F_a(r,Y_a).
  \label{en:app-piecewise-ODE}
\end{equation}
The center conditions therefore leave exactly the \(K\) enthalpy values
free: \(m_I(0)=0\), spherical regularity fixes the other center data, and
exterior normalization fixes the lapse constant.

Fix \(v\in T_\sigma U\) and start with the raw field tuple
\(B_\sigma v\) defined in
\eqref{en:eq-raw-equilibrium-derivative}; no lift through \(\iota_\sigma\) is
being assumed.  On every compact subinterval of an open active cell, its
components are the fixed-\(r\) derivatives along
\(\sigma+\tau v\).  Differentiating the exact equilibrium equations there
therefore gives the linearized bulk equations in \conditionref{S1}.
Differentiating the
center-regular Volterra system
\eqref{en:app-center-mass}--\eqref{en:app-center-pressure} gives the center
orders in \eqref{en:eq-static-center-regularity}.

For a matter field \(f_I\in\{n_I,\epsilon_I,p_I,g_I\}\), let
\(\delta_vf_I\) be its fixed-\(r\) interior derivative, with the linearized
  EoS and metric identity used for the derived fields.  The
  uniqueness of the \(L^1_{\mathrm{fd}}\) derivative, together with
  Lemma~\statementpartref{en:lem-equilibrium-package}
  {en:item-crossing-regularity}, gives
\begin{equation}
 D(f_I^0)_\sigma[v]=(\delta_vf_I)^0
 \quad\hbox{in }L^1_{\mathrm{fd}}(0,R_*).
 \label{en:app-zero-extension-differential}
\end{equation}
In particular, differentiating the exact integral formula for the extended
component mass yields
\begin{equation}
 D_v \widetilde m_I(r)
 =4\pi\int_0^r s^2(\delta_v\epsilon_I)^0(s)\,\dd s,
 \qquad 0\le r\le R_*.
 \label{en:app-component-mass-differential}
\end{equation}
Hence \conditionref{S2} holds, including the matching one-sided values and the
absence of a
surface measure in the weak derivative.  Differentiating
\(H_I(R_I(\sigma);\sigma)=0\) gives \conditionref{S3}.  At a moving interface a continuous background field \(F\) has the
linearized matching condition \([\delta F]+[F^{\prime}]D_vR_I=0\).
Here \(m_{\rm tot}'=4\pi r^2\epsilon_{\rm tot}\) and \(\Phi'\) are
continuous, since all ending densities and pressures vanish. Hence
\([\delta m_{\rm tot}]=[\delta\Phi]=0\). Every surviving enthalpy has
\(\delta H_J=v^J-D_v\psi\) on both sides. These facts give
\conditionref{S4}, also when initially coincident surfaces separate.
Differentiating the normalized exterior gives \conditionref{S5}.  Finally, the
  endpoint estimates in Lemma~
  \statementpartrange{en:lem-equilibrium-package}
  {en:item-eos-consequences}{en:item-crossing-regularity}
  give \conditionref{S6}.  Together, these verifications establish
\begin{equation}
 B_\sigma v\in\mathcal S_\sigma^{\mathrm{stat}}.
 \label{en:app-raw-derivative-membership}
\end{equation}

It remains to identify this field tuple with the ambient Fr\'echet
differential.  By
\eqref{en:eq-raw-eulerian-enthalpy},
\begin{equation}
 D_v H_I^{\mathrm{pb}}(s)
 =\delta_vH_I(R_Is)+sH_I'(R_Is)D_v R_I.
 \label{en:app-pulled-enthalpy-differential}
\end{equation}
This is precisely the enthalpy component of \(\iota_\sigma\); the mass,
lapse, radius, and total mass components agree by
\eqref{en:eq-raw-chart-derivatives}, and the matter components agree by
\eqref{en:app-zero-extension-differential}.  Therefore
\begin{equation}
 \iota_\sigma(B_\sigma v)
 =D(J_{\mathrm{fd}}\circ X)_\sigma[v].
 \label{en:app-static-commuting-lift}
\end{equation}
This proves existence of the lift; injectivity of \(\iota_\sigma\) proves
its uniqueness.  The linearity of \(B_\sigma\) follows from the linearity of
the Fr\'echet derivatives in
\eqref{en:eq-raw-chart-derivatives}.

Conversely, for \(\delta X\in\mathcal S_\sigma^{\mathrm{stat}}\), set
\begin{equation}
  v^I=\delta H_I(0),\qquad
  v=v^I\frac{\partial}{\partial H_I^c},\qquad
  Z=\delta X-B_\sigma v .
  \label{en:app-static-center-data}
\end{equation}
Write
\begin{equation}
  z_I:=Z_{H_I},\qquad
  \nu_I:=Z_{m_I},\qquad
  \varphi:=Z_\Phi,\qquad
  \nu_{\mathrm{tot}}:=\sum_{I=1}^K\nu_I .
  \label{en:app-static-difference-components}
\end{equation}
The symbols \(Z_{R_I}\) and \(Z_M\) denote, respectively, the labeled radius
and total mass components of the difference tuple \(Z\).
Since \(\pi_{\mathrm{cen}}\circ X=\operatorname{Id}_U\) and \(m_I(0)=0\),
\begin{equation}
  z_I(0)=\delta H_I(0)-D(H_I^c)_\sigma[v]
  =v^I-v^I=0,\qquad
  \nu_I(0)=0 .
  \label{en:app-static-zero-center-data}
\end{equation}
Put \(C_\Phi:=\varphi(0)\) and
\(\widehat\varphi:=\varphi-C_\Phi\).  The
linearized enthalpy equation gives
\begin{equation}
  z_I'=-\widehat\varphi',\qquad
  z_I(0)=\widehat\varphi(0)=0
  \quad\Longrightarrow\quad
  z_I=-\widehat\varphi
  \quad\text{on every active component interval}.
  \label{en:app-static-H-phi-relation}
\end{equation}

\emph{Center.}
Choose \(r_c\in(0,\mathcal R_1)\) in the central active cell and put
\(\epsilon_{I,H}:=\partial_{H_I}\epsilon_I\).  The mass and lapse variations satisfy
\begin{align}
  \nu_I(r)
  &=4\pi\int_0^r s^2\epsilon_{I,H}(s)z_I(s)\,\dd s,
  \label{en:app-static-center-mass-difference}\\
  \widehat\varphi'(r)
  &=\frac{1+2r\Phi'(r)}{r^2A(r)}\,\nu_{\mathrm{tot}}(r)
    +\frac{4\pi r}{A(r)}
      \sum_{J=1}^K h_J(r)z_J(r).
  \label{en:app-static-center-lapse-difference}
\end{align}
Here \(\epsilon_{I,H},h_I\) are bounded, \(A=1+O(r^2)\), and
\(\Phi'=O(r)\) at the center.  Define
\begin{equation}
  z_{\mathrm{sup}}(r):=\max_{1\le I\le K}\sup_{0\le s\le r}|z_I(s)|,
  \label{en:app-static-center-supremum}
\end{equation}
The preceding bounds give
\begin{align}
  |\nu_{\mathrm{tot}}(r)|&\le Cr^3z_{\mathrm{sup}}(r),&
  \left|\frac{1+2r\Phi'}{r^2A}\right|&\le Cr^{-2},&
  \left|\frac{4\pi rh_I}{A}\right|&\le Cr,
  \label{en:app-static-center-coefficient-estimates}\\
  |\widehat\varphi'(r)|
  &\le Crz_{\mathrm{sup}}(r),&
  z_{\mathrm{sup}}(r)&\le C\int_0^r sz_{\mathrm{sup}}(s)\,\dd s .
  \label{en:app-static-center-Volterra-bound}
\end{align}
For \(z_{\mathrm{int}}(r):=\int_0^r sz_{\mathrm{sup}}(s)\,\dd s\),
\begin{equation}
  z_{\mathrm{int}}'(r)\le Crz_{\mathrm{int}}(r),\qquad
  z_{\mathrm{int}}(0)=0
  \quad\Longrightarrow\quad
  z_{\mathrm{int}}(r)=z_{\mathrm{sup}}(r)=0\quad(0\le r\le r_c),
  \label{en:app-static-Gronwall-conclusion}
\end{equation}
by the zero-initial-value case of Gronwall's inequality
\cite{WalterODE}.  Hence
\begin{equation}
  z_I=\nu_I=0,\qquad \widehat\varphi'=0
  \quad\text{on }[0,r_c].
  \label{en:app-static-center-conclusion}
\end{equation}

\emph{Open active cells.}
For \(a=1,\ldots,q\), put
\begin{equation}
  e_a:=(1)_{I\in\mathcal K_a}\in\mathbb R^{|\mathcal K_a|},
  \qquad
  e_K:=(1)_{I=1}^K\in\mathbb R^K .
  \label{en:app-static-cell-ones}
\end{equation}
Define the closed difference state
\begin{equation}
  Z_a^{\mathrm{cell}}(r):=
  \begin{pmatrix}
    (z_I(r))_{I\in\mathcal K_a}\\
    (\nu_I(r))_{I=1}^K
  \end{pmatrix}.
  \label{en:app-static-cell-state}
\end{equation}
All component masses are retained in \(Z_a^{\mathrm{cell}}\); if \(I\notin\mathcal K_a\),
then \(\nu_I'=0\).  On \(\mathcal I_a\), the linearized equations give
\begin{equation}
  (Z_a^{\mathrm{cell}})'=A_a^{\mathrm{sys}}(r)Z_a^{\mathrm{cell}},\qquad
  A_a^{\mathrm{sys}}(r):=
  \begin{pmatrix}
    -e_a(\beta_a^{\mathrm{cell}})^{\mathsf T}
      &-\gamma_a^{\mathrm{cell}} e_a e_K^{\mathsf T}\\
    A_a^{\mathrm{mass}}&0
  \end{pmatrix},
  \label{en:app-static-cell-IVP}
\end{equation}
where
\begin{equation}
  \gamma_a^{\mathrm{cell}}:=\frac{1+2r\Phi'}{r^2A},\qquad
  (\beta_a^{\mathrm{cell}})_I:=\frac{4\pi rh_I}{A},\qquad
  (A_a^{\mathrm{mass}})_{IJ}:=
  \begin{cases}
    4\pi r^2\epsilon_{I,H},&I=J\in\mathcal K_a,\\
    0,&\text{otherwise}.
  \end{cases}
  \label{en:app-static-cell-coefficients}
\end{equation}
At the right endpoint of \(\mathcal I_a\),
\begin{equation}
  \epsilon_{I,H}(r)=
  \begin{cases}
    O((\mathcal R_a-r)^{\alpha_I-1}),&I\in\mathcal B_a,\\
    O(1),&I\in\mathcal K_a\setminus\mathcal B_a,
  \end{cases}
  \qquad \alpha_I>0,
  \label{en:app-static-cell-E-bound}
\end{equation}
and the left endpoint is regular for \(a\ge2\).  Therefore
\begin{equation}
  A_1^{\mathrm{sys}}\in L^1((r_c,\mathcal R_1)),\qquad
  A_a^{\mathrm{sys}}\in L^1(\mathcal I_a)
  \quad(a=2,\ldots,q).
  \label{en:app-static-cell-L1}
\end{equation}
Put
\begin{equation}
  s_1:=r_c,\qquad
  s_a:=\mathcal R_{a-1}\quad(a=2,\ldots,q).
  \label{en:app-static-cell-left-points}
\end{equation}
For \(r\in(s_a,\mathcal R_a)\),
\begin{align}
  Z_a^{\mathrm{cell}}(r)
  &=Z_a^{\mathrm{cell}}(s_a^+)
    +\int_{s_a}^rA_a^{\mathrm{sys}}(s)Z_a^{\mathrm{cell}}(s)\,\dd s,
  \label{en:app-static-cell-integral}\\
  |Z_a^{\mathrm{cell}}(r)|
  &\le |Z_a^{\mathrm{cell}}(s_a^+)|
      \exp\!\left(\int_{s_a}^r\|A_a^{\mathrm{sys}}(s)\|\,\dd s\right).
  \label{en:app-static-cell-Gronwall}
\end{align}
Taking \(s_1=r_c\), the center conclusion gives
\(Z_1^{\mathrm{cell}}=0\) on \((r_c,\mathcal R_1)\).

\emph{Surface nodes.}
If \(Z_a^{\mathrm{cell}}(\mathcal R_a^-)=0\), the mass equation and \conditionrefp{S2} give
\begin{equation}
  \nu_I(\mathcal R_a^-)=0\qquad(I=1,\ldots,K).
  \label{en:app-static-node-left-mass}
\end{equation}
For every \(a=1,\ldots,q\) and every ending label,
\begin{equation}
  Z_{R_I}
  =-\frac{z_I(\mathcal R_a^-)}
          {H_I'(\mathcal R_a)}
  =0,
  \qquad I\in\mathcal B_a,
  \label{en:app-block-surface-variations}
\end{equation}
by \conditionrefp{S3} and \(H_I'(\mathcal R_a)<0\).  If \(a<q\), then
\begin{equation}
  \mathcal K_{a+1}=\mathcal K_a\setminus\mathcal B_a
  =\mathcal A_a^+ .
  \label{en:app-static-node-active-set}
\end{equation}
The mass and no-layer conditions give
\begin{equation}
  \nu_I(\mathcal R_a^+)=0\quad(I=1,\ldots,K),\qquad
  z_J(\mathcal R_a^+)=z_J(\mathcal R_a^-)=0
  \quad(J\in\mathcal K_{a+1}),
  \label{en:app-static-node-data}
\end{equation}
whenever \(Z_a^{\mathrm{cell}}(\mathcal R_a^-)=0\).  Moreover,
\begin{equation}
  \varphi(\mathcal R_a^+)=\varphi(\mathcal R_a^-)=C_\Phi,
  \qquad
  Z_{a+1}^{\mathrm{cell}}(\mathcal R_a^+)=0 .
  \label{en:app-static-node-induction}
\end{equation}
The integral formula \eqref{en:app-static-cell-integral} and
\eqref{en:app-static-cell-Gronwall} therefore give
\(Z_{a+1}^{\mathrm{cell}}=0\) on
\(\mathcal I_{a+1}\).  Induction over all surface nodes yields
\begin{equation}
  z_I=0,\qquad \nu_I=0
  \quad(0<r<R_I,\ I=1,\ldots,K),\qquad
  \widehat\varphi'=0.
  \label{en:app-static-global-nongauge-zero}
\end{equation}
The constitutive and metric identities then give
\begin{equation}
\begin{aligned}
  \delta\epsilon_I-(B_\sigma v)_{\epsilon_I}&=0,&
  \delta p_I-(B_\sigma v)_{p_I}&=0,\\
  \delta\Lambda-(B_\sigma v)_\Lambda
  &=\frac{\nu_{\mathrm{tot}}}{rA}=0,&
  Z_M&=\nu_{\mathrm{tot}}(R_{\mathrm{out}}^-)=0 .
\end{aligned}
  \label{en:app-static-derived-zero}
\end{equation}
\emph{Exterior and lapse constant.}
By \eqref{en:app-static-global-nongauge-zero},
\(\varphi=C_\Phi\) throughout the star.  In the vacuum exterior,
\begin{equation}
  \nu_{\mathrm{tot}}=Z_M,\qquad
  \delta\Phi-(B_\sigma v)_\Phi
  =-\frac{Z_M}{rA}+C_{\mathrm{vac}}.
  \label{en:app-static-vacuum-lapse}
\end{equation}
The outer matching condition \eqref{en:eq-static-outer-matching} and
\(\delta\Phi(\infty)=0\) give
\begin{equation}
  C_\Phi=C_{\mathrm{vac}},\qquad
  Z_M=0,\qquad
  C_{\mathrm{vac}}=0 .
  \label{en:app-static-gauge-fixing}
\end{equation}
Hence \(Z=0\), so every static perturbation is \(B_\sigma v\).  Finally,
the central enthalpy components of \(B_\sigma v\) are \(v^I\); therefore
\(B_\sigma v=0\) implies \(v=0\), and \(B_\sigma\) is an isomorphism.
\end{proof}

\subsection{Particle conservation and the material displacement}
\label{en:app-proof-particle-package}

\begin{lemma}[Particle conservation and displacement augmentation]
\label{en:lem-particle-package}
Here a regular radial material perturbation is a \(C^1\) one-parameter family
of spherically symmetric configurations equipped, for each component \(I\),
with shell maps \(F_{I,\tau}^{\mathrm{mat}}\) that are orientation-preserving
\(C^1\) diffeomorphisms between the unperturbed and perturbed radial
intervals, depend \(C^1\) on \(\tau\), satisfy
\(F_{I,0}^{\mathrm{mat}}=\operatorname{Id}\), and carry every unperturbed
shell to the perturbed shell containing the same material particles; the
infinitesimal generator of this Lagrangian shell map is \(\xi_I\).  Writing
\(g_{I,\tau}(r)\,\dd r\) for the shell particle number form in the perturbed
configuration, separate particle conservation gives
\begin{equation}
  (F_{I,\tau}^{\mathrm{mat}})^*\bigl(g_{I,\tau}(r)\,\dd r\bigr)
  =g_I(r)\,\dd r,
  \qquad
  F_{I,\tau}^{\mathrm{mat}}(r)=r+\tau\xi_I(r)+o(\tau).
  \label{en:eq-material-number-pullback}
\end{equation}
Then:
\begin{enumerate}
\renewcommand{\theenumi}{\alph{enumi}}
\renewcommand{\labelenumi}{(\theenumi)}
  \item \label{en:item-particle-conservation}
  \begin{equation}
    \delta g_I+(g_I\xi_I)'=0,
    \qquad
    \delta_E\mathcal N_I+g_I\xi_I=0;
    \label{en:eq-local-particle-conservation}
  \end{equation}
  finite surface limits imply \(\delta N_I=0\);
  \item \label{en:item-displacement-augmentation}
  if \(\delta X=B_\sigma v\) and \(DN_\sigma v=0\), then
  \begin{equation}
    \xi_{I,v}(r)=-\frac{\delta_{E,v}\mathcal N_I(r)}{g_I(r)}
    \label{en:eq-xi-construction}
  \end{equation}
  is the unique center-regular displacement satisfying part (a), and
  \begin{equation}
    \xi_{I,v}(r)\longrightarrow\delta_vR_I
  \qquad(r\to R_I^-).
    \label{en:eq-xi-surface-limit}
  \end{equation}
  It defines the unique augmentation
  \(A_\sigma:\mathcal S_{\sigma,N}^{\mathrm{stat}}
  \to\mathcal S_{\sigma,0,N}^{\mathrm{full}}\).  With the forgetful map
  \(F_\sigma\) defined in \eqref{en:eq-forgetful-map},
  the two maps satisfy
  \begin{equation}
    F_\sigma A_\sigma=\operatorname{Id},
    \qquad
    A_\sigma F_\sigma=\operatorname{Id}.
    \label{en:eq-augmentation-forgetful-inverses}
  \end{equation}
The identities in \eqref{en:eq-augmentation-forgetful-inverses} show that
\(A_\sigma\) and \(F_\sigma\) are inverse linear isomorphisms.
\end{enumerate}
\end{lemma}

\setproofparent{en:lem-particle-package}

\begin{proof}[Proof of part (a)]
On a time slice with future unit normal \(\nu^\mu\), the shell charge is
\[
  g_{I,\tau}\,\dd r=4\pi r^2e^\Lambda n_I(-u_I^\mu\nu_\mu)\,\dd r .
\]
The relative three-velocity vanishes in equilibrium, so the Lorentz factor
is one there and its first variation is zero. Thus the first variation
of \(4\pi r^2e^\Lambda n_I\) gives the linear shell density even for a
time-dependent displacement.  The material map
\(F_{I,\tau}^{\mathrm{mat}}\) in \eqref{en:eq-material-number-pullback} compares the same
fluid particle labels.  Differentiating its pullback identity at \(\tau=0\)
gives
\begin{equation}
  \left.\frac{\dd}{\dd\tau}\right|_{\tau=0}
  (F_{I,\tau}^{\mathrm{mat}})^*\bigl(g_{I,\tau}\,\dd r\bigr)
  =\{\delta g_I+\xi_Ig_I'+g_I\xi_I'\}\dd r
  =\{\delta g_I+(g_I\xi_I)'\}\dd r.
\end{equation}
The left side is zero by \eqref{en:eq-material-number-pullback}, so the
bracket vanishes.  Separate current conservation propagates this material
identity in time; its zero-frequency restriction is the particle label
condition.
Integrating from the regular center gives
\begin{equation}
  \delta_E\mathcal N_I(r)=-g_I(r)\xi_I(r).
  \label{en:app-particle-inner-variation}
\end{equation}
Differentiation recovers the local conservation law.  This shell identity expresses Lagrangian particle conservation. For its classical fluid formulation see Ref.~\cite{FriedmanSchutz1978}, and for the relativistic multifluid perturbation equations see Refs.~\cite{Kain2020,CaballeroRipleyYunes2024}.
At an ordinary vacuum surface,
\(g_I=O((R_I-r)^{\alpha_I})\); hence a finite displacement limit gives
\(g_I\xi_I\to0\) and \(\delta_E\mathcal N_I(R_I)=0\).  The moving-endpoint
formula is therefore
\begin{equation}
  \delta N_I
  =\delta_E\mathcal N_I(R_I)+g_I(R_I)\delta R_I=0.
  \label{en:app-total-particle-conservation}
\end{equation}

\end{proof}

\begin{proof}[Proof of part (b)]
Fix \(\delta X\in\mathcal S_{\sigma,N}^{\mathrm{stat}}\).  Surjectivity in
Lemma~\ref{en:lem-static-representation} gives the unique family
direction \(v\) with
\(\delta X=B_\sigma v\), and \eqref{en:eq-xi-construction} gives local
conservation.  At the center put \(s=r^2\).  Rewriting the center equations
in this variable and using \conditionref{H1} gives
\[
  g_I(r;\sigma)=r^2g_{I,\mathrm{cen}}(r^2;\sigma),
\]
where \(g_{I,\mathrm{cen}}\) and its family derivative are \(C^1\) in \(s\)
near zero and \(g_{I,\mathrm{cen}}(0;\sigma)>0\).  Therefore
\begin{equation}
  \delta_E\mathcal N_I(r)
  =r^3\int_0^1t^2\delta g_{I,\mathrm{cen}}(r^2t^2)\,\dd t,
  \qquad
  \xi_I(r)
  =-r\frac{\int_0^1t^2\delta g_{I,\mathrm{cen}}(r^2t^2)\,\dd t}
             {g_{I,\mathrm{cen}}(r^2)}.
\end{equation}
This quotient is \(rF_I(r^2)\), with \(F_I\in C^1\).  Hence, setting
\(a_I^{\xi,\mathrm{cen}}
=-\delta g_{I,\mathrm{cen}}(0)/(3g_{I,\mathrm{cen}}(0))\),
\begin{equation}
  \xi_I=a_I^{\xi,\mathrm{cen}}r+O(r^3),\qquad
  \xi_I'=a_I^{\xi,\mathrm{cen}}+O(r^2).
\end{equation}
If \(\xi_I\) and \(\widetilde\xi_I\) satisfy the same local equation and
center condition,
then \([g_I(\xi_I-\widetilde\xi_I)]'=0\); its center constant is zero, so
positivity of \(g_I\) gives uniqueness.

The quotient is \(C^2\) on every open active cell.  At an internal ending block,
\(\delta_E\mathcal N_I(r)=\int_0^r\delta g_I(s)\,\dd s\) has matching one-sided
limits while \(g_I(\mathcal R_a)>0\), so each surviving \(\xi_I\) is
continuous.  The required one-sided limits of the pressure and metric fields
are already part of the definition of \(\mathcal S_\sigma^{\mathrm{stat}}\).

On each open active cell,
\begin{equation}
  g_I=4\pi r^2e^\Lambda n_I,
  \qquad
  \frac{\delta g_I}{g_I}
  =\frac{\delta n_I}{n_I}+\delta\Lambda,
  \label{en:app-augmentation-g-variation}
\end{equation}
and a direct calculation gives
\begin{equation}
  \frac{(g_I\xi_I)'}{g_I}
  =\frac{e^\Phi}{n_Ir^2}
    (n_Ir^2e^{-\Phi}\xi_I)'
   +(\Lambda'+\Phi')\xi_I.
  \label{en:app-augmentation-flux-identity}
\end{equation}
Substitution of \eqref{en:app-augmentation-flux-identity} into
\eqref{en:eq-local-particle-conservation} gives precisely the particle number
equation required in \conditionrefp{F3}:
\begin{equation}
  \frac{\delta n_I}{n_I}
  =-\frac{e^\Phi}{n_Ir^2}
    (n_Ir^2e^{-\Phi}\xi_I)'
   -\delta\Lambda-(\Lambda'+\Phi')\xi_I.
  \label{en:app-augmentation-number-equation}
\end{equation}
The linearized EoS relations hold by construction.  The derived metric
identity \(rA\delta\Lambda=\delta m_{\mathrm{tot}}\), together with
\conditionrefp{S2}, gives the Hamiltonian equation.  The linearization of
\eqref{en:eq-tov-phi} gives the static \(rr\) equation, and combining it
with the linearization of \eqref{en:eq-tov-pressure} gives the Euler equation;
both are part of \conditionrefp{S1}.  These identities complete the
verification of \conditionrefp{F3}.

The tail estimate in
Lemma~\statementpartref{en:lem-equilibrium-package}{en:item-surface-tail},
together
with \(\delta N_I=0\), gives
\(\delta_E\mathcal N_I=-g_I\delta R_I+o(g_I)\).  Since
\(\xi_I=-\delta_E\mathcal N_I/g_I\), we obtain
\(\xi_I\to\delta R_I\).  Differentiating \(H_I(R_I)=0\) gives
\(\Delta H_I(R_I)=0\), so \conditionrefp{F5} holds.

Together with the center, transmission, endpoint, and exterior properties
established above or inherited from \(\delta X\), this proves
\begin{equation}
  A_\sigma\delta X
  :=\bigl(\delta X,(\xi_I)_{I=1}^K\bigr)
  \in\mathcal S_{\sigma,0,N}^{\mathrm{full}}.
  \label{en:app-augmentation-membership}
\end{equation}

Forgetting the augmented displacements gives
\(F_\sigma A_\sigma=\operatorname{Id}\).  Conversely, start with any full
perturbation.  Its particle number conservation equation and center regularity give
\([g_I(\xi_I-\xi_{I,A})]'=0\) with zero center constant, where
\(\xi_{I,A}\) is reconstructed by \eqref{en:eq-xi-construction}.
Positivity of \(g_I\) and continuity through internal ending blocks give
equality everywhere, which proves
\(A_\sigma F_\sigma=\operatorname{Id}\) and completes part~\textup{(b)}.
\end{proof}

\subsection{Recovery of the fields and matching at the surfaces}
\label{en:app-proof-reduction-package}

\begin{theorem}[Bulk recovery and boundary condition equivalence]
\label{en:thm-reduction-package}
Let \(\widehat R_\sigma\), \(\widehat{\mathcal C}_\sigma\), and
\(\widehat{\mathcal D}_\sigma\) be the recovery map, ambient class, and
reduced domain defined above.  Then:
\begin{enumerate}
\renewcommand{\theenumi}{\alph{enumi}}
\renewcommand{\labelenumi}{(\theenumi)}
  \item \label{en:item-residual-factorization}
  on every active cell, for every
  \(\zeta\in\widehat{\mathcal C}_\sigma\),
  all particle number conservation, EoS, Hamiltonian,
  component mass, and \(rr\) residuals vanish.  At this ambient level the
  mass--metric compatibility quantity $C_\zeta^{m\Lambda}$ satisfies the cellwise identity
  \begin{equation}
    (C_\zeta^{m\Lambda})'=0,
    \qquad
    C_\zeta^{m\Lambda}
    :=rA\delta\Lambda[\zeta]-\delta m_{\mathrm{tot}}[\zeta].
    \label{en:eq-ambient-metric-compatibility}
  \end{equation}
  Once \conditionref{D1} and \conditionref{D3} are imposed, the center and
  transmission conditions give
  \(C_\zeta^{m\Lambda}=0\) throughout the star, hence
  \(\delta m_{\mathrm{tot}}=rA\delta\Lambda\).  The remaining Euler residual is
  \begin{equation}
    \mathcal E_I(\widehat R_\sigma\zeta)
    =e^{-\Lambda-2\Phi}(\mathcal L_\sigma\zeta)_I.
    \label{en:eq-residual-factorization}
  \end{equation}
  Since \(e^{-\Lambda-2\Phi}>0\), the multiplier is pointwise invertible, and
  \(\mathcal L_\sigma\) is the radial zero-frequency pulsation expression.
  None of the recovery formulas divides by \(\omega\), so they remain
  well defined at \(\omega=0\);
  \item \label{en:item-boundary-equivalence}
  \emph{Boundary condition equivalence.}
  For every \(\zeta\in\widehat{\mathcal C}_\sigma\),
  \begin{equation}
    \widehat R_\sigma\zeta\in\mathcal X_{\mathrm{full}}^{\mathrm{tr}}
    \ \text{and}\
    \operatorname{BC}_{\mathrm{full}}(\widehat R_\sigma\zeta)=0
    \quad\Longleftrightarrow\quad
    \zeta\in\widehat{\mathcal D}_\sigma,
    \label{en:eq-BC-equivalence}
  \end{equation}
  In other words, this equivalence translates the one-sided limit conditions
  built into \(\widehat{\mathcal D}_\sigma\) into the corresponding conditions
  on the recovered full field tuple.
  When these conditions hold, the matching one-sided limits also remove all
  junction-supported delta terms, so the cellwise residual identities in part (a)
  agree with their distributional versions.
  All full boundary blocks and their trace spaces are listed in
  \eqref{en:eq-full-center-block}--\eqref{en:eq-full-boundary-operator}.
\end{enumerate}
\end{theorem}

\setproofparent{en:thm-reduction-package}

\subsubsection{Static system and recovery formulas}

The background identities are
\begin{equation}
  H_I'=\frac{p_I'}{h_I}=-\Phi',
  \qquad
  p_I'=h_IH_I',
  \qquad
  \Lambda'+\Phi'=4\pi rA^{-1}h_{\mathrm{tot}}.
  \label{en:app-background-identities}
\end{equation}
On each fixed open active cell, the static system is
\begin{align}
  \frac{\delta n_I}{n_I}
  &=-\frac{e^\Phi}{n_Ir^2}
  (n_Ir^2e^{-\Phi}\xi_I)'
  -\delta\Lambda-(\Lambda'+\Phi')\xi_I,
  \label{en:app-zf-number}\\
  \delta\epsilon_I
  &=\frac{h_I}{n_I}\delta n_I,
  \qquad
  \delta p_I=c_{s,I}^2\delta\epsilon_I,
  \label{en:app-zf-eos}\\
  (rA\delta\Lambda)'
  &=4\pi r^2\sum_I\delta\epsilon_I,
  \label{en:app-zf-Hamiltonian}\\
  \delta\Phi'
  &=4\pi rA^{-1}\sum_I\delta p_I
   +\left(2\Phi'+\frac1r\right)\delta\Lambda,
  \label{en:app-zf-rr}\\
  0&=\delta p_I'+h_I\delta\Phi'
  +(\delta\epsilon_I+\delta p_I)\Phi'.
  \label{en:app-zf-Euler}
\end{align}
The center conditions are
\begin{equation}
  \lim_{r\to0^+}rA\delta\Lambda=0,
  \qquad \xi_I=O(r).
  \label{en:app-zf-center}
\end{equation}
\subsubsection{Metric constraint identity}
Every piecewise solution of \eqref{en:app-zf-number}--
\eqref{en:app-zf-Hamiltonian} satisfying
\eqref{en:app-zf-center} and the internal no-layer transmission
conditions in \conditionref{F4}, with \(h_I\xi_I\to0\) for every
label ending at an internal node, obeys
\begin{equation}
  \delta\Lambda=-4\pi rA^{-1}\sum_Ih_I\xi_I.
  \label{en:app-zf-lambda}
\end{equation}

\begin{proof}[Proof of the metric constraint identity]
Set
\begin{equation}
  C_{\mathrm{con}}:=rA\delta\Lambda+4\pi r^2\sum_Ih_I\xi_I.
  \label{en:app-C-definition}
\end{equation}
Equations \eqref{en:app-zf-number}, \eqref{en:app-zf-eos}, and
\eqref{en:app-background-identities} give
\begin{equation}
  r^2\delta\epsilon_I+(r^2h_I\xi_I)'
  =-r^2h_I[\delta\Lambda+(\Lambda'+\Phi')\xi_I].
  \label{en:app-cancellation}
\end{equation}
Summing and using \eqref{en:app-zf-Hamiltonian} gives
\begin{equation}
  C_{\mathrm{con}}'=-\frac{4\pi rh_{\mathrm{tot}}}{A}C_{\mathrm{con}}.
  \label{en:app-C-ODE}
\end{equation}
Since \(C_{\mathrm{con}}(0)=0\), the same Carath\'eodory--Gronwall
uniqueness theorem \cite{WalterODE} gives
\(C_{\mathrm{con}}=0\) on the innermost active cell.  At every internal
ending block \(\mathcal R_a\), the background junction regularity and
\conditionref{F4} imply
\[
  [C_{\mathrm{con}}]_{\mathcal R_a}=0.
\]
Indeed, \(\delta\Lambda\) and every surviving displacement have
matching one-sided limits, while \(h_I\xi_I\to0\) for every ending
label.  Reapplying the same uniqueness theorem on each successive
active cell therefore propagates \(C_{\mathrm{con}}=0\) throughout the star.
\end{proof}

On each active cell, use the enthalpy-weighted displacement \(\bar\zeta\)
defined in Remark~\ref{en:rem-material-surface-condition} and put
\begin{equation}
  \xi_I=\frac{e^\Phi}{r^2}\zeta_I,
  \label{en:app-zf-xi}
\end{equation}
define the recovered fields by
\begin{align}
  \delta\Lambda[\zeta]
  &=-\frac{4\pi e^\Phi}{rA}\sum_Jh_J\zeta_J,
  \label{en:app-recover-lambda}\\
  \delta p_I[\zeta]
  &=-\frac{e^\Phi}{r^2}
  [b_I\zeta_I'+p_I'\zeta_I
  +b_I(\Lambda'+\Phi')(\zeta_I-\bar\zeta)],
  \label{en:app-recover-p}\\
  \delta H_I[\zeta]
  &:=\frac{\delta p_I[\zeta]}{h_I}
  =-\frac{e^\Phi}{r^2}
  [c_{s,I}^2\zeta_I'+H_I'\zeta_I
  +c_{s,I}^2(\Lambda'+\Phi')(\zeta_I-\bar\zeta)],
  \label{en:app-recover-H}\\
  \delta\epsilon_I[\zeta]
  &=c_{s,I}^{-2}\delta p_I[\zeta],
  \qquad
  \delta n_I[\zeta]=\frac{n_I}{h_I}\delta\epsilon_I[\zeta],
  \label{en:app-recover-matter}\\
  \delta m_I[\zeta](r)
  &=4\pi\int_0^r s^2\bigl(\delta\epsilon_I[\zeta]\bigr)^0(s)\,\dd s,
  \qquad
  \delta m_{\mathrm{tot}}[\zeta]=\sum_I\delta m_I[\zeta],
  \label{en:app-recover-mass}\\
  \delta\Phi'_{\mathrm{int}}[\zeta]
  &=4\pi rA^{-1}\sum_J\delta p_J[\zeta]
  +\left(2\Phi'+\frac1r\right)\delta\Lambda[\zeta],
  \qquad 0<r<R_{\mathrm{out}}.
  \label{en:app-recover-phi-prime}
\end{align}
On active cells we henceforth abbreviate
\(\delta\Phi'_{\mathrm{int}}[\zeta]\) to \(\delta\Phi'[\zeta]\).
For a vector satisfying \conditionref{C1}--\conditionref{C3}, the
integrability in \conditionref{C2} and the finite displacement trace in
\conditionref{C3} make the following global data well defined.  Set
\begin{equation}
  \delta R_I
  :=\frac{e^{\Phi(R_I)}}{R_I^2}\zeta_I(R_I^-)
  =\lim_{r\to R_I^-}\xi_I(r),
  \qquad
  \delta M:=\delta m_{\mathrm{tot}}(R_{\mathrm{out}}^-).
  \label{en:app-recover-global-data}
\end{equation}
If \conditionref{D2} holds, then the limit in
\eqref{en:app-B-surface} exists and vanishes, and hence
\begin{equation}
  \lim_{r\to R_I^-}\delta H_I[\zeta](r)
  =-H_I'(R_I)\xi_I(R_I^-)
  =-H_I'(R_I)\delta R_I.
  \label{en:eq-recovered-enthalpy-surface-trace}
\end{equation}
The quotient \(\delta p_I/h_I\) is used only on the active interval
\(r<R_I\).  Recover the
lapse inward from the normalized Schwarzschild boundary value,
\begin{equation}
  \delta\Phi[\zeta](r)
  :=-\frac{\delta M}{R_{\mathrm{out}}A(R_{\mathrm{out}})}
    -\int_r^{R_{\mathrm{out}}}\delta\Phi'_{\mathrm{int}}[\zeta](s)\,\dd s,
  \qquad 0<r\le R_{\mathrm{out}}.
  \label{en:app-recover-phi-interior}
\end{equation}
For every vector satisfying \conditionref{C1}--\conditionref{C3}, this
formula defines a locally absolutely continuous
(\(\mathrm{AC}_{\mathrm{loc}}\)) lapse on \((0,R_{\mathrm{out}}]\).
Condition~\conditionref{D1} supplies its
regular limit at \(r=0\).
For \(r\ge R_{\mathrm{out}}\), the recovery is defined explicitly by
\begin{equation}
  \delta m_{\mathrm{tot}}[\zeta](r)=\delta M,
  \qquad
  \delta\Lambda[\zeta](r)=\frac{\delta M}{rA(r)},
  \qquad
  \delta\Phi[\zeta](r)=-\frac{\delta M}{rA(r)}.
  \label{en:app-recover-exterior}
\end{equation}
The exterior formula builds in \(\delta\Phi(\infty)=0\).  Together, these
formulas realize the recovery map \(\widehat R_\sigma\) defined in
\eqref{en:eq-recovery-map-tuple}.
The reduced differential expression is
\begin{equation}
  (\mathcal L_\sigma\zeta)_I
  =e^{\Lambda+2\Phi}
  [\delta p_I[\zeta]'+h_I\delta\Phi'[\zeta]
  +(\delta\epsilon_I[\zeta]+\delta p_I[\zeta])\Phi'].
  \label{en:app-reduced-operator}
\end{equation}

\begin{sublemma}[Zero-frequency recovery identities]
\label{en:lem-canonical-zero-recovery}
Let \(\zeta\) be a real vector satisfying the preliminary requirements
\textup{(C1)--(C3)} in Definition~\ref{en:def-reduced-domain}, set
\(\xi_I=e^\Phi\zeta_I/r^2\), and define
\(\widehat R_\sigma\zeta\) by
\eqref{en:app-recover-lambda}--\eqref{en:app-recover-exterior}.  The
cellwise conclusions below use only these preliminary requirements; each
global conclusion states its additional domain conditions explicitly.  Then:
\begin{enumerate}
\renewcommand{\theenumi}{\alph{enumi}}
\renewcommand{\labelenumi}{(\theenumi)}
  \item \label{en:item-canonical-bulk}
  On every open active cell, the particle number, EoS,
  Hamiltonian, component mass, and static \(rr\) residuals vanish, and
  \((C_\zeta^{m\Lambda})'=0\), with \(C_\zeta^{m\Lambda}\) as in
  \eqref{en:eq-ambient-metric-compatibility}.  The remaining Euler residual
  factors as
  \begin{equation}
    \mathcal E_I(\widehat R_\sigma\zeta)
    =e^{-\Lambda-2\Phi}(\mathcal L_\sigma\zeta)_I.
    \label{en:app-Euler-factorization}
  \end{equation}

  \item \label{en:item-canonical-global-metric}
  If \conditionref{D1} and \conditionref{D3} hold, then
  \(C_\zeta^{m\Lambda}=0\) throughout the star.

  \item \label{en:item-canonical-surface}
  At a labeled surface, whenever the one-sided limit in
  \eqref{en:app-B-surface} exists,
  \begin{equation}
    \lim_{r\to R_I^-}
      \bigl(\delta H_I[\zeta]+H_I'\xi_I\bigr)
    =\widehat B_{I,\mathrm{s}}(\zeta).
    \label{en:eq-canonical-surface-identity}
  \end{equation}
  In particular, \conditionref{D2} is equivalent to the recovered material
  surface condition \(\Delta H_I(R_I)=0\).

  \item \label{en:item-canonical-particle}
  On an open active cell \(\mathcal I_a\subset(0,R_I)\), choose any
  \(r_{I,a}\in\mathcal I_a\) and set
  \(\delta\mathcal N_{I,a}^{\mathrm{loc}}(r)
    :=\int_{r_{I,a}}^r\delta g_I(s)\,\dd s\).  The recovered number equation
  gives
  \begin{equation}
    \bigl(\delta\mathcal N_{I,a}^{\mathrm{loc}}+g_I\xi_I\bigr)'=0,
    \qquad
    \delta\mathcal N_{I,a}^{\mathrm{loc}}(r)+g_I(r)\xi_I(r)=C_{I,a},
    \quad r\in\mathcal I_a,
    \label{en:app-recovered-number-cell-constant}
  \end{equation}
  for a cellwise constant \(C_{I,a}\).  If \conditionref{D1} holds, the
  center behavior and the endpoint weights make \(\delta g_I\) integrable on
  \((0,R_I)\), so the physical primitive
  \(\delta_E\mathcal N_I(r)=\int_0^r\delta g_I(s)\,\dd s\) is defined.  If
  \conditionref{D3} also holds, the cellwise constants of
  \(\delta_E\mathcal N_I+g_I\xi_I\) agree across all active cells and vanish,
  so
  \begin{equation}
    \delta_E\mathcal N_I(r)=-g_I(r)\xi_I(r).
    \label{en:app-recovered-enclosed-number}
  \end{equation}
  throughout \(0<r<R_I\).  The finite labeled surface limit in
  \conditionref{C3} then gives
  \begin{equation}
    \delta N_I=0.
    \label{en:app-recovered-total-number}
  \end{equation}

\end{enumerate}
\end{sublemma}

\begin{proof}
The algebraic metric constraint calculation
\eqref{en:app-C-definition}--\eqref{en:app-C-ODE} first gives
  \eqref{en:app-zf-lambda}; the recovered Hamiltonian equation follows in the
  next step.
For the recovered fields, division of
\eqref{en:app-recover-p} by \(b_I\), together with
\(n_I'/n_I=p_I'/b_I\), gives
\[
 \frac{\delta n_I}{n_I}
 =-\frac{e^\Phi}{n_Ir^2}(n_I\zeta_I)'
   -\delta\Lambda-(\Lambda'+\Phi')\xi_I,
\]
which is \eqref{en:app-zf-number}; the constitutive, component mass, and
\(rr\) equations hold by definition.  If \(C_{\mathrm{Ham}}\) is the
Hamiltonian residual, then
\(C_{\mathrm{con}}'=C_{\mathrm{Ham}}
 -(4\pi rh_{\mathrm{tot}}/A)C_{\mathrm{con}}\).
The definition of the recovered \(\delta\Lambda\) gives
\(C_{\mathrm{con}}=0\), hence \(C_{\mathrm{Ham}}=0\), and the component mass
equations give \((C_\zeta^{m\Lambda})'=0\).  Equation
~\eqref{en:app-reduced-operator} is exactly
the Euler factorization \eqref{en:app-Euler-factorization}.

Condition~\conditionref{D1} gives \(C_\zeta^{m\Lambda}(0)=0\), while the integral
mass recovery and \conditionref{D3} ensure continuity at every internal node;
hence its cellwise constant is
zero globally.  Adding \(H_I'\xi_I\) to \eqref{en:app-recover-H} gives
\eqref{en:app-B-surface} after taking the inner limit specified above.

Finally, using \(g_I=4\pi r^2e^\Lambda n_I\), the recovered number equation
is equivalent on every open active cell to
\begin{equation}
  \delta g_I+(g_I\xi_I)'=0.
  \label{en:app-recovered-local-number-form}
\end{equation}
Integrating from an arbitrary base point in that cell gives
\eqref{en:app-recovered-number-cell-constant} directly from the cellwise
equation.  Under \conditionref{D1}, the recovered center estimates give
\(\delta g_I=O(r^2)\), while the ordinary vacuum endpoint weights are
integrable because \(\alpha_I>0\).  Hence
\(\delta_E\mathcal N_I(r)=\int_0^r\delta g_I(s)\,\dd s\) is well defined, and
integration on the cell containing the center gives
\begin{equation}
  \delta_E\mathcal N_I(r)+g_I(r)\xi_I(r)=C_I^{\mathrm{cen}},
  \qquad
  C_I^{\mathrm{cen}}:=\lim_{s\to0^+}g_I(s)\xi_I(s).
  \label{en:app-recovered-number-center-constant}
\end{equation}
Condition~\conditionref{D1} gives \(\xi_I=O(r)\) and \(g_I=O(r^2)\), hence
\(C_I^{\mathrm{cen}}=0\).  On each active cell, denote the constant value of
\(\delta_E\mathcal N_I+g_I\xi_I\) by \(\widetilde C_{I,a}\).  At an internal
ending block at which component \(I\) survives,
\(\delta_E\mathcal N_I\) is continuous by its integral definition, and
\conditionref{D3} gives \([\xi_I]=0\).  Accordingly, the adjacent cell
constants satisfy
\[
  \widetilde C_{I,a+1}-\widetilde C_{I,a}
  =g_I(\mathcal R_a)[\xi_I]_{\mathcal R_a}=0,
\]
which proves \eqref{en:app-recovered-enclosed-number} throughout
\((0,R_I)\).  At the labeled surface, \conditionref{C3} gives a finite inner
limit of \(\xi_I\), while
\(g_I=O((R_I-r)^{\alpha_I})\) gives \(g_I\xi_I\to0\).  The moving-endpoint
formula then gives \(\delta N_I=0\), proving
\eqref{en:app-recovered-total-number}.
\end{proof}

\begin{sublemma}[Full-tuple recovery uniqueness]
\label{en:lem-full-tuple-recovery-uniqueness}
Let a piecewise full tuple satisfying the center expansions in
\conditionref{F2} and whose matter fields obey the labeled endpoint weights
\eqref{en:eq-static-weighted-limits} satisfy the
particle number and EoS equations, the global component mass
condition
\eqref{en:eq-component-mass-global-domain}, the Hamiltonian and static
\(rr\) equations, the internal no-layer transmission conditions in
\conditionref{F4}, the labeled surface identities
\(\delta R_I=\xi_I(R_I^-)\), and normalized exterior matching.  Suppose its
metric fields also satisfy the metric constraint \eqref{en:app-zf-lambda}.  If
\(\zeta_I=r^2e^{-\Phi}\xi_I\), then every algebraic, integral, exterior,
and trace formula in \eqref{en:app-zf-xi} and
\eqref{en:app-recover-lambda}--\eqref{en:app-recover-exterior} has a
well-defined right-hand side and reproduces the corresponding field or
surface datum of the given tuple.
\end{sublemma}

\begin{proof}
The particle number equation, the metric constraint, and the equation of
state give the pressure and enthalpy recovery formulas
\eqref{en:app-recover-p}--\eqref{en:app-recover-H} by direct algebra.
The assumed cellwise regularity and endpoint weights, global mass condition,
and finite transmission, surface, and exterior traces show that all
algebraic, integral, exterior, and trace formulas just listed are well
defined.
The global component mass condition gives
\[
  \delta m_I(r)=4\pi\int_0^r s^2\delta\epsilon_I(s)\,\dd s,
\]
so there are no independent integration constants or jumps at internal
nodes.  The Hamiltonian and static \(rr\) equations then give the recovered
metric derivatives.  The internal transmission conditions propagate these
identities across all active cells, while normalized exterior matching fixes
the remaining lapse constant.  The labeled surface identities also show that
\(\delta R_I\) is reproduced by the corresponding trace formula.  Taken
together, these observations show that each recovery formula listed in the
sublemma reproduces the corresponding entry of the original full tuple on
its entire domain.
\end{proof}

\begin{proof}[Proof of Theorem~\ref{en:thm-reduction-package}(a)]
The cellwise residual identities, the identity \((C_\zeta^{m\Lambda})'=0\), and
the Euler factorization are given by
Lemma~\statementpartref{en:lem-canonical-zero-recovery}
{en:item-canonical-bulk}.  The global metric conclusion in
Lemma~\statementpartref{en:lem-canonical-zero-recovery}
{en:item-canonical-global-metric} gives \(C_\zeta^{m\Lambda}=0\) once
\conditionref{D1} and \conditionref{D3} are imposed.  It remains only to
identify the frequency-dependent equation.
The standard radial linearization of fluid conservation, in our sign and
lapse convention, is~\cite{Kain2020,CaballeroRipleyYunes2024}
\begin{equation}
  e^{2(\Lambda-\Phi)}h_I\ddot\xi_I
  +\mathcal E_I(\delta X)=0.
  \label{en:app-full-dynamical-Euler}
\end{equation}
With \(e^{i\omega t}\), \(\ddot\xi_I=-\omega^2\xi_I\),
\(\xi_I=e^\Phi\zeta_I/r^2\), and
\eqref{en:app-Euler-factorization}, the equation becomes
\begin{equation}
  (\mathcal L_\sigma\zeta)_I
  =\omega^2\frac{e^{3\Lambda+\Phi}h_I}{r^2}\zeta_I.
  \label{en:app-dynamic-generalized-mode}
\end{equation}
This is \eqref{en:eq-generalized-mode}.  Since it follows directly from the
time domain equation \eqref{en:app-full-dynamical-Euler}, the reduction remains
valid at zero frequency.
\end{proof}

For a piecewise field, write \([f]_R=f(R^+)-f(R^-)\), and use the surface
blocks \(\mathcal B_a\) and surviving sets \(\mathcal A_a^+\) of
\eqref{en:eq-surface-blocks}.  Define one labeled surface operator for every
\(I\in\mathcal B_a\), one common transmission operator for every internal
block, and one exterior operator:
\begin{align}
  \widehat B_{I,\mathrm{s}}(\zeta)
  &:=\lim_{r\to R_I^-}
  \left[\delta H_I[\zeta]+H_I'\xi_I\right]
  \notag\\
  &=\lim_{r\to R_I^-}
  \frac{\delta p_I[\zeta]+p_I'\xi_I}{h_I}
  \notag\\
  &=-\lim_{r\to R_I^-}\frac{e^\Phi}{r^2}c_{s,I}^2
  \left[\zeta_I'+(\Lambda'+\Phi')
  (\zeta_I-\bar\zeta)\right],
  \label{en:app-B-surface}\\
  B_{a,\mathrm{int}}(\zeta)
  &:=\left(( [\zeta_J]_{\mathcal R_a},[\delta p_J]_{\mathcal R_a})
  _{J\in\mathcal A_a^+},
  [\delta\Lambda]_{\mathcal R_a},[\delta\Phi]_{\mathcal R_a}\right),
  \qquad a<q,
  \label{en:app-B-internal}\\
  B_\infty(\zeta)
  &:=\left(
  \delta\Lambda(R_{\mathrm{out}}^-)-\frac{\delta M}{R_{\mathrm{out}}A(R_{\mathrm{out}})},
  \delta\Phi(R_{\mathrm{out}}^-)+\frac{\delta M}{R_{\mathrm{out}}A(R_{\mathrm{out}})}
  \right),
  \label{en:app-B-exterior}
\end{align}
where \(\delta M=\delta m_{\mathrm{tot}}(R_{\mathrm{out}}^-)\).  The cancellation in
\eqref{en:app-B-surface} follows from \(p_I'=h_IH_I'\) and
\eqref{en:app-recover-H}.  Since \(e^\Phi/r^2\) is finite and nonzero at a
finite horizonless surface, the component surface condition is equivalently
\begin{equation}
  \lim_{r\to R_I^-}c_{s,I}^2
  \left[\zeta_I'+(\Lambda'+\Phi')
  (\zeta_I-\bar\zeta)\right]=0.
  \label{en:app-reduced-enthalpy-surface}
\end{equation}
In particular, \(c_{s,I}^2\sim
  \kappa_I(R_I-r)/\alpha_I\), so its leading derivative condition is
\((R_I-r)\zeta_I'\to0\).  The trace requirements used here are those in
Definition~\ref{en:def-solution-classes}(IV).  For a vector satisfying
\conditionref{C1}--\conditionref{C3}, the internal and exterior traces in
\eqref{en:app-B-internal}--\eqref{en:app-B-exterior} exist.

Definition~\ref{en:def-solution-classes}(IV) imposes
\conditionref{D1}--\conditionref{D4} on the recovered
fields in this piecewise classical weighted class.

The piecewise qualification is essential.  If \(I\) ends at an internal
block, \(J\) survives, and \(x=\mathcal R_a-r\), write
\(\bigl((\mathcal L_\sigma\zeta)_J\bigr)_{\mathrm{sing}}\) for the
contribution generated by the ending label \(I\).  The coefficients in
\eqref{en:eq-symmetric-cross-coefficients}--
\eqref{en:eq-Q-derivative-splitting} give
\begin{equation}
  f_{JI}'=O(x^{\alpha_I-1})\in L^1,
  \qquad
  \bigl((\mathcal L_\sigma\zeta)_J\bigr)_{\mathrm{sing}}
  =f_{JI}'(\zeta_I-\zeta_J).
\end{equation}
Each fluid carries its own displacement.  Since
\(P_J(\mathcal R_a)>0\), a classical solution may therefore have
\(\zeta_J''=O(x^{\alpha_I-1})\): it is integrable but unbounded for
\(0<\alpha_I<1\), and may jump for \(\alpha_I=1\).  Global
\(C^2(0,R_J)\) would exclude admissible solutions.

\begin{sublemma}[Piecewise traces and distributional gluing]
\label{en:lem-piecewise-distributional-gluing}
Let \(0=r_0<r_1<\cdots<r_\ell=R\).  If
\(f|_{(r_{j-1},r_j)}\in W^{1,1}\) and all one-sided traces are finite,
then
\begin{equation}
  \partial_r f=(f')_{\mathrm{pw}}
  +\sum_{j=1}^{\ell-1}[f]_{r_j}\delta_{r_j}
  \quad\hbox{in }\mathcal D'(0,R).
  \label{en:eq-piecewise-first-derivative}
\end{equation}
If \(u\) is piecewise \(W^{2,1}\) with finite traces of \(u,u'\), then
\begin{equation}
  \partial_r^2u=(u'')_{\mathrm{pw}}
  +\sum_{j=1}^{\ell-1}[u']_{r_j}\delta_{r_j}
  +\sum_{j=1}^{\ell-1}[u]_{r_j}\delta_{r_j}'.
  \label{en:eq-piecewise-second-derivative}
\end{equation}
Equation~\eqref{en:eq-piecewise-first-derivative} shows that a piecewise
\(W^{1,1}\) field is globally \(W^{1,1}\) exactly when all of its jumps
vanish.  A zero-extended field whose inner trace is zero creates
no atom at its ending point.  Finally, if \(F\in L^1(0,R)\), then
\(c+\int_{r_0}^rF(s)\,\dd s\) is absolutely continuous and has matching
one-sided traces at every \(r_j\).
\end{sublemma}

\begin{proof}
For \(\phi\in C_c^\infty(0,R)\), integration by parts on each open
interval gives
\[
  -\int_0^R f\phi'\,\dd r
  =\int_0^R(f')_{\mathrm{pw}}\phi\,\dd r
   +\sum_{j=1}^{\ell-1}[f]_{r_j}\phi(r_j),
\]
which is \eqref{en:eq-piecewise-first-derivative}.  Applying this identity
first to \(u\) and then to its piecewise derivative gives
\eqref{en:eq-piecewise-second-derivative}.  The global \(W^{1,1}\) criterion
follows because an \(L^1\) weak derivative has no atomic part.  The
zero extension statement is the same formula with exterior trace zero, and
the final assertion is the fundamental theorem for \(W^{1,1}\).
\end{proof}

\begin{sublemma}[Boundary blocks of the recovered fields]
\label{en:lem-recovered-boundary-blocks}
Let \(\zeta\in\widehat{\mathcal C}_\sigma\) and
\(\delta X=\widehat R_\sigma\zeta\).  Then:
\begin{enumerate}
\renewcommand{\theenumi}{\alph{enumi}}
\renewcommand{\labelenumi}{(\theenumi)}
  \item The integral definitions in
  \eqref{en:app-recover-mass} and \eqref{en:app-recover-phi-interior}
  give \(\mathcal B_{\mathrm{mass}}=0\),
  \(C_{m,I}=C_{\Phi,\infty}=C_M=0\), and matching traces of
  \(\delta m_{\mathrm{tot}}\) and \(\delta\Phi\) at every internal node.

  \item The center block \(\mathcal B_{\mathrm{cen}}\) is well defined and
  vanishes if and only if \conditionref{D1} holds.
  Under this condition \(C_\Lambda=C_{N,I}=0\) for every \(I\).

  \item For each label \(I\),
  the labeled surface block \(\mathcal B_{I,\mathrm{s}}\) is well defined
  and vanishes if and only if \conditionref{D2} holds.
  If \conditionref{D1}--\conditionref{D3} hold, then
  \(\mathcal B_N=0\).

  \item At an internal node \(\mathcal R_a\),
  \(\mathcal B_{a,\mathrm{int}}=0\) if and only if
  \conditionref{D3} holds.  When these conditions hold at every internal
  node, the cellwise recovery equations are global distributional equations
  and their distributional right-hand sides have no node-supported
  \(\delta\)- or \(\delta'\)-terms.

  \item \(\mathcal B_\infty=0\) if and only if \conditionref{D4} holds.
  Simultaneous termination of several labels contributes one surface block
  for each ending label and one common metric block, with no further trace
  datum.
\end{enumerate}
Combining parts~\textup{(a)}--\textup{(e)} gives
\[
  \widehat R_\sigma\zeta\in\mathcal X_{\mathrm{full}}^{\mathrm{tr}}
  \ \text{and}\
  \operatorname{BC}_{\mathrm{full}}(\widehat R_\sigma\zeta)=0
  \quad\Longleftrightarrow\quad
  \conditionref{D1}\text{--}\conditionref{D4}\text{ hold}.
\]
\end{sublemma}

\begin{proof}
The component mass formula gives
\[
  \delta m_I(r)=4\pi\int_0^r
     s^2\delta\epsilon_I(s)\,\dd s.
\]
Hence \(\delta m_I\in W^{1,1}(0,R_*)\),
  \(\partial_r\delta m_I=4\pi r^2(\delta\epsilon_I)^0\), and
\(\delta m_I(0)=0\).  The sum \(\delta m_{\mathrm{tot}}=\sum_I\delta m_I\) is therefore
continuous at every internal node, including a node at which several labels
  end.  Likewise,
  \(\delta\Phi'_{\mathrm{int}}\in L^1_{\mathrm{loc}}((0,R_{\mathrm{out}}])\) by the
  endpoint weights and \eqref{en:app-recover-phi-prime}; formula
  \eqref{en:app-recover-phi-interior} gives
  \(\delta\Phi\in W^{1,1}_{\mathrm{loc}}((0,R_{\mathrm{out}}])\), with matching traces
  at every positive-radius internal node.  Its outer value and
\eqref{en:app-recover-exterior} give
\(C_{\Phi,\infty}=0\), while
\eqref{en:app-recover-global-data} gives \(C_M=0\).  This proves part (a)
and identifies all mass and lapse integration constants.

At the regular background center,
\(e^\Phi=e^{\Phi(0)}[1+O(r^2)]\).  Therefore
\[
  \zeta_I=a_I^{\mathrm{cen}}r^3+O(r^5),\quad
  \zeta_I'=3a_I^{\mathrm{cen}}r^2+O(r^4)
\]
is equivalent to
\[
  \xi_I=a_I^{\xi,\mathrm{cen}}r+O(r^3),\quad
  \xi_I'=a_I^{\xi,\mathrm{cen}}+O(r^2),
  \qquad a_I^{\xi,\mathrm{cen}}=e^{\Phi(0)}a_I^{\mathrm{cen}}.
\]
Substitution in \eqref{en:app-recover-lambda}--
\eqref{en:app-recover-phi-prime} gives the remaining center estimates in
\eqref{en:eq-full-center-block}; conversely, its displacement estimates give
\conditionref{D1}.  The same substitution gives
\(rA\delta\Lambda\to0\) and \(\delta\Phi'=O(r)\), so the locally absolutely
continuous lapse extends regularly through the center.  Moreover,
\eqref{en:app-recovered-number-center-constant} and
\(g_I\xi_I=O(r^3)\) give \(C_{N,I}=0\), proving part (b).

Equations~\eqref{en:app-recover-global-data} and
\eqref{en:app-B-surface} give
\(\mathcal B_{I,\mathrm{s}}=0\) if and only if \conditionref{D2} holds.
For particle number, \conditionref{D1} sets the center
constant to zero, \conditionref{D3} propagates it across every internal node,
and \conditionref{C3} gives the finite labeled surface trace.  Lemma
~\statementpartref{en:lem-canonical-zero-recovery}
{en:item-canonical-particle} then gives
\(\delta N_I=0\), proving part (c).

At a positive-radius node, continuity of the background gives
\[
  [\xi_J]_{\mathcal R_a}
  =\frac{e^{\Phi(\mathcal R_a)}}{\mathcal R_a^2}
    [\zeta_J]_{\mathcal R_a}
  \qquad(J\in\mathcal A_a^+).
\]
The \(\delta m_{\mathrm{tot}}\) and \(\delta\Phi\) entries of the full internal block
already vanish by part (a); its remaining entries are exactly those in
\eqref{en:app-B-internal}.  This proves the equivalence with
\conditionref{D3}.

We next verify the hypotheses needed for distributional gluing.  Fix a
positive-radius internal node and a surviving label \(J\).  By
\conditionref{C4} and \eqref{en:app-Euler-factorization},
\(\mathcal E_J\in L^1\) on each one-sided neighborhood of the node.  The
recovery formulas and the endpoint weights give
\[
  h_J\delta\Phi'\in L^1,
  \qquad
  (\delta\epsilon_J+\delta p_J)\Phi'\in L^1,
\]
there, and hence the Euler identity gives \(\delta p_J'\in L^1\).  Since
\(b_J(\mathcal R_a)>0\), the pressure recovery formula can be solved for
\(\zeta_J'\):
\[
  \zeta_J'
  =-\frac{r^2e^{-\Phi}}{b_J}\delta p_J
   -\frac{p_J'}{b_J}\zeta_J
   -(\Lambda'+\Phi')(\zeta_J-\bar\zeta).
\]
Its finite one-sided data give finite traces of \(\zeta_J'\);
differentiating that formula, using \(\delta p_J'\in L^1\) and the
ordinary vacuum orders
\(O(|r-\mathcal R_a|^{\alpha_I-1})\), gives piecewise
\(\zeta_J\in W^{2,1}\).  These estimates also imply that, on the same
one-sided neighborhoods, \(n_J\zeta_J\), \(rA\delta\Lambda\),
\(\delta\Phi\), and \(\delta p_J\) belong to \(W^{1,1}\).  Lemma
~\ref{en:lem-piecewise-distributional-gluing} therefore applies to each of
these fields.

The singular coefficients can be read off explicitly.  Put
\(\kappa_{J,a}:=e^{\Phi(\mathcal R_a)}b_J(\mathcal R_a)/\mathcal R_a^2\).
The distributional extension of the first-order pressure recovery expression
is
\begin{equation}
  \begin{aligned}
  (\delta p_J)_{\mathrm{dist}}
  &=(\delta p_J)_{\mathrm{pw}}
    -\sum_{a:\,\mathcal R_a<R_J}
      \kappa_{J,a}[\zeta_J]_{\mathcal R_a}
      \delta_{\mathcal R_a},\\
  \partial_r(\delta p_J)_{\mathrm{dist}}
  &=(\delta p_J')_{\mathrm{pw}}
    +\sum_{a:\,\mathcal R_a<R_J}
      [\delta p_J]_{\mathcal R_a}\delta_{\mathcal R_a}
    -\sum_{a:\,\mathcal R_a<R_J}
      \kappa_{J,a}[\zeta_J]_{\mathcal R_a}
      \delta'_{\mathcal R_a}.
  \end{aligned}
  \label{en:eq-recovered-pressure-distribution}
\end{equation}
Condition~\conditionref{D3} first removes the \(\delta'\)-coefficient through
\([\zeta_J]_{\mathcal R_a}=0\), and then removes the \(\delta\)-coefficient
through \([\delta p_J]_{\mathcal R_a}=0\).  The same condition removes the
jumps of \(rA\delta\Lambda\) and \(\delta\Phi\), and it removes the jump of
the surviving particle flux because \(n_J\) is continuous and positive at
the node.  For an ending label \(I\),
\(n_I\zeta_I\), \(h_I\zeta_I\), and \(b_I\zeta_I'\) have zero inner traces by
the ordinary vacuum powers and the finite endpoint conditions, so their zero
extensions create no atom.  These statements apply labelwise when an entire
block \(\mathcal B_a\) ends at the same radius.

All remaining coefficient terms are locally \(L^1\), including the
integrable orders allowed by \(\mathcal Y_\sigma\).  Accordingly,
\conditionref{C4} controls the cellwise \(L^1\) part, while
\conditionref{D3} removes every node-supported distribution.  This proves
part (d).  Conditions \conditionref{D1} and \conditionref{D3} also give
\(\delta m_{\mathrm{tot}}=rA\delta\Lambda\) throughout the star by Lemma
~\statementpartref{en:lem-canonical-zero-recovery}
{en:item-canonical-global-metric}.

In the vacuum exterior, direct integration of the spherical Hamiltonian and
\(rr\) equations gives
\[
  \delta m_{\mathrm{tot}}=C_M^{\mathrm{vac}},\qquad
  \delta\Lambda=\frac{C_M^{\mathrm{vac}}}{rA},\qquad
  \delta\Phi=-\frac{C_M^{\mathrm{vac}}}{rA}+C_\Phi^{\mathrm{vac}}.
\]
The inner mass trace fixes \(C_M^{\mathrm{vac}}=\delta M\), and
\(\delta\Phi(\infty)=0\) fixes \(C_\Phi^{\mathrm{vac}}=0\).  Hence the full
exterior block is zero exactly when \eqref{en:app-B-exterior}, namely
\conditionref{D4}, holds.  This proves part (e) and the final equivalence.
\end{proof}

\begin{proof}[Proof of Theorem~\ref{en:thm-reduction-package}(b)]
Lemma~\ref{en:lem-recovered-boundary-blocks} gives the componentwise
equivalence in \eqref{en:eq-BC-equivalence}.  Its distributional conclusion,
together with \conditionref{C4}, proves that the cellwise residual identities
in part~\textup{(a)} are their global distributional counterparts.
\end{proof}

\subsection{Inverse map on the physical zero modes}
\label{en:app-proof-zero-kernel-package}

\begin{lemma}[Full and reduced zero mode correspondence]
\label{en:lem-zero-kernel-package}
For the full zero-frequency space whose element type is specified in
\eqref{en:eq-full-element-type} and the reduced domain of
Definition~\ref{en:def-solution-classes}(IV), the projection
\begin{equation}
  \Pi_\sigma^{\mathrm{red}}:\mathcal S_{\sigma,0,N}^{\mathrm{full}}
  \longrightarrow\ker\mathcal L_\sigma,
  \qquad \Pi_\sigma^{\mathrm{red}}(\delta X)_I=r^2e^{-\Phi}\xi_I,
  \label{en:eq-projection}
\end{equation}
is a linear isomorphism with inverse \(\widehat R_\sigma\).  More precisely,
\begin{equation}
  \Pi_\sigma^{\mathrm{red}}\widehat R_\sigma
    =\operatorname{Id}_{\ker\mathcal L_\sigma},
  \qquad
  \widehat R_\sigma\Pi_\sigma^{\mathrm{red}}
    =\operatorname{Id}_{\mathcal S_{\sigma,0,N}^{\mathrm{full}}}.
  \label{en:eq-PR-RP}
\end{equation}
\end{lemma}

\setproofparent{en:lem-zero-kernel-package}

\begin{proof}
Let \(\delta X\in\mathcal S_{\sigma,0,N}^{\mathrm{full}}\) and set
\(\zeta_I=r^2e^{-\Phi}\xi_I\).  By \conditionref{F1}, the tuple
satisfies the global component mass condition
\eqref{en:eq-component-mass-global-domain}; \conditionref{F4} and
\conditionref{F5} supply the internal transmission, labeled surface, and
normalized exterior hypotheses.  The number, EoS, Hamiltonian,
and static \(rr\) equations in \conditionref{F3}, together with the center
regularity in \conditionref{F2}, allow the metric constraint calculation
\eqref{en:app-C-definition}--\eqref{en:app-C-ODE} to be applied across all
active cells.  Hence the full tuple satisfies
\eqref{en:app-zf-lambda}.  The full tuple now satisfies all hypotheses of the
independent full-tuple recovery uniqueness lemma.  Its uniqueness conclusion,
together with \conditionref{F1}, \conditionref{F2},
\conditionref{F4}, and \conditionref{F5}, first verifies
\conditionref{C1}--\conditionref{C3}: the local recovery expressions agree
with the original full fields; \conditionref{F2} and \conditionref{S1} give
their cellwise regularity; \conditionref{S2} gives the mass integrability;
the static \(rr\) formula together with \conditionref{S4}--\conditionref{S6}
gives the local lapse integrability; \conditionref{S6} gives the endpoint
weights; and the remaining traces are supplied by \conditionref{F2},
\conditionref{F4}, and \conditionref{F5}.  Hence
\(\widehat R_\sigma\zeta\) is well defined, and
the same formula identities give
\[
  \widehat R_\sigma\zeta=\delta X.
\]
This equality identifies the recovered fields with the original full fields.
The conditions listed as \conditionref{D1}--\conditionref{D4} follow from
\conditionref{F2}, \conditionref{F4}, and \conditionref{F5}.  We may
therefore apply the
cellwise recovery identities in
Lemma~\statementpartref{en:lem-canonical-zero-recovery}
{en:item-canonical-bulk}.  The Euler equation in \conditionref{F3},
together with its residual factorization, gives
\(\mathcal L_\sigma\zeta=0\) on every active cell.  The zero function
belongs to \(\mathcal Y_\sigma\), so \conditionref{C4} now holds and
\(\zeta\in\widehat{\mathcal C}_\sigma\).  Hence
\(\zeta\in\widehat{\mathcal D}_\sigma\).
\begin{equation}
  \widehat R_\sigma\Pi_\sigma^{\mathrm{red}}(\delta X)=\delta X.
  \label{en:app-RP}
\end{equation}

Conversely, let \(\zeta\in\ker\mathcal L_\sigma\).  By the definition of
this kernel, \(\zeta\in\widehat{\mathcal D}_\sigma\); in particular,
\conditionref{D1}--\conditionref{D4} hold.
Lemma~\statementpartref{en:lem-canonical-zero-recovery}
{en:item-canonical-bulk} gives all bulk equations, and
Theorem~\ref{en:thm-reduction-package}\textup{(b)} gives their endpoint and
transmission conditions.  The angular Einstein equation, not used in the
recovery, follows from the contracted Bianchi identity.  Set
\begin{equation}
  \mathcal G^\mu{}_\nu
  :=\delta G^\mu{}_\nu-8\pi\sum_I\delta T_I^\mu{}_\nu.
\end{equation}
The linearized identity~\cite{WaldGR} and separate
fluid conservation give
\(\nabla_\mu\mathcal G^\mu{}_\nu=0\) distributionally.  On each active cell
of the static, spherically symmetric background, its radial component is
\begin{equation}
  (\mathcal G^r{}_r)'
  +\Phi'(\mathcal G^r{}_r-\mathcal G^t{}_t)
  +\frac2r(\mathcal G^r{}_r-\mathcal G^\theta{}_\theta)=0.
\end{equation}
The Hamiltonian and \(rr\) residuals vanish, so the angular residual does too;
the transmission conditions remove junction-supported delta terms, extending
this conclusion globally.  For particle number, \conditionref{D1} sets
\(C_I^{\mathrm{cen}}=0\) in
\eqref{en:app-recovered-number-center-constant}, \conditionref{D3}
propagates this value across every internal node, and \conditionref{C3}
provides the finite labeled surface limit.  Lemma
~\statementpartref{en:lem-canonical-zero-recovery}
{en:item-canonical-particle} therefore gives
\eqref{en:app-recovered-enclosed-number} and
\(\delta N_I=0\) for every \(I\).  These particle number conclusions,
together with the field equation and boundary checks above, give
\(\widehat R_\sigma\zeta\in\mathcal S_{\sigma,0,N}^{\mathrm{full}}\), and
\begin{equation}
  \Pi_\sigma^{\mathrm{red}}\widehat R_\sigma\zeta
  =r^2e^{-\Phi}\left(\frac{e^\Phi}{r^2}\zeta\right)
  =\zeta.
  \label{en:app-PR}
\end{equation}
The two composition identities show that \(\widehat R_\sigma\) and
\(\Pi_\sigma^{\mathrm{red}}\) are inverse.
\end{proof}

\section{Equivalence of the static and dynamical kernels}
\label{en:sec-main-proof}

At a fixed equilibrium \(\sigma\), \(\ker DN_\sigma\) consists of
equilibrium family tangents along which all particle numbers have vanishing
first variation, whereas \(\ker\mathcal L_\sigma\) consists of reduced radial
zero-frequency solutions satisfying the physical regularity and boundary
conditions. In this argument perturbations are real. The radial equations
and boundary conditions have real coefficients, so complexifying the kernel
preserves its dimension over the corresponding scalar field.
We relate the two spaces through
\begin{equation}
  \ker DN_\sigma
  \xrightarrow{\ B_\sigma\ }
  \mathcal S_{\sigma,N}^{\mathrm{stat}}
  \xrightarrow{\ A_\sigma\ }
  \mathcal S_{\sigma,0,N}^{\mathrm{full}}
  \xrightarrow{\ \Pi_\sigma^{\mathrm{red}}\ }
  \ker\mathcal L_\sigma.
  \label{en:eq-main-roadmap}
\end{equation}
Here \(B_\sigma\) identifies a particle-preserving equilibrium tangent with
its static Eulerian field perturbation, \(A_\sigma\) adds the unique
displacements determined by particle conservation, and
\(\Pi_\sigma^{\mathrm{red}}\) extracts
the reduced displacement.  Lemma~\ref{en:lem-static-representation},
Lemma~\ref{en:lem-particle-package}, and
Lemma~\ref{en:lem-zero-kernel-package} show that these maps are linear
isomorphisms.  Their composition identifies the two kernels, and
rank--nullity gives the multiplicity formula.

\begin{theorem}[Static criticality if and only if a radial zero mode exists]
\label{en:thm-main}
Under Assumption~\ref{en:ass-regular}, let
\(X:U\to\mathcal X_{\mathrm{eq}}\), \(U\subset\mathbb R^K\), be the \(C^1\)
equilibrium embedding established in
Lemma~\statementpartref{en:lem-equilibrium-package}{en:item-map-regularity}.
Then, for every \(\sigma\in U\),
\begin{equation}
  \boxed{
  \rankop DN_\sigma<K
  \quad\Longleftrightarrow\quad
  \exists\,\zeta\in\widehat{\mathcal D}_\sigma\setminus\{0\}:
  \mathcal L_\sigma\zeta=0.}
  \label{en:eq-main-equivalence}
\end{equation}
The multiplicity is
\begin{equation}
  \dim\ker\mathcal L_\sigma
  =K-\rankop DN_\sigma.
  \label{en:eq-nullity}
\end{equation}
\end{theorem}

\begin{proof}
Compose the three isomorphisms established above.
Lemma~\ref{en:lem-static-representation} gives
\(B_\sigma:T_\sigma U\cong\mathcal S_\sigma^{\mathrm{stat}}\).  The chain rule
and the defining constraint \(\delta N_I=0\) show that its restriction maps
\(\ker DN_\sigma\) onto
\(\mathcal S_{\sigma,N}^{\mathrm{stat}}\).  Lemma~\ref{en:lem-particle-package}
then adds the unique displacements
\[
  \xi_I=-\frac{\delta_E\mathcal N_I}{g_I},
\]
and identifies this space with
\(\mathcal S_{\sigma,0,N}^{\mathrm{full}}\).  Finally,
Lemma~\ref{en:lem-zero-kernel-package} identifies the latter space with
\(\ker\mathcal L_\sigma\) through \(\Pi_\sigma^{\mathrm{red}}\), with inverse
  \(\widehat R_\sigma\).  Composing these three identifications yields
\begin{equation}
  \Pi_\sigma^{\mathrm{red}}A_\sigma B_\sigma
       \big|_{\ker DN_\sigma}
  :\ker DN_\sigma\xrightarrow{\cong}\ker\mathcal L_\sigma.
\end{equation}
Since \(\dim T_\sigma U=K\), rank--nullity and the above isomorphism give
\begin{equation}
  \rankop DN_\sigma<K
  \Longleftrightarrow
  \ker\mathcal L_\sigma\ne\{0\},
  \qquad
  \dim\ker\mathcal L_\sigma=K-\rankop DN_\sigma.
\end{equation}
\end{proof}

For \(K=2\), rank deficiency is equivalent to
\(\det D(N_1,N_2)=0\); if both covectors are nonzero, their gradients are
parallel. For \(K>2\), requiring every pair of particle number gradients to be parallel
would incorrectly force \(DN_\sigma\) to have rank at most one.  For every
finite \(K\), the correct parameterization-independent condition is
\[
  \rankop DN_\sigma<K,
\]
as stated in Theorem~\ref{en:thm-main}.

\section{Spectral realization and the local stability boundary}
\label{en:sec-spectral}

The zero mode correspondence becomes a stability statement after the radial
equations are placed in a self-adjoint spectral framework.  The classical variational principle for stellar oscillations~\cite{chandrasekhar_general_1964}, its relativistic radial form~\cite{Chandrasekhar1964}, and the self-adjoint reduction of spherical perturbations~\cite{seifert_general_2007} provide the foundation for this construction.  The construction
has two stages.  At a fixed equilibrium, a closed semibounded form defines the
radial operator, compactness gives a discrete spectrum, and finite-energy
endpoint selection identifies its eigenspaces with the physical radial modes.
The resulting forms are then transported to fixed weighted spaces to establish
continuity along the equilibrium family.

\subsection{Weighted radial spaces and endpoint structure}
\label{en:sec-spectral-setting}

Using the matter coefficients in \eqref{en:eq-total-matter-notation}, put
\begin{equation}
  W_I:=\frac{e^{3\Lambda+\Phi}h_I}{r^2},
  \qquad
  P_I:=\frac{e^{\Lambda+3\Phi}b_I}{r^2},
  \label{en:eq-spectral-WP}
\end{equation}
\begin{nameddefinition}[Dynamical Hilbert space]
\phantomsection
\label{en:def-dynamical-Hilbert-space}
At a fixed equilibrium \(\sigma\), the dynamical Hilbert space is
\begin{equation}
  \mathcal H_\sigma
  :=\bigoplus_{I=1}^K L^2((0,R_I),W_I\,\dd r).
  \label{en:eq-dynamical-Hilbert-space}
\end{equation}
Its elements are complex vectors \(u=(u_I)_{I=1}^K\), defined up to weighted
almost everywhere equality, with
\begin{equation}
  \|u\|_{\mathcal H_\sigma}^2
  =\sum_{I=1}^K\int_0^{R_I}W_I|u_I|^2\,\dd r<\infty.
\end{equation}
\end{nameddefinition}
Sturm--Liouville operators in direct sum spaces provide the operator-theoretic
background for the direct sum, multi-interval aspect of this realization
~\cite{everitt_sturmliouville_1986}.
The weight in \eqref{en:eq-spectral-WP} is the component weight of
\(\mathcal W_\sigma\) in \eqref{en:eq-generalized-mode}.  Reference~\cite{CaballeroRipleyYunes2024} gives the symmetric radial equations for \(K\) fluids in Sec.~V and Appendix~A. Here we use the associated form to treat the singular free surfaces, including changes in their ordering along the equilibrium family.

\begin{remark}[Material surface condition]
\label{en:rem-material-surface-condition}
The labeled surface of component \(I\) is the material level set \(H_I=0\).
Its linearized motion therefore satisfies
\begin{equation}
  \Delta H_I(R_I)=0.
  \label{en:eq-material-surface-condition}
\end{equation}
Since \(\Delta p_I=h_I\Delta H_I\) and \(h_I\to0\) at an ordinary vacuum
surface, the normalized enthalpy condition retains the first-order motion of
the material boundary.  On each active cell \(\mathcal I_a\), define the
enthalpy-weighted displacement
\(\bar\zeta:=h_{\mathrm{tot}}^{-1}\sum_{J\in\mathcal K_a}h_J\zeta_J\).
In reduced variables the condition is
\begin{equation}
 \lim_{r\to R_I^-}c_{s,I}^2
 \left[\zeta_I'+(\Lambda'+\Phi')(\zeta_I-\bar\zeta)\right]=0.
 \label{en:eq-main-reduced-material-surface}
\end{equation}
The endpoint asymptotics below give \(c_{s,I}^2\asymp R_I-r\), and the
finite-energy endpoint analysis yields
\((R_I-r)\zeta_I'\to0\) for every eigenfunction of the form realization.
Together, these endpoint properties show that the spectral and material
surface conditions select the same radial modes.
\end{remark}

\begin{nameddefinition}[Fixed radius order sector]
\phantomsection
\label{en:def-smooth-equilibrium-sector}
A \emph{fixed radius order sector} is a \(C^1\) parameter submanifold
\(\mathcal U\subset U\) for which constants
\(s_{IJ}\in\{-1,0,1\}\) exist such that
\begin{equation}
  \operatorname{sgn}\!\bigl(R_I(\sigma)-R_J(\sigma)\bigr)=s_{IJ},
  \qquad \sigma\in\mathcal U,\quad I,J=1,\ldots,K.
  \label{en:eq-smooth-equilibrium-sector}
\end{equation}
On every relatively compact neighborhood in \(\mathcal U\), distinct surface
nodes have a positive uniform separation.  Labels with \(s_{IJ}=0\) form one
surface block throughout the sector.  A fixed radius order sector is also
called a \emph{smooth equilibrium sector}.
\end{nameddefinition}

\begin{lemma}[Endpoint structure and form control]
\label{en:lem-spectral-endpoint-structure}
Under Assumption~\ref{en:ass-regular}, fix an equilibrium in \(U\).
At the center,
\begin{equation}
\begin{aligned}
  W_I&=w_{I,c}r^{-2}[1+O(r^2)],&
  P_I&=p_{I,c}r^{-2}[1+O(r^2)],\\
  f_{IJ}&=O(r^{-1}),&
  \widetilde Q_I,S_{IJ}&=O(r^{-2}),
\end{aligned}
  \label{en:eq-derived-Fuchs-center}
\end{equation}
where \(w_{I,c},p_{I,c}>0\). The relative remainders in the
principal weights are \(O(r^2)\), with first radial derivatives \(O(r)\).
Indeed \(H_I=H_I^c+O(r^2)\), \(H_I'=O(r)\), and \(h_I,b_I\) are
\(C^1\) functions of \(H_I\) at positive central density. Here \(f_{IJ}\), \(S_{IJ}\), and \(\widetilde Q_I\) are defined later in \eqref{en:eq-symmetric-cross-coefficients} and \eqref{en:eq-complete-Qtilde}, with the relation between \(Q_I\) and \(\widetilde Q_I\) given in \eqref{en:eq-Q-derivative-splitting}. At a distinct component radius \(R\), put
\begin{equation}
  x:=R-r,\qquad
  \mathcal B_R:=\{I:R_I=R\},\qquad
  \mathcal A_R^+:=\{J:R_J>R\}.
  \label{en:eq-Fuchs-ending-block}
\end{equation}
For \(I\in\mathcal B_R\), there are \(p_{I,0},w_{I,0}>0\) and
\(\eta_{\mathrm{sp}}>0\)
such that
\begin{align}
  P_I&=p_{I,0}x^{\alpha_I+1}(1+\pi_I),&
  W_I&=w_{I,0}x^{\alpha_I}(1+\varpi_I),
  \label{en:eq-Fuchs-principal-expansions}\\
  |\pi_I|+|\varpi_I|+|x\pi_I'|+|x\varpi_I'|
  &\le Cx^{\eta_{\mathrm{sp}}} .
  \label{en:eq-Fuchs-principal-remainders}
\end{align}
Every surviving \(P_J,W_J\), \(J\in\mathcal A_R^+\), extends positively and
\(C^1\) through \(R\).  The lower-order coefficients factor as
\begin{align}
  \widetilde Q_I&=x^{\alpha_I}O_{\mathrm{con}}(1),
  \notag\\
  J\in\mathcal A_R^+:\quad
  S_{IJ}&=x^{\alpha_I}O_{\mathrm{con}}(1),&
  f_{IJ}&=x^{\alpha_I+1}O_{\mathrm{con}}(1),&
  f_{JI}&=x^{\alpha_I}O_{\mathrm{con}}(1),
  \notag\\
  J\in\mathcal B_R:\quad
  S_{IJ}&=x^{\alpha_I+\alpha_J}O_{\mathrm{con}}(1),&
  f_{IJ}&=x^{\alpha_I+\alpha_J+1}O_{\mathrm{con}}(1),
  \label{en:eq-Fuchs-coupling-factorization}
\end{align}
where each \(O_{\mathrm{con}}(1)\) factor and its \(x\partial_x\) derivative
is bounded. These expansions concern one fixed surface node. The bounds
are locally uniform within a fixed radius order sector when the collar is
chosen disjoint from the other distinct nodes.
For positive quantities \(a,b\), the notation \(a\asymp b\) means that
\[
  cb\le a\le Cb
\]
on the stated region for some constants \(0<c\le C<\infty\).  Uniform
comparisons use constants independent of \(\sigma\).  After the labelwise
rescaling
\(r=R_I(\sigma)s\), take
\begin{equation}
  w_I^*(s):=s^{-2}(1-s)^{\alpha_I},\qquad
  p_I^*(s):=s^{-2}(1-s)^{\alpha_I+1}.
  \label{en:eq-model-spectral-weights}
\end{equation}
After the labelwise pullback, set
\begin{equation}
\begin{aligned}
 \widehat W_{I,\sigma}(s)
 &:=R_I(\sigma)W_I(R_I(\sigma)s;\sigma),\\
 \widehat P_{I,\sigma}(s)
 &:=\frac{P_I(R_I(\sigma)s;\sigma)}{R_I(\sigma)}.
\end{aligned}
\label{en:eq-pulled-back-spectral-coefficients}
\end{equation}
The four normalized quantities
\[
 \frac{\widehat W_{I,\sigma}}{w_I^*},\qquad
 \frac{w_I^*}{\widehat W_{I,\sigma}},\qquad
 \frac{\widehat P_{I,\sigma}}{p_I^*},\qquad
 \frac{p_I^*}{\widehat P_{I,\sigma}}
\]
depend continuously on \(\sigma\) in \(L^\infty(0,1)\).  On a smooth
equilibrium sector, a \emph{fixed-node pullback} is a family of continuous,
cellwise \(C^2\) diffeomorphisms \(r=\Psi_\sigma(s)\) from a fixed reference
partition, depending \(C^1\) on \(\sigma\) in the cellwise \(C^2\) topology,
with locally uniform bounds
\(0<c\le\partial_s\Psi_\sigma\le C\), and sending each reference node to the
corresponding radius \(\mathcal R_a(\sigma)\). Under this pullback the
normalized principal and Hilbert weights are \(C^1\) on the sector.
The four lower-order form control ratios are
\begin{equation}
 \frac{\widetilde Q_I}{W_I},\qquad
 \frac{S_{IJ}}{(W_IW_J)^{1/2}},\qquad
 \frac{f_{IJ}}{(P_IW_J)^{1/2}},\qquad
 \frac{f_{IJ}}{(P_IW_I)^{1/2}}.
 \label{en:eq-form-control-ratios}
\end{equation}
On the full chart these ratios are locally uniformly bounded on their
active cells and converge uniformly on compact intervals away from the
limiting center and surface nodes. Within each smooth equilibrium sector
they are \(C^1\) in the cellwise \(L^\infty\) norms after the fixed-node
pullback.
\end{lemma}

\subsection{Closed form, compactness, and physical eigenspaces}
\label{en:sec-self-adjoint-realization}

On the test space
\(C_{c,\sigma}^{\infty}=\bigoplus_I C_c^\infty(0,R_I)\), define
\begin{equation}
\begin{aligned}
 \mathfrak a_\sigma[u,v]
 &:=\sum_I\int_0^{R_I}\overline{u_I}(\mathcal L_\sigma v)_I\,\dd r,\\
 \mathfrak a_{0,\sigma}[u,v]
 &:=\sum_I\int_0^{R_I}P_I\overline{u_I'}v_I'\,\dd r,\\
 \mathfrak a_{0,\sigma}[u]&:=\mathfrak a_{0,\sigma}[u,u].
\end{aligned}
 \label{en:eq-main-form-definition}
\end{equation}
Define the principal form norm and its completion by
\begin{equation}
\begin{aligned}
 \|u\|_{\mathcal V_\sigma}^2
 &:=\mathfrak a_{0,\sigma}[u]+\|u\|_{\mathcal H_\sigma}^2,
 &&u\in C_{c,\sigma}^{\infty},\\
 \mathcal V_\sigma
 &:=\overline{C_{c,\sigma}^{\infty}}^{\,\|\cdot\|_{\mathcal V_\sigma}}.
\end{aligned}
 \label{en:eq-main-closed-form-domain}
\end{equation}
Friedrichs extensions and domain and endpoint characterizations for singular
Sturm--Liouville expressions provide standard background for this closure
construction~\cite{niessen_singular_1992,kaper_characterizations_1986}.
The symmetric expression in Section~\ref{en:app-spectral-realization}
extends the difference
\(\mathfrak r_\sigma:=\mathfrak a_\sigma-\mathfrak a_{0,\sigma}\)
continuously to the principal form completion.

\begin{theorem}[Self-adjoint form realization, compact resolvent, and physical eigenspaces]
\label{en:thm-fixed-spectral-package}
Under Assumption~\ref{en:ass-regular}, \(\mathfrak a_\sigma\) is closable in
\(\mathcal H_\sigma\), and its closure is a densely defined, closed,
lower-semibounded form.  Let \(T_\sigma\) be the self-adjoint operator associated
with this closed form by the first representation theorem.  Then:
\begin{enumerate}
\renewcommand{\theenumi}{\alph{enumi}}
\renewcommand{\labelenumi}{(\theenumi)}
  \item \label{en:item-spectral-lower-bound}
  For every \(\bar\sigma\in U\) there are a
  relatively compact neighborhood \(V\Subset U\) of \(\bar\sigma\) and
  \(C_V<\infty\) such
  that
  \begin{equation}
    \langle u,T_\sigma u\rangle_{\mathcal H_\sigma}
    \ge-C_V\|u\|_{\mathcal H_\sigma}^2,
    \qquad \sigma\in V,\quad u\in\operatorname{Dom}(T_\sigma).
    \label{en:eq-uniform-lower-bound}
  \end{equation}
  \item \label{en:item-spectral-compact-resolvent}
  The embedding of the closed form domain into \(\mathcal H_\sigma\) is
  compact.  Hence \(T_\sigma\) has compact resolvent and a real discrete
  spectrum with finite multiplicities and no finite accumulation point;
  \item \label{en:item-spectral-kernel-compatibility}
  Every eigenfunction has a piecewise classical representative satisfying the
  conditions at the center and at each labeled component surface, together
  with the internal junction and normalized exterior conditions in
  Definition~\ref{en:def-solution-classes}(IV), and every
  piecewise classical physical mode in that domain belongs to the form
  realization.  For every real \(\lambda\),
  \begin{equation}
    \ker\!\left(T_\sigma-\lambda\operatorname{Id}_{\mathcal H_\sigma}\right)
    =\{\zeta\in\widehat{\mathcal D}_\sigma:
      \mathcal L_\sigma\zeta
      =\lambda\mathcal W_\sigma\zeta\},
    \qquad \lambda\in\mathbb R.
    \label{en:eq-form-realization-classical-eigenspaces}
  \end{equation}
  In particular, the operator kernel is the physical zero mode space:
  \begin{equation}
    \ker T_\sigma
    =\{\zeta\in\widehat{\mathcal D}_\sigma:
      \mathcal L_\sigma\zeta=0\}.
    \label{en:eq-form-realization-classical-kernel}
  \end{equation}
\end{enumerate}
Part~\hyperref[en:item-spectral-kernel-compatibility]{\textup{(\ref*{en:item-spectral-kernel-compatibility})}} characterizes the eigenspaces
inside the closed operator domain.  The surrounding weak domain is described
in Remark~\ref{en:rem-closed-versus-classical-domain}.
\end{theorem}

The two essential endpoint facts used in the proof are
\begin{equation}
  W_I\asymp P_I\asymp r^{-2}\quad(r\to0^+),
  \qquad
  W_I\asymp(R_I-r)^{\alpha_I},\quad
  P_I\asymp(R_I-r)^{\alpha_I+1}\quad(r\to R_I^-).
  \label{en:eq-spectral-endpoint-summary}
\end{equation}
The center powers select the \(r^3\) family, while the surface powers select
finite endpoint values with \((R_I-r)u_I'\to0\).  They also give the uniform
tail bounds
\begin{equation}
 \int_0^\delta W_I|u_I|^2\,\dd r
 \le C\delta^2\|u\|_{\mathcal V_\sigma}^2,
 \qquad
 \int_{R_I-\delta}^{R_I}W_I|u_I|^2\,\dd r
 \le C(\delta+\delta^{\alpha_I+1})\|u\|_{\mathcal V_\sigma}^2,
 \label{en:eq-main-spectral-tail-summary}
\end{equation}
Interior Rellich compactness together with
\eqref{en:eq-main-spectral-tail-summary} proves the
compact embedding.  The lower-order coefficients satisfy
locally uniformly on the equilibrium chart,
\begin{equation}
 |\widetilde Q_I|\le CW_I,\qquad
 |S_{IJ}|\le C(W_IW_J)^{1/2},\qquad
 |f_{IJ}|^2\le CP_IW_J,\qquad
 |f_{IJ}|^2\le CP_IW_I.
\end{equation}
These coefficient bounds imply
\begin{equation}
 |\mathfrak r_\sigma[u]|
 \le\varepsilon\mathfrak a_{0,\sigma}[u]
   +C_{\varepsilon,V}\|u\|_{\mathcal H_\sigma}^2,
 \label{en:eq-main-relative-form-bound}
\end{equation}
which gives closedness and the local uniform lower bound.

\subsection{Symmetric expression and coefficient bounds}
\label{en:app-spectral-realization}

Sections~\ref{en:app-spectral-realization}--\ref{en:sec-physical-endpoints}
prove Lemma~\ref{en:lem-spectral-endpoint-structure} and
Theorem~\ref{en:thm-fixed-spectral-package}.
Section~\ref{en:sec-spectral-parameter-dependence} proves
Theorem~\ref{en:thm-ordered-spectral-continuity}.  The fixed-equilibrium argument uses
the weak quadratic form of the radial expression.  At an ordinary vacuum
endpoint, coefficient derivatives are integrable and can be unbounded; the
weak formulation treats them as integrated terms controlled by uniform
relative-form estimates.  Transport to common weighted spaces and
endpoint tail estimates for changing overlaps then yield parameter continuity.

\setproofparent{en:lem-spectral-endpoint-structure}

Fix \(\sigma\in U\), suppress it from the notation, and retain
\(h_I,b_I,W_I,P_I\) from \eqref{en:eq-spectral-WP}.  All matter coefficients
are extended by zero outside their labeled active intervals.  A coefficient
with labels \(I,J\) is therefore supported in
\((0,R_I\wedge R_J)\).

\begin{sublemma}[Symmetric radial expression]
\label{en:lem-symmetric-spectral-form}
Introduce the background blocks
\begin{equation}
  \Xi:=\frac{e^{\Lambda+3\Phi}}{r^2},\qquad
  \Gamma:=4\pi rA^{-1},\qquad
  \Theta:=\Lambda'+\Phi'=\Gamma h_{\mathrm{tot}},
  \qquad \chi:=\Phi'.
  \label{en:eq-spectral-background-blocks}
\end{equation}
and recall that \(c_{s,I}^2=b_I/h_I\).  Define
\begin{equation}
  X_I:=b_I\zeta_I'+(-h_I\chi+b_I\Theta)\zeta_I
  -b_I\Gamma\sum_Jh_J\zeta_J.
  \label{en:eq-spectral-X-block}
\end{equation}
Then \(\delta p_I=-(e^\Phi/r^2)X_I\), and the reduced expression obeys the
exact identity
\begin{align}
  (\mathcal L_\sigma\zeta)_I
  ={}&-(\Xi X_I)'
  +\Xi\left(\Theta-\frac{\chi}{c_{s,I}^2}\right)X_I
  -\Xi\Gamma h_I\sum_{L=1}^KX_L
  \notag\\
  &-\Xi\Gamma\left(2\chi+\frac1r\right)
       h_I\sum_{L=1}^Kh_L\zeta_L.
  \label{en:eq-spectral-exact-X-identity}
\end{align}
On every piecewise classical test vector, the expression
\(\mathcal L_\sigma\) in \eqref{en:app-reduced-operator} can be written as
\begin{equation}
  (\mathcal L_\sigma\zeta)_I
  =-(P_I\zeta_I')'+Q_I\zeta_I
  +\sum_{J\ne I}
  \left\{(f_{IJ}\zeta_J)'-f_{JI}\zeta_J'+S_{IJ}\zeta_J\right\},
  \label{en:eq-symmetric-K-fluid-expression}
\end{equation}
where all coefficients are real,
\begin{equation}
  f_{IJ}=4\pi\frac{e^{3\Lambda+3\Phi}}r\,b_Ih_J,
  \qquad
  S_{IJ}=\Xi\Gamma h_Ih_J
  \left[\Gamma b_{\mathrm{tot}}-\frac1r
        -(c_{s,I}^2+c_{s,J}^2)\Theta\right]
  =S_{JI},
  \label{en:eq-symmetric-cross-coefficients}
\end{equation}
and
\begin{equation}
  Q_I=\widetilde Q_I-\sum_{J\ne I}f_{IJ}',
  \label{en:eq-Q-derivative-splitting}
\end{equation}
where the complete diagonal coefficient is
\begin{equation}
  \widetilde Q_I
  =\Xi h_I\left[
    \Phi''+\chi^2-\left(\Theta+\frac2r\right)\chi
    +c_{s,I}^2\Theta(\Theta-2\Gamma h_I)
    +\Gamma h_I\left(\Gamma b_{\mathrm{tot}}-\frac1r\right)
  \right].
  \label{en:eq-complete-Qtilde}
\end{equation}
The splitting places each differentiated endpoint coefficient in
\(-\sum_{J\ne I}f_{IJ}'\), while \(\widetilde Q_I\) consists of the algebraic
background terms collected in \eqref{en:eq-complete-Qtilde}.
\end{sublemma}

\begin{proof}
By \eqref{en:app-recover-p}, \(\Theta=\Gamma h_{\mathrm{tot}}\), and
\eqref{en:eq-spectral-X-block}, one has
\(\delta p_I=-(e^\Phi/r^2)X_I\).  Substitution of the recovery identities
into \eqref{en:app-reduced-operator}, followed by
\(\Xi'/\Xi=\Theta+2\chi-2/r\), gives
\eqref{en:eq-spectral-exact-X-identity}.  Expanding \(X_I\) and using
\[
\begin{aligned}
 \Xi b_I(\Theta-\Gamma h_I)
   &=\sum_{J\ne I}f_{IJ},\\
 -h_I\chi+b_I\Theta-\Gamma h_I b_{\mathrm{tot}}
   &=h_I(-\chi+c_{s,I}^2\Theta-\Gamma b_{\mathrm{tot}}),\\
 \frac{h_I'}{h_I}
   &=-(1+c_{s,I}^{-2})\chi .
\end{aligned}
\]
yields \eqref{en:eq-symmetric-K-fluid-expression}--
\eqref{en:eq-complete-Qtilde}; \(S_{IJ}=S_{JI}\) proves the asserted
symmetry.
\end{proof}

\phantomsection
\label{en:app-proof-spectral-endpoint}
\continueproofparent{en:lem-spectral-endpoint-structure}
\begin{proof}[Proof of Lemma~\ref{en:lem-spectral-endpoint-structure}]
At the center the even background expansions give
\eqref{en:eq-derived-Fuchs-center}.  Here
\(O_{\mathrm{con}}(x^{\eta_{\mathrm{sp}}})\) denotes a
term for which the finitely many \(x\partial_x\)-derivatives used below are
also \(O(x^{\eta_{\mathrm{sp}}})\).  At a component surface, with \(x=R_I-r\),
Lemma~\statementpartrange{en:lem-equilibrium-package}
{en:item-eos-consequences}{en:item-surface-tail} gives
\(H_I=\kappa_Ix[1+O_{\mathrm{con}}(x^{\eta_{\mathrm{sp}}})]\),
\(h_I\asymp x^{\alpha_I}\), and
\(b_I\asymp x^{\alpha_I+1}\), with the required conormal remainders.  The
metric multipliers have positive one-sided limits.  The background identities
\begin{equation}
  m_{\mathrm{tot}}'=4\pi r^2\epsilon_{\mathrm{tot}},\qquad
  \Phi'=\frac{m_{\mathrm{tot}}+4\pi r^3p_{\mathrm{tot}}}{r^2A},\qquad A=1-\frac{2m_{\mathrm{tot}}}{r}.
  \label{en:eq-app-background-algebraic-system}
\end{equation}
show that the endpoint-sensitive differentiated matter contribution satisfies
\(x\partial_x\epsilon_I=O(x^{\alpha_I})\).  Direct substitution in
\eqref{en:eq-symmetric-cross-coefficients} and
\eqref{en:eq-complete-Qtilde} therefore proves
\eqref{en:eq-Fuchs-principal-expansions}--
\eqref{en:eq-Fuchs-coupling-factorization}. Transversality and the
no-horizon bound control the metric and each component's own endpoint
weight locally uniformly. For the single-node conormal expansions, the
other distinct nodes stay outside the chosen collar. The endpoint
substitutions are detailed in
Section~\ref{sup:sec-conormal-endpoint-substitutions}.

For parameter dependence,
Lemma~\ref{en:lem-fixed-domain-parameter-regularity} and the labelwise identity
\(H_I(R_Is;\sigma)=(1-s)k_I(s;\sigma)\), \(\inf k_I>0\), give
\(L^\infty\) convergence of
\[
 \frac{\widehat W_{I,\sigma}}{w_I^*},\quad
 \frac{w_I^*}{\widehat W_{I,\sigma}},\quad
 \frac{\widehat P_{I,\sigma}}{p_I^*},\quad
 \frac{p_I^*}{\widehat P_{I,\sigma}}.
\]
On a smooth equilibrium sector, let \(r=\Psi_\sigma(s)\) denote a cellwise
fixed-node pullback.  For any background profile \(f\), write
\[
  D_v^{\mathrm{fn}}f:=D_v\bigl[f(\Psi_\sigma(\mathord\cdot);\sigma)\bigr]
  \quad\text{with \(s\) held fixed}.
\]
Because this pullback fixes every surface node,
\(D_v^{\mathrm{fn}}H_I=O(x)\), \(D_v^{\mathrm{fn}}h_I=O(x^{\alpha_I})\), and
\(D_v^{\mathrm{fn}}b_I=O(x^{\alpha_I+1})\).  Differentiating the algebraic formula for
\(\Phi'\) eliminates \(\Phi''\), and the conormal chain rule then applies
termwise to the four normalized coefficients in
\eqref{en:eq-form-control-ratios}. The sectorwise difference quotients
converge in the cellwise \(L^\infty\) norms by
Section~\ref{sup:sec-conormal-endpoint-substitutions}. Across changes in
the order of the radii, the algebraic formulas
\eqref{sup:eq-parameter-RQ}--\eqref{sup:eq-parameter-Rf-diagonal} keep the
ratios uniformly bounded up to their own endpoints. On compact intervals
avoiding the limiting nodes they converge uniformly by the background
dependence in Lemma~\ref{en:lem-equilibrium-package}. For overlaps near
meeting surfaces, each fluid retains its own distance \(x_I=R_I-r\) as in
\eqref{sup:eq-overlap-two-end}.
\end{proof}

\begin{sublemma}[Uniform form coefficient bounds]
\label{en:lem-spectral-coefficient-bounds}
For every \(\bar\sigma\in U\), there is a relatively compact neighborhood
\(V\Subset U\) of \(\bar\sigma\) and a constant \(C_V<\infty\), independent
of \(\sigma\in V\), such that almost
everywhere on the relevant active intervals,
\begin{align}
  |\widetilde Q_I|&\le C_VW_I,
  \label{en:eq-Qtilde-bound}\\
  |S_{IJ}|&\le C_V(W_IW_J)^{1/2},
  \label{en:eq-S-bound}\\
  |f_{IJ}|^2&\le C_VP_IW_J,
  \label{en:eq-f-cross-bound}\\
  |f_{IJ}|^2&\le C_VP_IW_I.
  \label{en:eq-f-diagonal-bound}
\end{align}
\end{sublemma}

\begin{proof}
Away from the finitely many distinguished radii, ordinary compact-interval
bounds give the stated estimates.  At the center and component surfaces, the
endpoint factorizations reduce them to the normalized identities
\begin{align}
  \frac{\widetilde Q_I}{W_I}
  &=e^{2(\Phi-\Lambda)}
  \left[
    \Phi''+\chi^2-\left(\Theta+\frac2r\right)\chi
    +c_{s,I}^2\Theta(\Theta-2\Gamma h_I)
    +\Gamma h_I\left(\Gamma b_{\mathrm{tot}}-\frac1r\right)
  \right],
  \label{en:eq-Qtilde-normalized}\\
  \frac{S_{IJ}}{(W_IW_J)^{1/2}}
  &=e^{2(\Phi-\Lambda)}\Gamma(h_Ih_J)^{1/2}
  \left[\Gamma b_{\mathrm{tot}}-\frac1r-(c_{s,I}^2+c_{s,J}^2)\Theta\right].
  \label{en:eq-S-normalized}
\end{align}
At the center, \(\Gamma=O(r)\), \(\Theta,\chi=O(r)\), and
\(\Phi''=O(1)\); hence the factors \(\Gamma/r\) and \(\chi/r\) are
bounded.  At a component surface, \(r>0\), the metric factors and the square
brackets have bounded one-sided limits, while the ending \(h_I\) factors
vanish.  Equations~\eqref{en:eq-Qtilde-normalized}--
\eqref{en:eq-S-normalized} therefore prove
\eqref{en:eq-Qtilde-bound}--\eqref{en:eq-S-bound}.

For the derivative coefficients, cancellation of the metric factors gives,
up to fixed numerical constants and uniformly bounded positive metric
factors,
\begin{equation}
  \frac{f_{IJ}^2}{P_IW_J}\asymp r^2b_Ih_J,
  \qquad
  \frac{f_{IJ}^2}{P_IW_I}\asymp r^2c_{s,I}^2h_J^2,
  \label{en:eq-f-ratios}
\end{equation}
Both ratios are bounded and prove
\eqref{en:eq-f-cross-bound}--\eqref{en:eq-f-diagonal-bound}.  Uniformity follows
from compactness of \(\overline V\),
Lemma~\ref{en:lem-spectral-endpoint-structure}, and the no-horizon lower bound
in \eqref{en:eq-derived-A-and-kernel}.
Section~\ref{sup:sec-coefficient-order-table} lists the coefficient orders,
including coincident endpoints with unequal exponents.

If \(J\) ends strictly inside the support of \(I\), then
\(f_{IJ}'=O((R_J-r)^{\alpha_J-1})\).  The integrated form identity represents
this locally integrable contribution through \(f_{IJ}\), to which
\eqref{en:eq-f-diagonal-bound} applies.
\end{proof}

\subsection{Closed form, lower bound, and compact resolvent}
\label{en:app-proof-fixed-spectral}

\setproofparent{en:thm-fixed-spectral-package}

Recall the test space, principal form, and principal form domain defined in
\eqref{en:eq-main-form-definition} and
\eqref{en:eq-main-closed-form-domain}.
\begin{sublemma}[Form completion representation and endpoint estimates]
\label{en:lem-form-completion-representation}
Let \(\mathcal V_\sigma\) be the completion in
\eqref{en:eq-main-closed-form-domain}.  The canonical map
\[
  \jmath_\sigma:\mathcal V_\sigma\longrightarrow\mathcal H_\sigma,
  \qquad
  \jmath_\sigma([u^{(n)}])
  :=\lim_{n\to\infty}u^{(n)}
  \quad\hbox{in }\mathcal H_\sigma,
\]
is well defined, continuous, and injective, with
\(\|\jmath_\sigma u\|_{\mathcal H_\sigma}\le\|u\|_{\mathcal V_\sigma}\).
Its image is dense in \(\mathcal H_\sigma\), since
\(C_{c,\sigma}^{\infty}\) is dense there.  We identify \(\mathcal V_\sigma\) with
its image.  Every
\(u=(u_I)\in\mathcal V_\sigma\) has a unique componentwise
representative \(u_I\in H^1_{\mathrm{loc}}(0,R_I)\), with
\[
  \|u\|_{\mathcal V_\sigma}^2
  =\sum_I\int_0^{R_I}
    \bigl(P_I|u_I'|^2+W_I|u_I|^2\bigr)\,\dd r .
\]
This construction proves that \(\mathfrak a_{0,\sigma}\) is closable in
\(\mathcal H_\sigma\), with closed form domain \(\mathcal V_\sigma\).
For sufficiently small endpoint collars,
\[
  |u_I(r)|^2\le Cr^3\|u\|_{\mathcal V_\sigma}^2
  \quad(r\to0^+),\qquad
  |u_I(R_I-x)|^2
  \le C(1+x^{-\alpha_I})\|u\|_{\mathcal V_\sigma}^2
  \quad(x\to0^+).
\]
In particular, \(u_I(r)\to0\) at the center, and the corresponding weighted
\(L^2\) tails are uniformly controlled on form-bounded sets.
\end{sublemma}

\begin{proof}
Take a \(\mathcal V_\sigma\)-Cauchy sequence
\(u^{(n)}\in C_{c,\sigma}^{\infty}\).  It is Cauchy in
\(\mathcal H_\sigma\), while \(P_I^{1/2}u_I^{(n)\prime}\) is Cauchy in
\(L^2(0,R_I)\).  On every compact \(K\Subset(0,R_I)\), the endpoint
asymptotics make \(P_I\) and \(W_I\) bounded above and below by positive
constants.  The \(\mathcal H_\sigma\)-limit and the local derivative limit
therefore satisfy
\[
  \int_K u_I\phi'\,\dd r
  =-\int_K u_I'\phi\,\dd r
  \qquad(\phi\in C_c^\infty(K)),
\]
so \(u_I\in H^1_{\mathrm{loc}}(0,R_I)\).  If the \(\mathcal H_\sigma\)-limit is
zero, this identity forces every weak derivative limit to vanish; hence the
completion norm is zero.  This proves injectivity and the norm identity.

For test functions, weighted Cauchy--Schwarz and
\(P_I^{-1}\lesssim r^2\) at the center give the first estimate.  At a surface,
fix an interior point \(x_0>0\); a local Sobolev estimate controls
\(u_I(R_I-x_0)\) by the form norm, and
\(P_I^{-1}\lesssim x^{-\alpha_I-1}\) gives the second estimate.  Local
\(H^1\) convergence passes both estimates to the completion, and integration
against \(W_I\lesssim r^{-2}\) and \(W_I\lesssim x^{\alpha_I}\) gives the
corresponding center and surface tail bounds.
\end{proof}

Recall the full pre-form \(\mathfrak a_\sigma\) from
\eqref{en:eq-main-form-definition}.
Integration by parts in \eqref{en:eq-symmetric-K-fluid-expression} gives
\begin{align}
  \mathfrak a_\sigma[u,v]
  ={}&\sum_I\int P_I\overline{u_I'}v_I'
      +\sum_I\int \widetilde Q_I\overline{u_I}v_I
  \notag\\
  &+\sum_{I\ne J}\int
  \left[
        f_{IJ}(\overline{u_I'}v_I+\overline{u_I}v_I')
        -f_{IJ}\overline{u_I'}v_J
        -f_{JI}\overline{u_I}v_J'
        +S_{IJ}\overline{u_I}v_J
  \right].
  \label{en:eq-full-spectral-form-expanded}
\end{align}
Each two-label integral is over \((0,R_I\wedge R_J)\).  The diagonal
first-order pair is Hermitian, and pairing the ordered terms \(I,J\) and
\(J,I\), together with \(S_{IJ}=S_{JI}\), shows that the full form is
Hermitian.  The conormal and divergence fluxes associated with this integrated
form are
\begin{equation}
\begin{aligned}
  \mathcal F_I^{\mathrm{con}}[u]
  &:=P_Iu_I'+\sum_{J\ne I}f_{IJ}(u_I-u_J),\\
  \mathcal F_I^{\mathrm{div}}[u]
  &:=P_Iu_I'-\sum_{J\ne I}f_{IJ}u_J.
\end{aligned}
  \label{en:eq-form-conormal-flux}
\end{equation}
They differ by \(u_I\sum_{J\ne I}f_{IJ}\).  This difference vanishes at the
center and at a labeled endpoint and has the same one-sided limit from both
sides of an internal ending block.  Their endpoint and jump conditions
therefore coincide, while their interior expressions remain distinct.

\begin{sublemma}[Infinitesimal form bound]
\label{en:lem-infinitesimal-form-bound}
Let \(\bar\sigma\in U\) and let \(V\Subset U\) be a relatively compact
neighborhood as in Lemma~\ref{en:lem-spectral-coefficient-bounds}.  For every
\(\varepsilon>0\), after replacing \(V\), if necessary, by a smaller
relatively compact neighborhood of \(\bar\sigma\), there is
\(C_{\varepsilon,V}<\infty\) such that
\begin{equation}
  |\mathfrak a_\sigma[u]-\mathfrak a_{0,\sigma}[u]|
  \le\varepsilon\mathfrak a_{0,\sigma}[u]
  +C_{\varepsilon,V}\|u\|_{\mathcal H_\sigma}^2,
  \qquad \sigma\in V.
  \label{en:eq-infinitesimal-form-bound}
\end{equation}
\end{sublemma}

\begin{proof}
The term involving \(f_{IJ}'\) is defined on the form completion by
integration by parts,
\begin{equation}
  -\int f_{IJ}'|u_I|^2\,\dd r
  =2\operatorname{Re}\int f_{IJ}\overline{u_I}u_I'\,\dd r.
  \label{en:eq-fprime-form-integration}
\end{equation}
The right side defines the continuous form extension of this derivative
contribution. After exchanging indices or taking a complex conjugate where
necessary, the first-order terms obey
\[
 \left|\int f_{IJ}\overline{u_I'}u_L\,\dd r\right|
 \le\left(\int P_I|u_I'|^2\,\dd r\right)^{1/2}
    \left(\int f_{IJ}^2P_I^{-1}|u_L|^2\,\dd r\right)^{1/2},
 \qquad L\in\{I,J\}.
\]
For \(L=I\) use \eqref{en:eq-f-diagonal-bound}, and for \(L=J\)
use \eqref{en:eq-f-cross-bound}. Young's inequality then gives an
arbitrarily small principal form part, while \(\widetilde Q_I\) and \(S_{IJ}\)
are Hilbert-form bounded.  Summation over the finite label set proves
\eqref{en:eq-infinitesimal-form-bound}.
\end{proof}

For the form difference \(\mathfrak r_\sigma\) introduced after
\eqref{en:eq-main-closed-form-domain}, the concrete realization from
Lemma~\ref{en:lem-form-completion-representation}, the preceding proof gives
a unique continuous
extension of \(\mathfrak r_\sigma\) to \(\mathcal V_\sigma\), with
relative \(\mathfrak a_{0,\sigma}\)-bound zero.  The
Kato--Lax--Milgram--Nelson theorem
\cite{Kato} therefore implies that
\(\mathfrak a_{0,\sigma}+\mathfrak r_\sigma\) is a dense, closed,
lower-semibounded form on \(\mathcal V_\sigma\).  For fixed
\(\varepsilon<1\), \eqref{en:eq-infinitesimal-form-bound} makes the shifted
form norm equivalent to
\((\mathfrak a_{0,\sigma}+\|\cdot\|_{\mathcal H_\sigma}^2)^{1/2}\).
The norm equivalence makes the original test space
\(C_{c,\sigma}^{\infty}\), already a core for the principal form by
definition, a form core for the total form as well.  The resulting closed
form is the unique closure of the symmetric semibounded pre-form
\(\mathfrak a_\sigma\) on \(C_{c,\sigma}^{\infty}\).  The first
representation theorem associates a unique self-adjoint operator \(T_\sigma\)
with the closed form~\cite{Kato}.
Its distributional action on each open active cell is
\begin{equation}
  T_\sigma u=\mathcal W_\sigma^{-1}\mathcal L_\sigma u.
  \label{en:eq-Friedrichs-operator-action}
\end{equation}
Choosing \(\varepsilon<1\) gives the uniform lower bound
\eqref{en:eq-uniform-lower-bound}, proving
Theorem~\statementpartref{en:thm-fixed-spectral-package}
{en:item-spectral-lower-bound}.

\begin{sublemma}[Weighted Rellich compactness]
\label{en:lem-weighted-Rellich}
The embedding
\begin{equation}
  \mathcal V_\sigma\hookrightarrow\mathcal H_\sigma
  \label{en:eq-weighted-compact-embedding}
\end{equation}
is compact.
\end{sublemma}

\begin{proof}
It suffices to treat one component.  On each compact
\(K\Subset(0,R_I)\), Lemma~\ref{en:lem-form-completion-representation} gives
\(H^1(K)\)-bounded restrictions, so ordinary one-dimensional Rellich
compactness applies~\cite{BrezisFA}.

It remains to make both endpoint tails uniformly small on form-bounded sets.
At the center, the first estimate in
Lemma~\ref{en:lem-form-completion-representation} gives
\begin{align}
  |u_I(r)|^2
  &\le\left(\int_0^rP_I|u_I'|^2\,\dd s\right)
       \left(\int_0^rP_I^{-1}\,\dd s\right)
  \le Cr^3\mathfrak a_{0,\sigma}[u],
  \notag\\
  \int_0^\delta W_I|u_I|^2\,\dd r
  &\le C\delta^2\mathfrak a_{0,\sigma}[u].
  \label{en:eq-center-form-tail}
\end{align}

At the vacuum surface set \(x=R_I-r\), and choose a fixed point \(x_0\)
in an interior collar.  The second estimate in
Lemma~\ref{en:lem-form-completion-representation}, obtained from the local
Sobolev bound and \(P_I^{-1}\asymp x^{-\alpha_I-1}\), gives
\begin{align}
  |u_I(x)|^2
  &\le2|u_I(x_0)|^2
  +2\left(\int_x^{x_0}P_I|u_{I,x}|^2\,\dd t\right)
     \left(\int_x^{x_0}P_I^{-1}\,\dd t\right)
  \notag\\
  &\le C\left(\mathfrak a_{0,\sigma}[u]
    +\|u\|_{\mathcal H_\sigma}^2\right)(1+x^{-\alpha_I}).
  \label{en:eq-surface-pointwise-form-bound}
\end{align}
Multiplication by \(W_I\asymp x^{\alpha_I}\) yields
\begin{equation}
  \int_0^\delta W_I|u_I|^2\,\dd x
  \le C(\delta+\delta^{\alpha_I+1})
  \left(\mathfrak a_{0,\sigma}[u]
    +\|u\|_{\mathcal H_\sigma}^2\right).
  \label{en:eq-surface-form-tail}
\end{equation}
The same pointwise bound gives \(W_I|u_I|^2\le
C\|u\|_{\mathcal V_\sigma}^2\) throughout a fixed small surface collar.
The two tail estimates, interior Rellich compactness, and a finite diagonal
extraction over the \(K\) components prove the result.
\end{proof}

After adding a sufficiently large positive shift \(c_*\), the shifted resolvent factors
as
\begin{equation}
  \left(T_\sigma+c_*\operatorname{Id}_{\mathcal H_\sigma}\right)^{-1}
  :\mathcal H_\sigma
  \longrightarrow\mathcal V_\sigma
  \hookrightarrow\mathcal H_\sigma.
\end{equation}
The first arrow is bounded and the second is compact by
Lemma~\ref{en:lem-weighted-Rellich}.  This bounded--compact factorization
proves that the resolvent is compact, establishing
Theorem~\statementpartref{en:thm-fixed-spectral-package}
{en:item-spectral-compact-resolvent}.  The spectral theorem for a
self-adjoint compact-resolvent operator then gives the real discrete spectrum
and finite multiplicities asserted there~\cite{ReedSimonIV}.

\subsection{Endpoint regularity and physical eigenspaces}
\label{en:sec-physical-endpoints}

\begin{sublemma}[Finite-energy endpoint selection]
\label{en:lem-finite-energy-Fuchs-selection}
Let \(D_\alpha=\operatorname{diag}(\alpha_1,\ldots,\alpha_m)\), with every
\(\alpha_i>0\), and put
\begin{equation}
  A_0:=
  \begin{pmatrix}
    0&\operatorname{Id}_m\\
    0&-D_\alpha
  \end{pmatrix}.
\end{equation}
Suppose that
\(Y=(u,v)\in\mathrm{AC}_{\mathrm{loc}}((0,\delta))\) satisfies
\begin{equation}
  xY'(x)=[A_0+R_{\mathrm{rem}}(x)]Y(x)+q_{\mathrm{F}}(x),
  \qquad
  \|R_{\mathrm{rem}}(x)\|+\|q_{\mathrm{F}}(x)\|
  \le Cx^{\eta_{\mathrm{F}}}
  \label{en:eq-regular-Fuchs-system}
\end{equation}
for some \(\eta_{\mathrm{F}}>0\), and
\begin{equation}
  \sum_{i=1}^m\int_0^\delta
  \left(x^{\alpha_i}|u_i|^2+
        x^{\alpha_i-1}|v_i|^2\right)\,\dd x<\infty.
  \label{en:eq-Fuchs-finite-energy}
\end{equation}
Then every \(u_i\) has a finite limit at \(x=0\), and \(v_i(x)\to0\).
The conclusion includes repeated exponents, resonant blocks, and regular
corrections \(x^\gamma(\log x)^k\), \(\gamma>0\).
\end{sublemma}

\noindent\emph{Proof.}
The substitution \(t=-\log x\) converts
\eqref{en:eq-regular-Fuchs-system} into an exponentially perturbed constant
system.  Its Volterra decomposition consists of a bounded center family and
positive exponential families.  Condition
\eqref{en:eq-Fuchs-finite-energy} selects the center family.  The complete
spectral block construction is given in
Section~\ref{sup:sec-fuchs-construction}.

\begin{sublemma}[Weak-to-Fuchs regularity at a component radius]
\label{en:lem-weak-to-Fuchs-regularity}
Let \(u\in\operatorname{Dom}(T_\sigma)\) and \(T_\sigma u=\lambda u\).  Fix a distinct
component radius \(R\), choose \(\delta>0\) so that \((R-\delta,R)\) contains
no other surface node, put \(x=R-r\), and set
\(\mathcal K_R:=\mathcal B_R\cup\mathcal A_R^+\).  On this collar, for
\(L\in\mathcal K_R\), the fluxes in
\eqref{en:eq-form-conormal-flux} take the restricted form
\begin{equation}
  \mathcal F_L^{\mathrm{div}}[u]
  =P_Lu_L'-\sum_{J\in\mathcal K_R\setminus\{L\}}f_{LJ}u_J,
  \qquad
  \mathcal F_L^{\mathrm{con}}[u]
  =\mathcal F_L^{\mathrm{div}}[u]+u_L
    \sum_{J\in\mathcal K_R\setminus\{L\}}f_{LJ}.
  \label{en:eq-weak-Fuchs-fluxes}
\end{equation}
Then \(\mathcal F_L^{\mathrm{div}}[u],
\mathcal F_L^{\mathrm{con}}[u]\in W^{1,1}(R-\delta,R)\).  If
\(J\in\mathcal A_R^+\), the two fluxes with \(L=J\) extend through \(R\) as
\(W^{1,1}\) functions, and \(u_J,u_J'\) have finite one-sided limits there.
Moreover, with
\begin{equation}
  v_I:=xu_{I,x},
  \qquad
  Y_R:=\bigl((u_I)_{I\in\mathcal B_R},
             (v_I)_{I\in\mathcal B_R}\bigr),
  \label{en:eq-weak-Fuchs-vector}
\end{equation}
the vector \(Y_R\) belongs to
\(\mathrm{AC}_{\mathrm{loc}}((0,\delta))\) and satisfies
\begin{equation}
  xY_R'
  =\left[
  \begin{pmatrix}
    0&\operatorname{Id}_{|\mathcal B_R|}\\
    0&-\operatorname{diag}(\alpha_I)_{I\in\mathcal B_R}
  \end{pmatrix}
  +R_{\mathrm{rem},R}(x)\right]Y_R+q_{\mathrm{F},R}(x),
  \label{en:eq-endpoint-Fuchs-reduction}
\end{equation}
where, for \(\eta_0:=\min\{\eta_{\mathrm{sp}},1\}>0\),
\begin{equation}
  \|R_{\mathrm{rem},R}(x)\|+\|q_{\mathrm{F},R}(x)\|\le Cx^{\eta_0}.
  \label{en:eq-endpoint-Fuchs-remainder}
\end{equation}
It also obeys
\begin{equation}
  \sum_{I\in\mathcal B_R}\int_0^\delta
  \left(x^{\alpha_I}|u_I|^2+x^{\alpha_I-1}|v_I|^2\right)\,\dd x<\infty.
  \label{en:eq-endpoint-Fuchs-energy}
\end{equation}
\end{sublemma}

\begin{proof}
Since \(\operatorname{Dom}(T_\sigma)\subset\mathcal V_\sigma\), local ellipticity
gives \(u_L\in H^1_{\mathrm{loc}}\), while
\eqref{en:eq-surface-pointwise-form-bound} gives
\begin{equation}
  |u_I(x)|\le C(1+x^{-\alpha_I/2}),
  \qquad I\in\mathcal B_R.
  \label{en:eq-weak-Fuchs-pointwise-bound}
\end{equation}
Equations~\eqref{en:eq-symmetric-K-fluid-expression} and
\eqref{en:eq-Q-derivative-splitting} put the distributional eigenequation in
the exact flux form
\begin{equation}
\begin{aligned}
  (\mathcal F_L^{\mathrm{div}})'
  ={}&(\widetilde Q_L-\lambda W_L)u_L
  -\sum_{J\in\mathcal K_R\setminus\{L\}}f_{LJ}'u_L\\
  &-\sum_{J\in\mathcal K_R\setminus\{L\}}f_{JL}u_J'
  +\sum_{J\in\mathcal K_R\setminus\{L\}}S_{LJ}u_J
  \quad\hbox{in }\mathcal D'(R-\delta,R).
\end{aligned}
  \label{en:eq-weak-flux-derivative}
\end{equation}
First, \(\mathcal F_L^{\mathrm{div}}\in L^1\): Cauchy--Schwarz and
\eqref{en:eq-f-cross-bound} give
\begin{equation}
\begin{aligned}
  \int|P_Lu_L'|
  &\le\left(\int P_L|u_L'|^2\right)^{1/2}
       \left(\int P_L\right)^{1/2},\\
  \int|f_{LJ}u_J|
  &\le\left(\int\frac{f_{LJ}^2}{W_J}\right)^{1/2}
       \left(\int W_J|u_J|^2\right)^{1/2}<\infty.
\end{aligned}
  \label{en:eq-weak-flux-L1}
\end{equation}
For an ending row \(I\), the factorizations
\eqref{en:eq-Fuchs-coupling-factorization} and
\eqref{en:eq-weak-Fuchs-pointwise-bound} give, in the worst
ending--surviving cases,
\[
 f_{IJ}'u_I=O(x^{\alpha_I/2}),\qquad
 f_{JI}'u_J=O(x^{\alpha_I-1}).
\]
The latter is integrable because \(\alpha_I>0\).  The remaining derivative
terms are controlled by the same factorizations and Cauchy--Schwarz.  For two
surviving labels, the derivative coefficients \(f_{LJ}\) extend as bounded
\(C^1\) functions across the collar.  The multiplication coefficients need
not be \(C^1\) when another component ends with \(0<\alpha_I<1\); their
boundedness and local integrability are sufficient here.
Section~\ref{sup:sec-weak-to-fuchs} lists the termwise orders.  The terms involving
\(\widetilde Q_L,W_L,S_{LJ}\) are controlled by
 \eqref{en:eq-Qtilde-bound}--\eqref{en:eq-S-bound}.  These bounds place the
 right side of \eqref{en:eq-weak-flux-derivative} in \(L^1\) and hence prove
 \(\mathcal F_L^{\mathrm{div}}\in W^{1,1}\).  The same estimates applied to
\begin{equation}
  (\mathcal F_L^{\mathrm{con}}-\mathcal F_L^{\mathrm{div}})'
  =\sum_{J\in\mathcal K_R\setminus\{L\}}
    (f_{LJ}'u_L+f_{LJ}u_L')
  \label{en:eq-form-flux-difference-derivative}
\end{equation}
prove \(\mathcal F_L^{\mathrm{con}}\in W^{1,1}\).  For a surviving row the zero extension
of every ending coefficient has no boundary delta and the same estimates
hold on both sides of \(R\).  Testing the global form equation with functions
supported across \(R\) identifies the two one-sided representatives; hence
both fluxes are \(W^{1,1}\) through the node.

If \(J\in\mathcal A_R^+\), then \(P_J,W_J\) remain positive, so
\(u_J\in H^1(R-\delta,R)\) and has a finite left limit.  The flux also has a
finite limit; for ending \(I\),
\(f_{JI}u_I=O(x^{\alpha_I/2})\to0\), while products with surviving labels
have finite limits.  Hence
\begin{equation}
  u_J'=P_J^{-1}\left(\mathcal F_J^{\mathrm{div}}+
  \sum_{L\in\mathcal K_R\setminus\{J\}}f_{JL}u_L\right)
  \label{en:eq-surviving-derivative-limit}
\end{equation}
has a finite left limit; the right limit follows from ordinary regularity on
the next cell.  On every punctured collar, the same identity and
\eqref{en:eq-weak-flux-derivative} show that \(u_I'\) is locally absolutely
continuous, so \(Y_R\in\mathrm{AC}_{\mathrm{loc}}((0,\delta))\).

Divide the ending row \(I\) by \(p_{I,0}x^{\alpha_I-1}\).
Equations~\eqref{en:eq-Fuchs-principal-expansions}--
\eqref{en:eq-Fuchs-coupling-factorization} give
\eqref{en:eq-endpoint-Fuchs-reduction}: the principal remainder is
\(O(x^{\eta_{\mathrm{sp}}})\), an ending label \(J\) contributes
\(O(x^{\alpha_J+1})(u_J+v_J)\), and a surviving label contributes
\(O(x)(u_J+u_J')\).  The finite surviving limits just proved yield
\eqref{en:eq-endpoint-Fuchs-remainder}.  Finally,
\(P_I\asymp x^{\alpha_I+1}\), \(W_I\asymp x^{\alpha_I}\), and
\(u\in\mathcal V_\sigma\) give
\eqref{en:eq-endpoint-Fuchs-energy}.  These form domain estimates supply the
endpoint bounds required by the Fuchs reduction, and
Lemma~\ref{en:lem-finite-energy-Fuchs-selection} supplies the endpoint trace.
\end{proof}

\begin{subproposition}[Piecewise classical regularity of form realization eigenfunctions]
\label{en:prop-Friedrichs-eigenfunction-regularity}
If \(u\in\operatorname{Dom}(T_\sigma)\) and \(T_\sigma u=\lambda u\), then \(u\) is
piecewise classical and satisfies:
\begin{enumerate}
\renewcommand{\theenumi}{\roman{enumi}}
\renewcommand{\labelenumi}{(\theenumi)}
  \item \(u_I(r)=a_Ir^3+O(r^5)\) and
        \(u_I'(r)=3a_Ir^2+O(r^4)\) at the center;
  \item \(u_I\) has a finite labeled surface limit and
  \begin{equation}
    (R_I-r)u_I'(r)\longrightarrow0,
    \label{en:eq-Friedrichs-surface-eigenfunction}
  \end{equation}
  equivalently the reduced component surface condition
  \eqref{en:app-reduced-enthalpy-surface};
  \item at every internal ending block, surviving \(u_J\) and
  \(\delta p_J[u]\) are continuous, and the recovered metric obeys the
  no-layer transmission conditions;
  \item the recovered exterior is the normalized Schwarzschild perturbation.
\end{enumerate}
\end{subproposition}

\begin{proof}
Ordinary one-dimensional regularity applies away from the center and the
finitely many component radii.

At the center, set \(v_I=ru_I'\).  The center orders in
\eqref{en:eq-derived-Fuchs-center}, including all derivative couplings,
reduce the eigenequation to
\begin{equation}
  r\frac{\dd}{\dd r}
  \begin{pmatrix}u\\v\end{pmatrix}
  =
  \left[
  \begin{pmatrix}0&I_K\\0&3I_K\end{pmatrix}
  +O(r^2)
  \right]
  \begin{pmatrix}u\\v\end{pmatrix}.
\end{equation}
The same Volterra construction on \([T,\infty)\) as in
Lemma~\ref{en:lem-finite-energy-Fuchs-selection} gives a constant family and
an \(r^3\) family, including the lower-order resonant corrections.  The
constant family
is not in \(L^2(r^{-2}\dd r)\), so the Hilbert condition selects the latter.
For this branch \(Z=r^{-3}(u,v)\) is bounded and tends to
\((a,3a)\). Its constant indicial matrix has eigenvalues \(-3,0\), and
the forcing from the remainder is \(O(r^2)\). On the \(-3\) block,
multiplication by \(r^3\) and integration from zero give a correction
\(r^{-3}\int_0^r O(s^4)\,\dd s=O(r^2)\). On the zero block the
correction is \(\int_0^r O(s)\,\dd s=O(r^2)\).
Thus \(u_I(r)=a_Ir^3+O(r^5)\) and \(u_I'(r)=3a_Ir^2+O(r^4)\).

Now fix a distinct component radius \(R\) and put \(x=R-r\).
Lemma~\ref{en:lem-weak-to-Fuchs-regularity} gives the locally absolutely
continuous system \eqref{en:eq-endpoint-Fuchs-reduction}, the remainder bound
\eqref{en:eq-endpoint-Fuchs-remainder}, and the finite-energy condition
\eqref{en:eq-endpoint-Fuchs-energy} for the entire ending block
\(\mathcal B_R\).  These estimates place the block in the finite-energy class,
and Lemma~\ref{en:lem-finite-energy-Fuchs-selection} applied to
\eqref{en:eq-endpoint-Fuchs-reduction} therefore proves
\begin{equation}
  u_I(R^-)\in\mathbb C,\qquad
  (R-r)u_I'(r)=-v_I(x)\longrightarrow0
  \quad(I\in\mathcal B_R).
\end{equation}
Since
\(c_{s,I}^2/x\) has a finite positive limit, the derivative limit is
equivalent to \eqref{en:app-reduced-enthalpy-surface}.  The recovery formulas
then give
\begin{equation}
  \delta H_I=O(1),\qquad
  \delta n_I,\delta\epsilon_I=O(x^{\alpha_I-1}),\qquad
  \delta p_I=O(x^{\alpha_I}).
\end{equation}

At an internal ending block, Lemma~\ref{en:lem-weak-to-Fuchs-regularity}
gives the globally \(W^{1,1}\) divergence flux \(\mathcal F_J^{\mathrm{div}}[u]\) in
\eqref{en:eq-weak-Fuchs-fluxes}, so \([\mathcal F_J^{\mathrm{div}}]=0\) for every surviving
component \(J\).  If component \(I\) ends there, then
\(f_{JI}u_I=O(x^{\alpha_I/2})\to0\); all terms with surviving labels have
 common one-sided limits.  We therefore obtain
 \([\mathcal F_J^{\mathrm{div}}]=[P_Ju_J']=0\).
In \eqref{en:app-recover-p}, every term other than \(b_Ju_J'\) is continuous
across the block, and
\begin{equation}
  -\frac{e^\Phi}{r^2}b_Ju_J'
  =-e^{-\Lambda-2\Phi}P_Ju_J'.
\end{equation}
The identity above converts continuity of the weak conormal flux into the
physical pressure matching condition \([\delta p_J[u]]=0\).
Equations~\eqref{en:app-recover-lambda} and
\eqref{en:app-recover-phi-prime} then give the matching one-sided metric
limits required by the no-layer condition.  Finally,
\eqref{en:app-recover-lambda} and the Hamiltonian constraint give
\begin{equation}
  \delta m_{\mathrm{tot}}=rA\delta\Lambda
  =-4\pi e^\Phi\sum_Ih_Iu_I.
\end{equation}
All eigenfunctions have finite labeled surface limits, and \(h_I\to0\) for
every component ending at the outer surface, so \(\delta M=0\).  The inward
lapse recovery fixes the remaining
constant by \(\delta\Phi(\infty)=0\), giving the normalized exterior.
Finally, \(\mathcal L_\sigma u=\lambda\mathcal W_\sigma u\).  The center
order \(u_I=O(r^3)\), the surface orders for \(W_I\), and the piecewise
interior regularity show that
\(\lambda\mathcal W_\sigma u\in\mathcal Y_\sigma\).  Together with the
weighted recovery estimates above, this proves
\(u\in\widehat{\mathcal D}_\sigma\).
\end{proof}

\begin{proof}[Proof of Theorem~\statementpartref
{en:thm-fixed-spectral-package}{en:item-spectral-kernel-compatibility}]
Fix \(\lambda\in\mathbb R\).  If
\(u\in\ker\!\left(T_\sigma-\lambda\operatorname{Id}_{\mathcal H_\sigma}\right)\), Proposition
~\ref{en:prop-Friedrichs-eigenfunction-regularity} gives all center,
component surface, transmission, and exterior conditions.  The recovered enthalpy
perturbation is bounded at each surface, and
\begin{equation}
  \frac{\dd\epsilon_I}{\dd H_I}
  =O((R_I-r)^{\alpha_I-1}),
  \qquad h_I=O((R_I-r)^{\alpha_I}),
\end{equation}
so the recovered matter fields obey the weighted estimates
\eqref{en:eq-static-weighted-limits}.  Ordinary interior regularity gives
\(C^2\) regularity on every open active cell, while the distributional identity
\(T_\sigma u=\lambda u\) identifies the cellwise expression with
\(\lambda\mathcal W_\sigma u\), which belongs to
\(\mathcal Y_\sigma\).  Hence
\(u\in\widehat{\mathcal D}_\sigma\) and
\(\mathcal L_\sigma u=\lambda\mathcal W_\sigma u\) on every active cell,
equivalently as a distribution.

Conversely, let
\(\zeta\in\widehat{\mathcal D}_\sigma\) satisfy
\(\mathcal L_\sigma\zeta=\lambda\mathcal W_\sigma\zeta\).  Its
\(O(r^3)\) center behavior gives finite
Hilbert and principal form norms there.  At a vacuum surface,
\(x\zeta_I'\to0\); hence, for each fixed \(\varepsilon>0\), sufficiently near
the endpoint \(|\zeta_I'|\le\varepsilon/x\), and
\begin{equation}
  \int_0^\delta P_I|\zeta_I'|^2\,\dd x
  \le C\varepsilon^2\int_0^\delta x^{\alpha_I-1}\,\dd x<\infty
\end{equation}
because \(\alpha_I>0\).  This finite-energy statement also places \(\zeta\)
in the closure defining \(\mathcal V_\sigma\).  Indeed, use an
endpoint cutoff that changes from zero to one on \(x\in(\delta,2\delta)\).
The term in which the derivative hits the cutoff is bounded by
\begin{equation}
  C\delta^{-2}\int_\delta^{2\delta}
  x^{\alpha_I+1}|\zeta_I|^2\,\dd x=O(\delta^{\alpha_I})\longrightarrow0,
\end{equation}
because \(\zeta_I\) has a finite surface limit; the removed derivative and Hilbert
tails tend to zero by absolute continuity of the corresponding integrals.
The analogous center cutoff converges because \(\zeta_I=O(r^3)\).
At an internal ending block, \conditionref{D3} gives \([\zeta_J]=0\) and
finite matching one-sided limits of
\(\delta p_J\) for every surviving \(J\).  Since \(b_J\) is strictly positive
there and every other term in the recovery formula
\eqref{en:app-recover-p} has finite one-sided limits, \(\zeta_J'\) has finite
one-sided limits as well.  Hence each surviving \(\zeta_J\) belongs to the
ordinary local \(H^1\) space across the block.  Interior smoothing across
the finitely many blocks then gives compactly supported smooth approximants, proving
\(\zeta\in\mathcal V_\sigma\).
Moreover, \conditionref{D1} and the finite labeled surface limits give
\begin{equation}
  W_I\zeta_I=O(r)\quad(r\to0^+),
  \qquad
  W_I\zeta_I=O((R_I-r)^{\alpha_I})
  \quad(r\to R_I^-).
  \label{en:eq-classical-mode-weight-orders}
\end{equation}
These endpoint orders imply
\(\lambda\mathcal W_\sigma\zeta\in\mathcal Y_\sigma\subset L^1\) and
\(\lambda\zeta\in\mathcal H_\sigma\).

For \(v\in C_{c,\sigma}^{\infty}\), split the form integrals over the active
cells.  Integration by parts in each cell gives
\begin{equation}
  \mathfrak a_\sigma[v,\zeta]
  =\sum_{a=1}^q\sum_{I\in\mathcal K_a}
    \int_{\mathcal I_a}\overline{v_I}
    (\mathcal L_\sigma\zeta)_I\,\dd r
    +\mathcal J_\sigma^{\mathrm{bd}}[v,\zeta].
  \label{en:eq-cellwise-Green-identity}
\end{equation}
\begin{equation}
\begin{aligned}
  \mathcal J_\sigma^{\mathrm{bd}}[v,\zeta]
  ={}&-\sum_I\lim_{r\to0^+}\overline{v_I}\mathcal F_I^{\mathrm{con}}[\zeta]
    +\sum_{a=1}^q\sum_{I\in\mathcal B_a}
      \lim_{r\to\mathcal R_a^-}\overline{v_I}\mathcal F_I^{\mathrm{con}}[\zeta]
  \\
  &\quad-\sum_{a=1}^{q-1}\sum_{J\in\mathcal A_a^+}
    \overline{v_J(\mathcal R_a)}
    [\mathcal F_J^{\mathrm{con}}[\zeta]]_{\mathcal R_a}.
\end{aligned}
\label{en:eq-cellwise-Green-boundary-sum}
\end{equation}
The signs follow from the cell boundary term
\([\sum_{I\in\mathcal K_a}\overline{v_I}\mathcal F_I^{\mathrm{con}}[\zeta]]
_{\mathcal R_{a-1}}^{\mathcal R_a}\).  The form conormal has the exact
recovery identity
\begin{align}
  \mathcal F_I^{\mathrm{con}}[\zeta]
  &=\Xi\{b_I\zeta_I'+b_I(\Lambda'+\Phi')
       (\zeta_I-\bar\zeta)\}
  \notag\\
  &=-e^{\Lambda+2\Phi}
    (\delta p_I[\zeta]+p_I'\xi_I)
  \notag\\
  &=-e^{\Lambda+2\Phi}h_I
    (\delta H_I[\zeta]+H_I'\xi_I).
  \label{en:eq-conormal-lagrangian-pressure}
\end{align}
At the center, \conditionref{D1} gives \(\mathcal F_I^{\mathrm{con}}[\zeta]=O(1)\), while a
core test function vanishes there.  If \(I\in\mathcal B_a\),
\conditionref{D2} and \(h_I\to0\)
force its labeled surface term to vanish.  For a surviving
\(J\in\mathcal A_a^+\), the background multipliers have common one-sided
limits, and \conditionref{D3} gives
\begin{equation}
\begin{aligned}
  [\mathcal F_J^{\mathrm{con}}[\zeta]]_{\mathcal R_a}
  ={}&-e^{\Lambda+2\Phi}\big|_{\mathcal R_a}
  \left(
    [\delta p_J[\zeta]]_{\mathcal R_a}
    +p_J'(\mathcal R_a)
     \frac{e^{\Phi(\mathcal R_a)}}{\mathcal R_a^2}
     [\zeta_J]_{\mathcal R_a}
  \right)=0.
\end{aligned}
  \label{en:eq-conormal-jump-cancellation}
\end{equation}
Moreover,
\(\mathcal F_J^{\mathrm{con}}-\mathcal F_J^{\mathrm{div}}
=\zeta_J\sum_{I\ne J}f_{JI}\) has common one-sided limits: terms whose
second label ends are \(O(h_I)\to0\), and all surviving-label terms are
continuous.  Hence \([\mathcal F_J^{\mathrm{div}}]=[\mathcal F_J^{\mathrm{con}}]=0\).

Every term in \eqref{en:eq-cellwise-Green-boundary-sum} therefore vanishes.
Since
\(\mathcal L_\sigma\zeta=\lambda\mathcal W_\sigma\zeta\) cellwise,
\begin{equation}
  \mathfrak a_\sigma[v,\zeta]
  =\lambda\langle v,\zeta\rangle_{\mathcal H_\sigma}
  \qquad(v\in C_{c,\sigma}^{\infty}).
  \label{en:eq-classical-mode-form-identity}
\end{equation}
Density and form continuity extend this identity to every
\(v\in\mathcal V_\sigma\).  Since
\(\lambda\zeta\in\mathcal H_\sigma\), the first representation theorem
\cite{Kato} gives
\(\zeta\in\operatorname{Dom}(T_\sigma)\) and \(T_\sigma\zeta=\lambda\zeta\).  This proves
\eqref{en:eq-form-realization-classical-eigenspaces}; its \(\lambda=0\) case is
\eqref{en:eq-form-realization-classical-kernel}.
\end{proof}

\begin{subremark}[Eigenvectors in the weak operator domain]
\label{en:rem-closed-versus-classical-domain}
The operator domain is the weak domain associated with the closed form.  Its
eigenvectors are precisely the piecewise classical physical modes characterized
by \eqref{en:eq-form-realization-classical-eigenspaces}.
\end{subremark}

\subsection{Parameter dependence of the ordered spectrum}
\label{en:sec-spectral-parameter-dependence}

The eigenspace equality
\eqref{en:eq-form-realization-classical-eigenspaces} identifies the spectrum
of \(T_\sigma\), including multiplicities, with the squared frequency spectrum
of the piecewise classical physical radial modes satisfying
\eqref{en:eq-generalized-mode} and the domain conditions in
Definition~\ref{en:def-solution-classes}(IV).  Denote the ordered eigenvalues
by
\begin{equation}
  \omega_0^2(\sigma)\le\omega_1^2(\sigma)\le\cdots,
  \qquad \omega_n^2(\sigma)\longrightarrow+\infty.
  \label{en:eq-ordered-radial-spectrum}
\end{equation}
Eigenvalue dependence on varying Sturm--Liouville boundary data, including the
case of one singular endpoint, has been studied systematically
~\cite{kong_dependence_1996,zhu_dependence_2017}.
The labelwise pullback
\begin{equation}
 (\mathcal P_\sigma u)_I(s)=u_I(R_I(\sigma)s),
 \qquad 0<s<1,
 \label{en:eq-main-spectral-label-pullback}
\end{equation}
places the Hilbert products and closed forms on a common weighted pair
\(\mathcal V_*\hookrightarrow\mathcal H_*\).  Coefficient convergence controls
the common interiors, while the endpoint tail estimates control the moving
overlaps when labeled surfaces meet or exchange order.  This gives the compact
sequential convergence required by the min--max principle.

\begin{theorem}[Continuity of the ordered radial spectrum]
\label{en:thm-ordered-spectral-continuity}
Under Assumption~\ref{en:ass-regular}, for \(\sigma\in U\), every ordered
eigenvalue
\[
  \sigma\longmapsto\omega_n^2(\sigma),\qquad n=0,1,\ldots,
\]
is continuous.  On each smooth equilibrium sector satisfying
\eqref{en:eq-smooth-equilibrium-sector}, use a common fixed-node pullback for all
components, as in Lemma~\ref{en:lem-spectral-endpoint-structure}. In these
coordinates the Hilbert products are \(C^1\) in operator norm and the closed
forms are \(C^1\) in form norm.  For any \(\bar\sigma\) in the sector, if
\(\omega_n^2(\bar\sigma)\) is spectrally simple, then a neighborhood of
\(\bar\sigma\) carries a unique \(C^1\) eigenvalue branch through
\(\omega_n^2(\bar\sigma)\).
\end{theorem}

\phantomsection\label{en:app-proof-spectral-continuity}

\setproofparent{en:thm-ordered-spectral-continuity}

Using the model weights in \eqref{en:eq-model-spectral-weights}, define the
fixed reference spaces
\begin{align}
  \mathcal H_*
  &:=\bigoplus_{I=1}^K
  L^2((0,1),w_I^*(s)\,\dd s),
  \label{en:eq-fixed-spectral-Hilbert-space}\\
  \mathcal V_*
  &:=\overline{\bigoplus_I C_c^\infty(0,1)}^{\,\|\cdot\|_{\mathcal V_*}},
  \notag\\
  \|v\|_{\mathcal V_*}^2
  &:=\sum_I\int_0^1
  \left[p_I^*(s)|v_I'|^2+w_I^*(s)|v_I|^2\right]\dd s .
  \label{en:eq-fixed-spectral-form-space}
\end{align}
Use the labelwise pullback defined in
\eqref{en:eq-main-spectral-label-pullback}.  The pulled-back coefficients
\(\widehat W_{I,\sigma}\) and \(\widehat P_{I,\sigma}\) are those defined in
\eqref{en:eq-pulled-back-spectral-coefficients}.  Transport the Hilbert product
and the closed form by
\begin{equation}
\begin{aligned}
  \mathfrak b_\sigma[v,w]
  &:=\sum_I\int_0^1\widehat W_{I,\sigma}
    \overline{v_I}w_I\,\dd s,\\
  \widehat{\mathfrak a}_\sigma[v,w]
  &:=\mathfrak a_\sigma[\mathcal P_\sigma^{-1}v,
                         \mathcal P_\sigma^{-1}w].
\end{aligned}
  \label{en:eq-transported-spectral-forms}
\end{equation}
The center and surface expansions give, locally uniformly in \(\sigma\),
\begin{equation}
  \widehat W_{I,\sigma}\asymp w_I^*,
  \qquad
  \widehat P_{I,\sigma}\asymp p_I^*.
  \label{en:eq-common-weight-equivalence}
\end{equation}
The equivalence \eqref{en:eq-common-weight-equivalence} identifies
\(\mathcal P_\sigma\mathcal V_\sigma\) with \(\mathcal V_*\).  Fix
\(\bar\sigma\in U\) and restrict \(\sigma\) to a sufficiently small
relatively compact neighborhood \(V\Subset U\) of \(\bar\sigma\).  There are constants
\(c_*>0\) and \(0<C_1\le C_2<\infty\) such that
\begin{equation}
  C_1\|v\|_{\mathcal V_*}^2
  \le \widehat{\mathfrak a}_\sigma[v]
      +c_*\mathfrak b_\sigma[v]
  \le C_2\|v\|_{\mathcal V_*}^2.
  \label{en:eq-common-form-coercivity}
\end{equation}

If
\(z_j\rightharpoonup z\) in \(H^1(K)\) on a compact interior interval and
\(\kappa_j\to\kappa>0\), then on every compact \(J\) for which the dilated
intervals stay in \(K\),
\begin{equation}
 z_j(\kappa_j\,\cdot)\to z(\kappa\,\cdot)\quad\hbox{in }L^2(J),
 \qquad
 \kappa_jz_j'(\kappa_j\,\cdot)\rightharpoonup
 \kappa z'(\kappa\,\cdot)\quad\hbox{in }L^2(J).
\end{equation}
This follows from one-dimensional Rellich compactness and a change of
variables in the weak derivative pairing~\cite{BrezisFA}.

\begin{sublemma}[Moving-overlap estimate]
\label{en:lem-moving-overlap-estimate}
Fix \(\bar\sigma\in U\).  For two sufficiently nearby parameters
\(\sigma,\tau\), let \(\mathfrak r_{IJ,\delta}^{\sigma,\tau}\) be the signed
contribution of the two ordered \(\{I,J\}\) lower-order brackets in
\eqref{en:eq-full-spectral-form-expanded} to
\(\widehat{\mathfrak a}_\sigma-\widehat{\mathfrak a}_\tau\), restricted to
the part of the symmetric difference of the transported \(I\)--\(J\) overlap
intervals lying in surface collars of width \(\delta\).
There are \(C,\delta_0>0\), independent of the pair and of nearby
\(\sigma,\tau\), such that
\begin{equation}
  |\mathfrak r_{IJ,\delta}^{\sigma,\tau}[v,w]|
  \le C\delta^{1/2}
       \|v\|_{\mathcal V_*}\|w\|_{\mathcal V_*},
  \qquad 0<\delta<\delta_0.
  \label{en:eq-moving-overlap-tail}
\end{equation}
The same constant works when several labeled surfaces enter one collar.
\end{sublemma}

\begin{proof}
Undo the labelwise pullbacks on each overlap contribution.  Their Jacobians
and radii are uniformly bounded above and below.  The surface pointwise and
tail bounds
\eqref{en:eq-surface-pointwise-form-bound}--
\eqref{en:eq-surface-form-tail}, ordinary \(H^1\) control for a uniformly
surviving label, and \eqref{en:eq-common-form-coercivity} give
\begin{equation}
  \int_E W_K|z_K|^2\,\dd r
  \le C\delta\|z\|_{\mathcal V_*}^2,
  \qquad
  \int_E P_K|z_K'|^2\,\dd r
  \le C\|z\|_{\mathcal V_*}^2
  \label{en:eq-moving-overlap-collar-norms}
\end{equation}
on every such collar piece \(E\), for either an ending or a surviving label
\(K\). Indeed, near its own surface,
\eqref{en:eq-surface-pointwise-form-bound} and \(W_K\asymp x_K^{\alpha_K}\)
give \(W_K|z_K|^2\le C\|z\|_{\mathcal V_*}^2\) pointwise. Hence integrating
over any interval of length \(O(\delta)\) costs \(O(\delta)\) in weighted
mass, independently of the distance to the other surface. A uniformly
bounded form control ratio and Cauchy--Schwarz then give
\(O(\delta^{1/2})\) for each first-order term. The refined exponents are
recorded in
Section~\ref{sup:sec-moving-overlap} and
Table~\ref{sup:tab-overlap-exponents}.  Every resulting power of \(\delta\)
is at least \(1/2\).  Summation over the finite label set proves
\eqref{en:eq-moving-overlap-tail}, including simultaneous surface crossings.
\end{proof}

\begin{proof}[Proof of Theorem~\ref{en:thm-ordered-spectral-continuity}]
Let \(\sigma_j\to\sigma\).  In the coordinates
\eqref{en:eq-main-spectral-label-pullback}, the normalized principal and Hilbert
coefficients converge on \([0,1]\) in their weighted \(L^\infty\) norms by
Lemma~\ref{en:lem-spectral-endpoint-structure} and the \(C^1\) equilibrium
dependence in Lemma~\ref{en:lem-equilibrium-package}.  These convergence
statements imply that
\(\mathfrak b_{\sigma_j}\to\mathfrak b_\sigma\) in operator norm on
\(\mathcal H_*\), while the principal parts converge on \(\mathcal V_*\).
For a cross term written in the \(I\)-coordinate, the other component occurs
as
\(v_J((R_I/R_J)s)\) on
\(0<s<\min\{1,R_J/R_I\}\), with the corresponding factor \(R_I/R_J\) when
its derivative occurs.  On compact subintervals away from the center and the
moving endpoint, the moving-coordinate fact above gives strong convergence
for every undifferentiated component and weak convergence for every
differentiated component.  Each first-order cross term therefore converges by
a weak--strong pairing, and each multiplication term by strong--strong
pairing.  The two endpoint collars are treated uniformly below.

The parameter-dependent overlap endpoints occur in the two-label lower-order
terms.  When two radii meet or exchange order,
Lemma~\ref{en:lem-moving-overlap-estimate} controls the signed contribution
on each symmetric difference by the uniform bound
\eqref{en:eq-moving-overlap-tail}, including simultaneous crossings.
The Hilbert mass in these collars is uniformly negligible on form-bounded
sets.  Indeed, \eqref{en:eq-surface-form-tail} applies to every ending label,
while an ordinary one-dimensional Sobolev estimate
\cite{BrezisFA} applies to every surviving label.  At an internal surface
\(R\), these two estimates give
\begin{equation}
 \sum_{I\in\mathcal B_R}\int_{R-\delta}^{R}W_I|u_I|^2\,\dd r
 +\sum_{J\in\mathcal A_R^+}\int_{R-\delta}^{\min(R+\delta,R_J)}W_J|u_J|^2\,\dd r
 \le C\delta\left(\mathfrak a_{0,\sigma}[u]
                  +\|u\|_{\mathcal H_\sigma}^2\right).
 \label{en:eq-uniform-internal-collar-mass}
\end{equation}
When another labeled endpoint approaches the collar, its pointwise
weighted-mass bound from \eqref{en:eq-surface-pointwise-form-bound}
keeps the integral \(O(\delta)\) for every ratio of endpoint separation to
collar width. Integrals are restricted to each label's active support.  At
an outer surface, the corresponding ending-label estimate gives the same
bound. These estimates control each lower-order term on the entire collar,
including the part common to the two overlaps. Indeed, on a collar piece \(E\)
of length \(O(\delta)\), for either set of coefficients,
\[
 \left|\int_E f_{IJ}\overline{v_I'}w_J\,\dd r\right|
 \le C\left(\int_E P_I|v_I'|^2\,\dd r\right)^{1/2}
       \left(\int_E W_J|w_J|^2\,\dd r\right)^{1/2}
 \le C\delta^{1/2}\|v\|_{\mathcal V_*}\|w\|_{\mathcal V_*}.
\]
Here the arguments on \(E\) are pulled back to physical radius for the
parameter in question. The diagonal first-order terms obey the same bound
by \eqref{en:eq-f-diagonal-bound}. The multiplication terms gain
\(O(\delta)\) from their two mass factors. The principal derivative energy
need not be small on these collars and is retained on the whole interval.

There is also a uniform center collar estimate.  Let
\(\mathfrak r_{0,\delta}\) denote all lower-order terms of the form supported
in \(0<r<\delta\).  The center orders give
\begin{equation}
 \frac{f_{IJ}^2}{P_IW_J}=O(r^2),\qquad
 \frac{f_{IJ}^2}{P_IW_I}=O(r^2),\qquad
 \frac{|S_{IJ}|}{(W_IW_J)^{1/2}}=O(1),\qquad
 \frac{|\widetilde Q_I|}{W_I}=O(1).
\end{equation}
Together with \eqref{en:eq-center-form-tail}, Cauchy--Schwarz gives, uniformly
for nearby \(\sigma\),
\begin{equation}
 |\mathfrak r_{0,\delta}[v,w]|
 \le \varepsilon_0(\delta)
      \|v\|_{\mathcal V_*}\|w\|_{\mathcal V_*},
 \qquad \varepsilon_0(\delta)\longrightarrow0.
 \label{en:eq-center-lower-order-tail}
\end{equation}
For example, a first-order term gains one factor \(O(\delta)\) from the
normalized \(f\)-ratio and another from the square root of the Hilbert tail;
the multiplication terms use two Hilbert tails.

By \emph{compact Mosco convergence} we mean the Mosco recovery and lower
conditions together with precompactness in \(\mathcal H_*\) of sequences with
uniformly bounded shifted energy.  The recovery and lower conditions are those of Mosco~\cite{Mosco1969}. Adding compactness gives the notion used by Kuwae and Shioya~\cite{KuwaeShioya2003}. With this definition, the shifted forms satisfy
\begin{equation}
  \mathfrak a_j^{(c_*)}:=\widehat{\mathfrak a}_{\sigma_j}
     +c_*\mathfrak b_{\sigma_j}
  \quad\xrightarrow{\ \mathrm{compact\ Mosco}\ }\quad
  \mathfrak a^{(c_*)}:=\widehat{\mathfrak a}_{\sigma}
     +c_*\mathfrak b_{\sigma}
\end{equation}
The convergence holds on the common pair
\(\mathcal V_*\hookrightarrow\mathcal H_*\).

For recovery take \(v_j=v\) for every \(v\in\mathcal V_*\). The normalized
principal weights converge uniformly relative to \(p_I^*\), so the principal
energies converge on the full intervals. On common interiors apply the
moving-coordinate result to the constant sequence, and then use the
lower-order estimates above as the collar widths tend to zero. The mass
forms converge in norm. Thus
\(\mathfrak a_j^{(c_*)}[v]\to\mathfrak a^{(c_*)}[v]\).
The same conclusion for mixed pairs follows by polarization.

For the lower condition take a subsequence realizing the finite lower limit
and suppose \(v_j\rightharpoonup v\) in \(\mathcal H_*\). Coercivity bounds
this subsequence in \(\mathcal V_*\). After extraction it converges weakly
there to \(v\) and, by the compact embedding, strongly in \(\mathcal H_*\).
Uniform convergence of the normalized principal weights lets us replace the
varying principal energy by the fixed one with error \(o(1)\);
the latter is weakly lower semicontinuous on the full form domain.
On the common interiors the lower-order terms converge by weak--strong
pairing, and on the complete center and surface collars their absolute
contributions tend uniformly to zero with the collar widths. Hence
\(\mathfrak a^{(c_*)}[v]\le \liminf_j\mathfrak a_j^{(c_*)}[v_j]\).
The compact embedding supplies the required precompactness. Because
\(\mathfrak b_{\sigma_j}\to\mathfrak b_\sigma\) in operator norm, the argument
simultaneously tracks the varying mass form.

The min--max principle for lower-semibounded self-adjoint operators
~\cite{ReedSimonIV} then gives
\begin{equation}
  \omega_n^2(\sigma)
  =\min_{\substack{L\subset\mathcal V_*\\\dim L=n+1}}
    \ \max_{0\ne v\in L}
    \frac{\widehat{\mathfrak a}_\sigma[v]}
         {\mathfrak b_\sigma[v]}.
  \label{en:eq-common-minmax}
\end{equation}
Let \(L_n\) be the span of the first \(n+1\)
\(\mathfrak b_\sigma\)-orthonormal eigenvectors at \(\sigma\).  By the fixed-vector recovery and polarization, the mass and energy matrices
on a basis of \(L_n\) converge entrywise. The min--max principle then gives
\[
  \limsup_{j\to\infty}\omega_n^2(\sigma_j)
  \le
  \max_{0\ne v\in L_n}
  \frac{\widehat{\mathfrak a}_\sigma[v]}{\mathfrak b_\sigma[v]}
  =\omega_n^2(\sigma).
\]
For the reverse inequality, pass to a subsequence realizing
\(\liminf_j\omega_n^2(\sigma_j)\) and choose the first \(n+1\)
\(\mathfrak b_{\sigma_j}\)-orthonormal eigenvectors.  Coercivity and compact
Mosco convergence give, after further extraction, \(n+1\)
\(\mathfrak b_\sigma\)-orthonormal limits; let \(L_\infty\) denote their span.
The Mosco lower bound for the corresponding linear combinations and the
min--max principle yield
\[
  \omega_n^2(\sigma)
  \le
  \max_{0\ne v\in L_\infty}
  \frac{\widehat{\mathfrak a}_\sigma[v]}{\mathfrak b_\sigma[v]}
  \le
  \liminf_{j\to\infty}\omega_n^2(\sigma_j).
\]
Together, the \(\limsup\) and \(\liminf\) inequalities prove
\(\omega_n^2(\sigma_j)\to\omega_n^2(\sigma)\) for every \(n\).

Finally restrict to a smooth equilibrium sector \(\mathcal U\).  By
\eqref{en:eq-smooth-equilibrium-sector}, the equality and order pattern of all
labeled radii is fixed.  Choose breakpoints
\(0=s_0<s_1<\cdots<s_q=1\) and use a common continuous, cellwise \(C^2\)
fixed-node diffeomorphism \(r=\Psi_\sigma(s)\) that maps \(s_a\) to the corresponding
distinct radius \(\mathcal R_a(\sigma)\).  The support of every component and
every two-label overlap is then a fixed union of cells, and the pullback has a
single fixed Hilbert/form pair.  Fixing \(\sigma_0\in\mathcal U\), define
\(\mathcal H_{\mathrm{str}}\) and \(\mathcal V_{\mathrm{str}}\) as the
fixed-node pullbacks of \(\mathcal H_{\sigma_0}\) and
\(\mathcal V_{\sigma_0}\), equipped with their transported norms; then
\(\mathcal V_{\mathrm{str}}\hookrightarrow\mathcal H_{\mathrm{str}}\).
Under this fixed-node pullback, we retain the symbols
\(\mathfrak b_\sigma\) and \(\widehat{\mathfrak a}_\sigma\) for the transported
mass and energy forms from \eqref{en:eq-transported-spectral-forms}; their
domains are now \(\mathcal H_{\mathrm{str}}\) and
\(\mathcal V_{\mathrm{str}}\), respectively.

Use fixed model weights with the center and surface powers in
\eqref{en:eq-derived-Fuchs-center}--
\eqref{en:eq-Fuchs-principal-expansions}.  The weighted coefficient topology
is the maximum of the normalized Hilbert and principal ratios and the four
lower-order ratios in \eqref{en:eq-form-control-ratios}; by the proof of
Lemma~\ref{en:lem-infinitesimal-form-bound}, it controls Hilbert-product
operator norm and form norm.  With the surface nodes fixed,
\begin{equation}
 D_v^{\mathrm{fn}}H_I=O(x),\qquad
 D_v^{\mathrm{fn}}h_I=O(x^{\alpha_I}),\qquad
 D_v^{\mathrm{fn}}b_I=O(x^{\alpha_I+1}).
\end{equation}
For the four ratios in \eqref{en:eq-form-control-ratios}, write
\(\varrho_\sigma\) for the corresponding finite coefficient vector after
the fixed-node pullback.  The parameter estimates in
Section~\ref{sup:sec-conormal-endpoint-substitutions} give, in local
coordinates on the sector and for every tangent \(v\),
\begin{equation}
 \left\|
  \frac{\varrho_{\sigma+tv}-\varrho_\sigma}{t}
  -D_v^{\mathrm{fn}}\varrho_\sigma
 \right\|_{(L^\infty)^4}\longrightarrow0.
 \label{en:eq-form-ratio-difference-quotients}
\end{equation}
The same statement holds for the four normalized Hilbert and principal
ratios.  Cauchy--Schwarz then gives
\begin{equation}
 |D_v \mathfrak b_\sigma[z,w]|
 \le C\|z\|_{\mathcal H_{\mathrm{str}}}\|w\|_{\mathcal H_{\mathrm{str}}},
 \qquad
 |D_v \widehat{\mathfrak a}_\sigma[z,w]|
 \le C\|z\|_{\mathcal V_{\mathrm{str}}}\|w\|_{\mathcal V_{\mathrm{str}}}.
\end{equation}
Hence the Hilbert products are \(C^1\) in operator norm and the closed forms
are \(C^1\) in form norm.  Fixed surface nodes make all support and overlap
intervals parameter independent.

After the uniform positive shift, let \(K_\sigma\) be the associated compact
solution operator and \(G_\sigma\) the positive Riesz operator of the mass
form.  Operator norm \(C^1\) dependence gives a \(C^1\) compact self-adjoint
family
\(\widetilde K_\sigma=G_\sigma^{1/2}K_\sigma G_\sigma^{-1/2}\).
The Riesz projection of a simple isolated eigenvalue is \(C^1\) and has rank
one, so its eigenvalue \(\mu_{\mathrm{op}}(\sigma)\), and hence
\(\omega^2=\mu_{\mathrm{op}}^{-1}-c_*\), is \(C^1\).  The functional calculus,
mass conjugation, and Riesz projection details are given in
Section~\ref{sup:sec-kato-riesz}.
\end{proof}

\subsection{The strictly stable set and its boundary}
\label{en:sec-boundary-proof}

\begin{nameddefinition}[Strict radial stability]
\phantomsection
\label{en:def-strict-stability}
An equilibrium is \emph{strictly radially stable} when the lowest squared
radial mode frequency is positive.  The strictly stable subset of the
equilibrium chart is therefore
\begin{equation}
  U_{\mathrm{st}}:=\{\sigma\in U:\omega_0^2(\sigma)>0\}.
  \label{en:eq-strictly-stable-set}
\end{equation}
Its boundary relative to \(U\) is denoted by
\(\partial_U U_{\mathrm{st}}\). 
\end{nameddefinition}

\begin{proposition}[Zero mode locus and the local stability boundary]
\label{en:prop-spectral-boundary}
Under Assumption~\ref{en:ass-regular}, the strictly stable set
\eqref{en:eq-strictly-stable-set} satisfies
\begin{equation}
  \partial_U U_{\mathrm{st}}
  \subseteq\{\omega_0^2=0\}
  \subseteq\{\rankop DN<K\},
  \label{en:eq-boundary-inclusions}
\end{equation}
At every equilibrium in \(U\),
\begin{equation}
  0\in\operatorname{spec}T_\sigma
  \quad\Longleftrightarrow\quad
  \rankop DN_\sigma<K.
  \label{en:eq-spectral-zero-rank-equivalence}
\end{equation}
Let \(\mathcal U\subset U\) be a smooth equilibrium sector satisfying
\eqref{en:eq-smooth-equilibrium-sector}.  If
\(\sigma_*\in\mathcal U\) satisfies \(\omega_0^2(\sigma_*)=0\) (a
\emph{fundamental zero}), this eigenvalue is spectrally simple, and its
tangential differential satisfies
\(d(\omega_0^2|_{\mathcal U})_{\sigma_*}\ne0\), then in a relative neighborhood
\(V\subset\mathcal U\) of \(\sigma_*\),
\begin{equation}
  V\cap\partial_{\mathcal U}(U_{\mathrm{st}}\cap\mathcal U)
  =V\cap\{\omega_0^2=0\}
  =V\cap\{\rankop DN<K\},
  \label{en:eq-local-stable-rank-equality}
\end{equation}
and this common set is a \(C^1\) hypersurface across which
\(\omega_0^2\) changes sign within \(\mathcal U\).
\end{proposition}

\begin{proof}
Let \(\sigma_*\in\partial_U U_{\mathrm{st}}\), and choose
\(\sigma_j\in U_{\mathrm{st}}\) with \(\sigma_j\to\sigma_*\).
Continuity gives \(\omega_0^2(\sigma_*)\ge0\), while openness of the positive
set gives \(\omega_0^2(\sigma_*)=0\).  This proves the first inclusion in
\eqref{en:eq-boundary-inclusions}.

Compact resolvent turns every spectral zero into an eigenfunction.  Combining
the physical kernel identity
\eqref{en:eq-form-realization-classical-kernel} with
Theorem~\ref{en:thm-main} gives
\eqref{en:eq-spectral-zero-rank-equivalence}, and hence the second inclusion
in \eqref{en:eq-boundary-inclusions}.

On \(\mathcal U\), Theorem~\ref{en:thm-ordered-spectral-continuity} gives a
\(C^1\) branch for the simple fundamental eigenvalue.  The implicit function
theorem makes its zero set a
relative \(C^1\) hypersurface across which \(\omega_0^2\) changes sign.
Simplicity and ordering give \(\omega_1^2(\sigma_*)>0\), and continuity yields
\(\omega_1^2\ge\delta>0\) on a smaller neighborhood.  The spectral zero set
therefore consists of the fundamental branch, and
\eqref{en:eq-spectral-zero-rank-equivalence} gives
\eqref{en:eq-local-stable-rank-equality}.
\end{proof}

\section{Endpoint and operator estimates}
\label{sup:sec-spectral-estimates}
The following estimates supply the coefficient bounds, endpoint
selection and parameter dependence used in Section~\ref{en:sec-spectral}.

\subsection{Conormal endpoint substitutions}
\label{sup:sec-conormal-endpoint-substitutions}

Fix a component endpoint \(R\), set \(x=R-r\), and let
\(I\in\mathcal B_R\).  The generated EoS and
\conditionref{H2} give
\begin{align}
 h_I&=\mu_{I,0}C_IH_I^{\alpha_I}(1+r_{h,I}),&
 b_I&=\frac{\mu_{I,0}C_I}{\alpha_I}
      H_I^{\alpha_I+1}(1+r_{b,I}),
 \label{sup:eq-conormal-hb}
\end{align}
where, for every fixed \(k\),
\begin{equation}
 |(H_I\partial_{H_I})^kr_{h,I}|
 +|(H_I\partial_{H_I})^kr_{b,I}|
 \le C_kH_I^{\beta_I},
 \qquad \beta_I=\min\{1,\eta_I\}.
 \label{sup:eq-conormal-hb-remainder}
\end{equation}
The hydrostatic equation \(H_I'=-\Phi'\) and endpoint transversality yield
\begin{equation}
 H_I=\kappa_Ix(1+\theta_I),\qquad
 |(x\partial_x)^j\theta_I|\le C_jx^{\eta_{\mathrm{sp}}},\qquad j=0,1,2,
 \label{sup:eq-conormal-Hx}
\end{equation}
for some \(\eta_{\mathrm{sp}}>0\).  Indeed, in a surface collar,
\begin{equation}
 m_{\mathrm{tot}}'=4\pi r^2\epsilon_{\mathrm{tot}},\qquad
 \Phi'=\frac{m_{\mathrm{tot}}+4\pi r^3p_{\mathrm{tot}}}{r^2A},\qquad
 A=1-\frac{2m_{\mathrm{tot}}}{r}.
 \label{sup:eq-background-algebraic-system}
\end{equation}
Every ending contribution is a positive power of \(x\) times a conormal factor.
Contributions of surviving labels are \(C^1\) up to the surface, while \(R>0\)
and \(A>0\) keep the algebraic denominators regular.  One integration of
\(H_I'=-\Phi'\) proves \eqref{sup:eq-conormal-Hx}.  Differentiating
\eqref{sup:eq-background-algebraic-system} also controls \(\Phi''\) and
\(x\partial_x\Phi''\).  The differentiated matter term satisfies
\(x\partial_x\epsilon_I=O(x^{\alpha_I})\).

Here \(O_{\mathrm{con}}(x^{\eta_{\mathrm{sp}}})\) means that the term and each
of the finitely many \(x\partial_x\) derivatives used below are
\(O(x^{\eta_{\mathrm{sp}}})\).  Substitution gives
\begin{equation}
 h_I=h_{I,0}x^{\alpha_I}(1+O_{\mathrm{con}}(x^{\eta_{\mathrm{sp}}})),\qquad
 b_I=b_{I,0}x^{\alpha_I+1}(1+O_{\mathrm{con}}(x^{\eta_{\mathrm{sp}}})),
 \label{sup:eq-hb-x}
\end{equation}
 with positive leading constants.  The definitions of \(P_I\) and \(W_I\)
then give \(P_I\asymp x^{\alpha_I+1}\) and
\(W_I\asymp x^{\alpha_I}\).  Substitution
into the coefficient formulas of Section~\ref{en:app-spectral-realization} gives, when
\(I\) ends and \(J\) survives,
\begin{equation}
 \widetilde Q_I=x^{\alpha_I}O_{\mathrm{con}}(1),\quad
 S_{IJ}=x^{\alpha_I}O_{\mathrm{con}}(1),\quad
 f_{IJ}=x^{\alpha_I+1}O_{\mathrm{con}}(1),\quad
 f_{JI}=x^{\alpha_I}O_{\mathrm{con}}(1),
 \label{sup:eq-ending-surviving-orders}
\end{equation}
whereas, if both labels end at \(R\),
\begin{equation}
 S_{IJ}=x^{\alpha_I+\alpha_J}O_{\mathrm{con}}(1),\qquad
 f_{IJ}=x^{\alpha_I+\alpha_J+1}O_{\mathrm{con}}(1).
 \label{sup:eq-two-ending-orders}
\end{equation}
These endpoint orders hold for arbitrary positive exponents
\(\alpha_I\) and \(\alpha_J\).

For parameter dependence, labelwise rescaling gives
\(H_I(R_Is;\sigma)=(1-s)k_I(s;\sigma)\) with \(\inf k_I>0\).
Lemma~\ref{en:lem-equilibrium-package} gives uniform continuity in
\(\sigma\) of the pulled-back enthalpy slopes, and hence of \(k_I\) by
integration from \(s=1\). The metric factors are continuous, so the four
normalized Hilbert and principal weights in
Lemma~\ref{en:lem-spectral-endpoint-structure} converge in \(L^\infty\)
on the full chart. Within a fixed radius order sector the fixed-node
Jacobian and the normalized weights also have continuous parameter
derivatives, by the same endpoint factorization and the following bounds.

Fix a smooth equilibrium sector and use the endpoint distance \(x\) in a
fixed-node collar.  For every sector tangent \(v\),
\begin{equation}
 D_v^{\mathrm{fn}}H_I=O(x),\qquad
 D_v^{\mathrm{fn}}h_I=O(x^{\alpha_I}),\qquad
 D_v^{\mathrm{fn}}b_I=O(x^{\alpha_I+1}).
 \label{sup:eq-parameter-hb-orders}
\end{equation}
The factorization \(\epsilon_{I,H}=h_I/c_{s,I}^2=O(x^{\alpha_I-1})\) gives
\begin{equation}
 D_v^{\mathrm{fn}}\epsilon_I
 =\epsilon_{I,H}D_v^{\mathrm{fn}}H_I=O(x^{\alpha_I}),
 \qquad
 D_v^{\mathrm{fn}}p_I
 =h_I D_v^{\mathrm{fn}}H_I=O(x^{\alpha_I+1}).
 \label{sup:eq-parameter-matter-orders}
\end{equation}
Moreover,
\begin{equation}
 D_v^{\mathrm{fn}}c_{s,I}^2
 =D_v^{\mathrm{fn}}\!\left(\frac{b_I}{h_I}\right)
 =\frac{h_I D_v^{\mathrm{fn}}b_I-b_I D_v^{\mathrm{fn}}h_I}{h_I^2}
 =O(x).
 \label{sup:eq-parameter-sound-speed-order}
\end{equation}

To control the diagonal coefficient, differentiate the algebraic identity
\(\chi=\Phi'=(m_{\mathrm{tot}}+4\pi r^3p_{\mathrm{tot}})/(r^2A)\), use
\(m_{\mathrm{tot}}'=4\pi r^2\epsilon_{\mathrm{tot}}\) and \(p_{\mathrm{tot}}'=-h_{\mathrm{tot}}\chi\), and obtain
\begin{equation}
 \Phi''
 =\frac{4\pi r^2(\epsilon_{\mathrm{tot}}+3p_{\mathrm{tot}}-rh_{\mathrm{tot}}\chi)}{r^2A}
 -\chi\frac{2r-2m_{\mathrm{tot}}-8\pi r^3\epsilon_{\mathrm{tot}}}{r^2A}.
 \label{sup:eq-Phi-second-algebraic}
\end{equation}
On a surface collar \(r\) and \(A\) are bounded away from zero.  Equations
\eqref{sup:eq-parameter-hb-orders}--
\eqref{sup:eq-parameter-sound-speed-order} therefore give a bounded,
continuously parameter-dependent \(D_v^{\mathrm{fn}}\Phi''\), with the differentiated matter
contribution controlled by \eqref{sup:eq-parameter-matter-orders}.  At the
center the parameter-dependent even expansions give the same conclusion.

Write the four normalized coefficients as
\begin{align}
 \varrho_{Q,I}
 &:=\frac{\widetilde Q_I}{W_I}
 =e^{2(\Phi-\Lambda)}
 \left[\begin{aligned}
  &\Phi''+\chi^2-\left(\Theta+\frac2r\right)\chi\\
  &+c_{s,I}^2\Theta(\Theta-2\Gamma h_I)
   +\Gamma h_I\left(\Gamma b_{\mathrm{tot}}-\frac1r\right)
 \end{aligned}\right],
 \label{sup:eq-parameter-RQ}\\
 \varrho_{S,IJ}
 &:=\frac{S_{IJ}}{(W_IW_J)^{1/2}}
 =e^{2(\Phi-\Lambda)}\Gamma(h_Ih_J)^{1/2}\notag\\
 &\quad\times
 \left[\Gamma b_{\mathrm{tot}}-\frac1r-(c_{s,I}^2+c_{s,J}^2)\Theta\right],
 \label{sup:eq-parameter-RS}\\
 \varrho_{f,IJ}^{(1)}
 &:=\frac{f_{IJ}}{(P_IW_J)^{1/2}}
 =4\pi r e^{\Lambda+\Phi}(b_Ih_J)^{1/2},
 \label{sup:eq-parameter-Rf-cross}\\
 \varrho_{f,IJ}^{(2)}
 &:=\frac{f_{IJ}}{(P_IW_I)^{1/2}}
 =4\pi r e^{\Lambda+\Phi}c_{s,I}h_J.
 \label{sup:eq-parameter-Rf-diagonal}
\end{align}
The conormal chain rule shows that
\(D_v^{\mathrm{fn}}(h_Ih_J)^{1/2}\) and
\(D_v^{\mathrm{fn}}(b_Ih_J)^{1/2}\) have the same endpoint
orders as the corresponding square roots.  Equation
\eqref{sup:eq-parameter-sound-speed-order} similarly gives
\(D_v^{\mathrm{fn}}c_{s,I}=O(x^{1/2})\) when \(I\) ends.  Together, these
estimates give continuous parameter derivatives of the four ratios in
\eqref{sup:eq-parameter-RQ}--\eqref{sup:eq-parameter-Rf-diagonal} in the
cellwise \(L^\infty\) norms.

In local coordinates on the sector, for a coefficient \(F\), set
\(\operatorname{DQ}_{t,v}F_\sigma=(F_{\sigma+tv}-F_\sigma)/t\).  Applying the
product, quotient, and square root difference quotient identities to these
four factorizations gives, on every fixed cell,
\begin{equation}
 \begin{aligned}
 &\|\operatorname{DQ}_{t,v}\varrho_{Q,I}
      -D_v^{\mathrm{fn}}\varrho_{Q,I}\|_\infty
 +\|\operatorname{DQ}_{t,v}\varrho_{S,IJ}
      -D_v^{\mathrm{fn}}\varrho_{S,IJ}\|_\infty\\
 &\quad
 +\|\operatorname{DQ}_{t,v}\varrho_{f,IJ}^{(1)}
       -D_v^{\mathrm{fn}}\varrho_{f,IJ}^{(1)}\|_\infty
 +\|\operatorname{DQ}_{t,v}\varrho_{f,IJ}^{(2)}
       -D_v^{\mathrm{fn}}\varrho_{f,IJ}^{(2)}\|_\infty
 \longrightarrow0.
 \end{aligned}
 \label{sup:eq-four-ratio-difference-quotients}
\end{equation}
The convergence is locally uniform in \(\sigma\) and in \(v\) on bounded
sets.  Compact interior cells use ordinary \(C^1\) dependence, while the
center and surface collars use the estimates above.  Since there are finitely
many cells and label pairs, \eqref{sup:eq-four-ratio-difference-quotients}
gives the sectorwise \(C^1\) coefficient estimate used in
Section~\ref{en:sec-spectral-parameter-dependence}.

\subsection{Coefficient order table}
\label{sup:sec-coefficient-order-table}

The preceding substitutions give
\begingroup
\small
\begin{equation}
\begin{array}{c|c|c|c}
\text{region}&\widetilde Q_I&f_{IJ}\ \text{(and }f_{JI}\text{)}&S_{IJ}\\ \hline
r\to0^+&O(r^{-2})&O(r^{-1})&O(r^{-2})\\
I\text{ ends},\ J\text{ survives}
&O(x^{\alpha_I})
&O(x^{\alpha_I+1})\ \text{(}O(x^{\alpha_I})\text{)}
&O(x^{\alpha_I})\\
I,J\text{ end together}
&O(x^{\alpha_I})
&O(x^{\alpha_I+\alpha_J+1})
&O(x^{\alpha_I+\alpha_J})
\end{array}
\label{sup:eq-coefficient-order-table}
\end{equation}
\endgroup
The last row remains valid for unequal exponents.  The normalized identities
\begin{align}
 \frac{\widetilde Q_I}{W_I}
 &=e^{2(\Phi-\Lambda)}
 \left[\begin{aligned}
 &\Phi''+\chi^2-\left(\Theta+\frac2r\right)\chi\\
 &+c_{s,I}^2\Theta(\Theta-2\Gamma h_I)
 +\Gamma h_I\left(\Gamma b_{\mathrm{tot}}-\frac1r\right)
 \end{aligned}\right],
 \label{sup:eq-Q-normalized}\\
 \frac{S_{IJ}}{(W_IW_J)^{1/2}}
 &=e^{2(\Phi-\Lambda)}\Gamma(h_Ih_J)^{1/2}
 \left[\Gamma b_{\mathrm{tot}}-\frac1r-(c_{s,I}^2+c_{s,J}^2)\Theta\right],
 \label{sup:eq-S-normalized}\\
 \frac{f_{IJ}^2}{P_IW_J}&\asymp r^2b_Ih_J,\qquad
 \frac{f_{IJ}^2}{P_IW_I}\asymp r^2c_{s,I}^2h_J^2
 \label{sup:eq-f-ratios}
\end{align}
prove the four uniform form coefficient bounds.  The contribution of
\(f_{IJ}'=O(x^{\alpha_J-1})\) is represented by the integrated form identity,
while these normalized estimates apply directly to \(f_{IJ}\).

\subsection{Volterra proof of finite-energy endpoint selection}
\label{sup:sec-fuchs-construction}

\begin{proof}[Detailed proof of
Lemma~\ref{en:lem-finite-energy-Fuchs-selection}]
Set \(t=-\log x\).  Then
\begin{equation}
 \dot Y=[B_{\mathrm{F}}+R_t(t)]Y+q_t(t),
 \qquad B_{\mathrm{F}}=-A_0,\qquad
 \|R_t(t)\|+\|q_t(t)\|
 \le Ce^{-\eta_{\mathrm{F}}t}.
 \label{sup:eq-t-system}
\end{equation}
Here \(R_t(t)=-R_{\mathrm{rem}}(e^{-t})\) and \(q_t(t)=-q_{\mathrm{F}}(e^{-t})\), since \(\dd/\dd t=-x\,\dd/\dd x\). The matrix \(B_{\mathrm{F}}\) is diagonalizable, with center space
\(\mathcal Z_0=\{(c,0):c\in\mathbb C^m\}\) and positive eigenvalues
\(\alpha_i\).  Repeated values are treated as full spectral blocks.  For
sufficiently large \(T\), the corresponding Volterra maps on \([T,\infty)\) are
contractions in the chosen weighted space
\cite{WalterODE}.  They give
a particular solution \(Y_0=(u_0,v_0)\) with
\(Y_0=O(e^{-\eta_{\mathrm{F},0}t})=o(1)\) for some
\(\eta_{\mathrm{F},0}>0\).  Hence \(Y_0\) satisfies
\eqref{en:eq-Fuchs-finite-energy}: the \(u\)-weight becomes
\(e^{-(\alpha_i+1)t}\) and the derivative weight is \(e^{-\alpha_it}\).

For each \(c\in\mathbb C^m\), the same construction gives
\(Y_c=(c,0)+o(1)\).  Subtracting \(Y_0\) yields the homogeneous solutions
\(Z_c=Y_c-Y_0=(c,0)+o(1)\), and a basis of \(\mathcal Z_0\) produces
\(m\) independent bounded center solutions.
For each distinct positive eigenvalue \(\beta\), let \(P_\beta\) be its full
spectral projection and let \(P_<,P_>\) denote the lower and higher spectral
projections.  For \(q\in\operatorname{Ran}P_\beta\), set
\(Z=e^{\beta t}(q+\eta)\).  Variation of constants uses forward integration
on \(P_<\) and terminal integration on \(P_\beta+P_>\).  By
\eqref{sup:eq-t-system} and the finite-dimensional spectral bounds, the
projected kernels in their respective integration directions are dominated,
for a uniform constant \(C_{\mathrm F}>0\), by
\(e_T(s):=C_{\mathrm F}e^{-\eta_{\mathrm F}s}\) on \(s\ge T\).  Hence
\(e_T\in L^1(T,\infty)\) and
\(\|e_T\|_{L^1(T,\infty)}\le
C_{\mathrm F}\eta_{\mathrm F}^{-1}e^{-\eta_{\mathrm F}T}\to0\).  The
Volterra map is therefore contractive for large \(T\), and \(\eta(t)\to0\).
A basis in each full
\(\operatorname{Ran}P_\beta\), together with the center solutions, is a
fundamental family.  Expansion in this family shows that every noncenter
homogeneous solution has
\begin{equation}
 e^{-\beta t}Z(t)\longrightarrow q_\beta\ne0,\qquad
 q_\beta\in\operatorname{Ran}P_\beta,\qquad \beta>0.
 \label{sup:eq-leading-positive-block}
\end{equation}
Repeated exponents are grouped into full spectral blocks, and the \(o(1)\)
remainder includes the regular conormal logarithmic corrections.

Write \(q_\beta=(q_u,q_v)\).  The eigenvalue equation gives
\(-q_v=\beta q_u\) and \(D_\alpha q_v=\beta q_v\), so \(q_v\ne0\) and is supported
on indices with \(\alpha_i=\beta\).  Choose an index \(i\) with \((q_v)_i\ne0\). For this index,
\eqref{sup:eq-leading-positive-block} shows that
\(e^{-\alpha_it}|v_i|^2\) is comparable to \(e^{\beta t}\), and the integral
diverges.  Finite energy removes all positive branches.  Subtracting the
finite-energy particular solution leaves only the center family, proving
\(u_i\to c_i\) and \(v_i\to0\).
\end{proof}

\subsection{Weak-to-Fuchs termwise integrability}
\label{sup:sec-weak-to-fuchs}

At a radius \(R\), put \(x=R-r\).  The form domain pointwise estimate is
\(|u_I(x)|\le C(1+x^{-\alpha_I/2})\) for an ending label.  For an ending row
\(I\), the coefficient factorizations give
\begin{equation}
\begin{array}{@{}c|cc@{}}
 &J\in\mathcal A_R^+&J\in\mathcal B_R\\ \hline
 f_{IJ}'u_I&O(x^{\alpha_I/2})&O(x^{\alpha_I/2+\alpha_J})\\
 S_{IJ}u_J&O(x^{\alpha_I})&O(x^{\alpha_I+\alpha_J/2})\\
 f_{JI}/\sqrt{P_J}&O(x^{\alpha_I})
 &O(x^{\alpha_I+\alpha_J/2+1/2})
\end{array}
\label{sup:eq-weak-ending-table}
\end{equation}
and the last row controls \(f_{JI}u_J'\) by Cauchy--Schwarz.  In a surviving
row \(J\), an ending label \(I\) gives
\begin{equation}
 f_{JI}'u_J=O(x^{\alpha_I-1}),\qquad
 S_{JI}u_I=O(x^{\alpha_I/2}),\qquad
 \frac{f_{IJ}^2}{P_I}=O(x^{\alpha_I+1}).
 \label{sup:eq-weak-surviving-orders}
\end{equation}
The last power is used in
\begin{equation}
 \int|f_{IJ}u_I'|
 \le\left(\int P_I|u_I'|^2\right)^{1/2}
      \left(\int\frac{f_{IJ}^2}{P_I}\right)^{1/2}.
\end{equation}
Every exponent in \eqref{sup:eq-weak-ending-table} and
\eqref{sup:eq-weak-surviving-orders} is greater than \(-1\) because
\(\alpha_I>0\), so the corresponding terms are integrable.  For two
surviving labels, the
derivative coefficients \(f_{JL}\) are bounded and \(C^1\) across the collar.
The potential coefficients \(\widetilde Q_J\) and \(S_{JL}\) may have only
bounded continuous-conormal regularity when another label ends with exponent
below one; the four uniform form coefficient bounds give exactly the control
needed here.  Therefore the divergence flux and form
conormal are in \(W^{1,1}\).  Surviving components have finite \(u_J,u_J'\)
limits.  Dividing an ending row by \(p_{I,0}x^{\alpha_I-1}\) gives the Fuchs
system derived in Section~\ref{en:sec-physical-endpoints}. The principal remainder is
\(O(x^{\eta_{\mathrm{sp}}})\), an
ending label contributes \(O(x^{\alpha_J+1})(u_J+v_J)\), and a surviving label
contributes \(O(x)(u_J+u_J')\).

\subsection{Moving-overlap exponent table}
\label{sup:sec-moving-overlap}

Let \(I\) end in a collar of width \(\delta\).  When \(J\) survives uniformly,
\begin{align}
 \frac{f_{IJ}^2}{P_IW_J}&=O(x_I^{\alpha_I+1}),&
 \frac{f_{JI}^2}{P_JW_I}&=O(x_I^{\alpha_I}),\\
 \frac{f_{IJ}^2}{P_IW_I}&=O(x_I),&
 \frac{f_{JI}^2}{P_JW_J}&=O(x_I^{2\alpha_I}),&
 \frac{|S_{IJ}|}{(W_IW_J)^{1/2}}&=O(x_I^{\alpha_I/2}).
 \label{sup:eq-overlap-one-end}
\end{align}
If both endpoints enter the collar, \(0<x_I,x_J\le C\delta\) and
\begin{equation}
 P_L\asymp x_L^{\alpha_L+1},\quad W_L\asymp x_L^{\alpha_L},\quad
 |f_{IJ}|\le Cx_I^{\alpha_I+1}x_J^{\alpha_J},\quad
 |S_{IJ}|\le Cx_I^{\alpha_I}x_J^{\alpha_J},
 \label{sup:eq-overlap-two-end}
\end{equation}
with the exchanged bound for \(f_{JI}\).  Cauchy--Schwarz and the common
collar energy bounds give Table~\ref{sup:tab-overlap-exponents}.  An entry
\(q\) means a bound by
\(C\delta^q\|v\|_{\mathcal V_*}\|w\|_{\mathcal V_*}\).
\begin{table}[ht]
\caption{Powers in the moving-overlap estimate.}
\label{sup:tab-overlap-exponents}
\centering
\begin{tabular}{@{}lcc@{}}
\toprule
integrand & \(J\) survives & \(J\) enters the collar \\
\midrule
\(f_{IJ}\overline{v_I'}w_J\)
  & \(1+\alpha_I/2\) & \(1+(\alpha_I+\alpha_J)/2\) \\
\(f_{JI}\overline{v_I}w_J'\)
  & \((1+\alpha_I)/2\) & \(1+(\alpha_I+\alpha_J)/2\) \\
\(f_{IJ}(\overline{v_I'}w_I+\overline{v_I}w_I')\)
  & \(1\) & \(1+\alpha_J\) \\
\(f_{JI}(\overline{v_J'}w_J+\overline{v_J}w_J')\)
  & \(\alpha_I+1/2\) & \(1+\alpha_I\) \\
\(S_{IJ}\overline{v_I}w_J\)
  & \(1+\alpha_I/2\) & \(1+(\alpha_I+\alpha_J)/2\) \\
\bottomrule
\end{tabular}
\end{table}
When the other endpoint approaches but lies outside the collar,
the first column still gives upper bounds: \(h_J,b_J,c_{s,J}\) are
bounded, and the weighted mass of either label on the collar is
\(O(\delta)\) by \eqref{en:eq-moving-overlap-collar-norms}.
Every exponent is at least \(1/2\), and simultaneous crossings are finite sums
of the same estimates.

\subsection{Fixed-sector Kato--Riesz construction}
\label{sup:sec-kato-riesz}

Fix a smooth equilibrium sector and a reference inner product
\(\langle\cdot,\cdot\rangle_*\) on \(\mathcal H_{\mathrm{str}}\).  Let \(G_\sigma\) be
the positive Riesz operator of the mass form,
\begin{equation}
 \mathfrak b_\sigma[u,v]=\langle G_\sigma u,v\rangle_*.
 \label{sup:eq-mass-Riesz}
\end{equation}
Uniform equivalence implies that \(G_\sigma\) is boundedly invertible, with
spectrum in a common compact subset of \((0,\infty)\).  Operator norm \(C^1\) dependence of the
mass form gives a \(C^1\) family \(G_\sigma\), and functional calculus yields
the following formula, where \(\gamma_G\subset\mathbb C\setminus(-\infty,0]\)
is a fixed positively oriented contour enclosing that common spectral set:
\begin{equation}
 G_\sigma^{\pm1/2}
 =\frac{1}{2\pi i}\int_{\gamma_G}
 z^{\pm1/2}(z-G_\sigma)^{-1}\,\dd z.
 \label{sup:eq-mass-square-root}
\end{equation}
The resolvent identity proves \(C^1\) dependence of both square roots.

Define \(L_\sigma^{(c_*)}:\mathcal V_{\mathrm{str}}\to\mathcal V_{\mathrm{str}}^*\) by
\begin{equation}
 \langle L_\sigma^{(c_*)}u,z\rangle
 =\widehat{\mathfrak a}_\sigma[z,u]+c_*\mathfrak b_\sigma[z,u].
\end{equation}
By coercivity and the Lax--Milgram theorem, \(L_\sigma^{(c_*)}\) is
invertible.  With
\(\iota:\mathcal V_{\mathrm{str}}\hookrightarrow\mathcal H_{\mathrm{str}}\) and its
adjoint \(\iota^*\), set
\begin{equation}
 K_\sigma=\iota(L_\sigma^{(c_*)})^{-1}\iota^*G_\sigma.
 \label{sup:eq-compact-solution-operator}
\end{equation}
The \(C^1\) form estimates and
\(D(L^{-1})=-L^{-1}(DL)L^{-1}\) show that
\(\sigma\mapsto K_\sigma\in\mathcal K(\mathcal H_{\mathrm{str}})\) is \(C^1\) in the
operator norm.  The variational identity is
\(G_\sigma K_\sigma=K_\sigma^*G_\sigma\); hence
\begin{equation}
 \widetilde K_\sigma
 =G_\sigma^{1/2}K_\sigma G_\sigma^{-1/2}
 \label{sup:eq-self-adjoint-solution-operator}
\end{equation}
is a \(C^1\) family of compact self-adjoint operators on the fixed reference
Hilbert space.

Fix \(\bar\sigma\) in the sector.  Let
\(\mu_{\mathrm{op},0}=(\omega^2(\bar\sigma)+c_*)^{-1}\ne0\) be simple,
and let \(\gamma_{\mathrm{R}}\) enclose \(\mu_{\mathrm{op},0}\) but no other spectral
point.  Norm continuity keeps \(\gamma_{\mathrm{R}}\)
in the nearby resolvent sets, and
\begin{equation}
 \Pi_\sigma^{\mathrm{R}}=\frac{1}{2\pi i}\int_{\gamma_{\mathrm{R}}}
 (z-\widetilde K_\sigma)^{-1}\,\dd z
 \label{sup:eq-Riesz-projection}
\end{equation}
is \(C^1\).  Stability of an isolated finite spectral subspace gives
\(\rankop\Pi_\sigma^{\mathrm{R}}=\rankop\Pi_{\bar\sigma}^{\mathrm{R}}=1\)
\cite{Kato}.  For a unit
\(e\in\operatorname{Ran}\Pi_{\bar\sigma}^{\mathrm{R}}\),
\begin{equation}
 v_\sigma=\frac{\Pi_\sigma^{\mathrm{R}}e}
                 {\|\Pi_\sigma^{\mathrm{R}}e\|_*},\qquad
 \mu_{\mathrm{op}}(\sigma)
 =\langle v_\sigma,\widetilde K_\sigma v_\sigma\rangle_*
\end{equation}
are \(C^1\).  Since \(\mu_{\mathrm{op}}=(\omega^2+c_*)^{-1}\), the simple
branch \(\omega^2=\mu_{\mathrm{op}}^{-1}-c_*\) is \(C^1\).

\PrintSupplementBibliography
\end{document}